\documentclass[pdflatex,oneside,sn-mathphys-num]{sn-jnl}% Math and Physical Sciences Numbered Reference Style
\usepackage{graphicx}%
\usepackage{multirow}%
\usepackage{amsmath,amssymb,amsfonts}%
\usepackage{amsthm}%
\usepackage{mathrsfs}%
\usepackage[title]{appendix}%
\usepackage{xcolor}%
\usepackage{textcomp}%
\usepackage{manyfoot}%
\usepackage{booktabs}%
\usepackage{algorithm}%
\usepackage{algorithmicx}%
\usepackage{algpseudocode}%
\usepackage{listings}%
\theoremstyle{thmstyleone}%
\theoremstyle{thmstyletwo}%

\theoremstyle{thmstylethree}%

\newcommand{\rev}[1]{\textcolor{red}{#1}}
\newcommand{\model}{{{UFO-MGen}}}

\begin{document}

\title[Article Title]{Topology-Stratified Materials Discovery with A Flow-Based Generative Model}


\author[1]{\fnm{Jingyi} \sur{Zhou}}%\email{iauthor@gmail.com}
\author[1]{\fnm{Oyshee} \sur{ Chowdhury}}%\email{iiauthor@gmail.com}
\author[1]{\fnm{Noah} \sur{Oyeniran}}
\author*[1,2]{\fnm{Chongze} \sur{Hu}}\email{hucz@ua.edu}

\affil*[1]{\orgdiv{Department of Aerospace Engineering and Mechanics}, \orgname{University of Alabama}, \orgaddress{\city{Tuscaloosa}, \postcode{35487}, \state{Alabama}, \country{United States}}}

%\address[UA_AMI]{Alabama Materials Institute, The University of Alabama, Tuscaloosa, AL, 35487, USA}

\affil[2]{\orgdiv{Alabama Materials Institute}, \orgname{The University of Alabama}, \city{Tuscaloosa}, \postcode{35487}, \state{Alabama}, \country{United States}}

%\affil[3]{\orgdiv{Department}, \orgname{Organization}, \orgaddress{\street{Street}, \city{City}, \postcode{610101}, \state{State}, \country{Country}}}


\abstract{
Accurate generation of crystal structures is the foundation to the discovery of high-performance materials for extreme-environment applications, such as aerospace, additive manufacturing, and fusion energy systems.  
Although generative modeling has emerged as a promising approach for crystal design, its performance remains limited by the complex crystal structures and diverse chemical compositions.
In this work, we develop {\model}, a universal flow-based generative model that learns topological features of Wyckoff representations and leverages this information to accurately generate crystals across vast structural and chemical spaces.
Compared with state-of-the-art generative models, {\model} achieves the highest crystal generation success rate under a rigorous multi-stability evaluation framework, the highest SUN (stable, unique, novel) rate, and a remarkable extrapolation capability that has not been reported by previous models.
Furthermore, a fine-tuning module is implemented to {\model} for property-constrained crystal generation, enabling the inverse materials design toward target properties.
%
%\rev{The predicted materials are further validated by the quantum accuracy first-principles calculations}.
%By training the stratified Wyckoff manifolds using a flow-matching model, {\model} achieves the highest performance among the state-of-the-art generative models in generating physically stable crystals and also achieve a highest SUN rate of \rev{approximately 31\%}.
%
%but also enable accurate generation of complex systems with fine-tuned properties toward extreme limits. 
%
%First-principles calculations further validate the predictions of {\model} through a rigorous stability assessments, \rev{including thermodynamic, lattice-dynamically, and thermal stability tests}.  
%
The {\model} opens a new avenue for accelerated materials discovery and providing a foundation for universal materials intelligence.
%
%The abstract serves both as a general introduction to the topic and as a brief, non-technical summary of the main results and their implications. Authors are advised to check the author instructions for the journal they are submitting to for word limits and if structural elements like subheadings, citations, or equations are permitted.
}

\keywords{Crystal structure generation, generative models, Wychoff Representation, Topology, Inverse Materials Design}

\maketitle

\clearpage

\section{Introduction}\label{sec1}
As humanity moves toward the middle of 21$^\textrm{st}$ century, the development of high-performance, low-cost, and sustainable materials has become a top research priority across nearly all advanced technologies, including hypersonic vehicles~\cite{peters2024nc}, smart manufacturing~\cite{ren2022nature, zhu2022nm}, and fusion energy~\cite{zinkle2014annualreview, zinkle2009mt}.    
The rise of artificial intelligence (AI) and data-driven techniques has transformed materials discovery from a traditional trial-and-error Edison explorations into an era of intelligent design~\cite{butler2018nature, horton2025nm, griesemer2023ncs}. 
Although hundreds of thousands of materials are predicted by AI-driven framework each year, with many experimentally validated~\cite{cheng2026nm, lin2026afm, pyzer2022npjCM}, these design strategies still concentrate on expanding the compositional and structural spaces of existing material systems. 
The vast landscape of unexplored material systems still hold enormous potential for developing next-generation materials toward superior stability, properties, and performance~\cite{merchant2023scaling, shen2022relections}.

%The stability, property, and performance of any materials heavily rely on their internal skeleton, i.e., crystal structures \rev{[refs]}.
%
Accurate generation of previously unknown crystal structures is the foundation for discovering innovative materials with exceptional properties~\cite{zeni2025generative, okabe2026nm, cheng2026enhancing, park2026guiding, luo2024npjCM, luo2025crystalflow}.
However, generating unknown crystals remains notoriously challenging because of the enormous number of crystallographic configurations, diverse chemical environments, and stringent physical constrains (e.g., crystal symmetry) in real material systems~\cite{kelvinius2025wydiff, jiao2023crystal}.
%
%makes it difficult to represent three-dimensional (3D) crystal structures \rev{[refs]}. 
%
Although generative models have emerged as a powerful technique for navigating the highly irregular and constrained crystallographic space, their ability to generate physically reliable crystals remains very limited~\cite{metni2026generative}, particularly for complex crystal structures with large unit cells.
%
%, even under a relatively simple criteria based on thermodynamic stability. 
%
Moreover, existing crystal generative models have primarily demonstrated strong interpolation capability for the crystal structures sampled in the training data, while their ability to extrapolate beyond the learned crystallographic space remains limited. 
This poor extrapolation capability fundamentally restricts their potential to discover truly novel materials beyond the existing domain knowledge. 
%lity directly limits the generative models' performance in generating innovative materials beyond the domain knowledge.

In this work, we present {\model} (Universal Flow Omni-Materials Generation), a flow matching-based generative model capable of accurately generating previously unknown crystals across broad crystallographic and chemical spaces (Fig.~\ref{fig:Fig1}).
To rigorously evaluate the quality of generated structures, we further propose a physically rigorous multi-stability evaluation (MSE) framework based on thermodynamic, lattice-dynamic, and thermal stabilities (Fig.~\ref{fig:Fig2}).
Benchmark results show that {\model} outperforms all existing crystal generative models under the MSE criteria while also achieving the highest SUN (stable, unique, novel) rate (Fig.~\ref{fig:Fig3}). 
%
%Furthermore, the {\model} achieves a SUN score of 33\%, far exceeding the performance of all existing generative models, including MatterGen \rev{[ref]}, whose best reported value is 22\% (\rev{Fig. 2}), but also demonstrate its extraoridinary capability for structural generation compared to traditional generative model (\rev{Fig. 3}). 
%
More importantly, {\model} has demonstrated superior extrapolation capability that has never been reported in any prior generative models (Fig.~\ref{fig:Fig3}).
%
%, \rev{which can be ascribed to its capability in generating physically reliable crystal structures and transmitting \rev{Wyckoff information} between different crystals for extrapolation prediction (Fig.~\ref{fig:Fig4}).}
%
Compared with traditional diffusion models, {\model} achieves higher accuracy and faster convergence speed due to two unique mechanisms: ($i$) unified Wyckoff representation and ($ii$) flow-matching generation engine (Fig.~\ref{fig:Fig4}).
%
%First-principles density functional theory (DFT) calculations validate the prediction of {\model} based on a strict stability tests (\rev{Fig. 4}).
%
Finally, a fine-tuning module is also introduced to {\model} for property-constrained generation, enabling the inverse design of materials with exceptional properties for targeted engineering applications (Fig.~\ref{fig:Fig5}).

%%%%%%%%%%%% results and discussion %%%%%%%%%%%%%
\section{Results}\label{sec2}

\subsection{Unified Wyckoff representation}
%
%\rev{Current atom-by-atom method for crystal generation shows limitation of learning deeper physics.}
%
Generative crystal modeling requires a precise representation of crystal structures in a low-dimensional yet physically unified space~\cite{court2020JCIM}.
Among existing methods, the Wyckoff representation~\cite{hahn2005} is one of the most widely used, as shown in Fig.~\ref{fig:Fig1}(a).
It provides a compact description of periodic crystal through its space group ($G$), lattice parameters ($\ell$), the number of Wyckoff orbits ($K$), Wyckoff letters ($\omega_{1:K}$) and chemical species ($s_{1:K}$) for each $K$, and Wyckoff coordinates ($x^{orb}$)~\cite{kazeev2025wyckoff}, see Supplementary Section 1.1. 
%
%ignoring many key information in the latent space. 
%traditional generative modeling always concadent and train these parameters together, which is less efficient. 
%
%\rev{Based on the manifold hypothesis [ref], physically valid crystals on a low-dimensional manifold.} 
%
For generation purpose, all Wyckoff information must be incorporated within a unified generative space.
However, the Wyckoff representation includes both discrete features (e.g., $G$, $K$, $\omega$, $s_{1:K}$) and continuous components (e.g., $\ell$ and $x^{orb}$), which belong to distinct mathematical domains and therefore require different modeling strategies.
Unfortunately, existing models~\cite{kazeev2025wyckoff, chang2026space, levy2025symmcd} overlook this fundamental difference and instead directly combine these heterogeneous features within a single generative framework. 
As such, a physically consistent unification of discrete and continuous Wyckoff information is essential for accurate crystal generation.
% making it difficult to treat as a unified representation for generative modeling. 
%

The mathematical unification of discrete and continuous information can be traced to the concept of fiber bundles in topology~\cite{steenrod1951FB}.
Inspired by this theory, we adopt the analogous terms ``scaffold" ($c$) and ``fiber" ($\mathcal{F}$) to denote the discrete and continuous features of the Wyckoff representation, respectively (Supplementary Section 1.2).
Accordingly, a scaffold is expressed as $c$ = ($G, K, \omega_{1:K}, s_{1:K}$), which encodes the key structural information (e.g., symmetry) and thus dominates the complexity of crystal generation.
We introduce two parameters to quantify the dimensionality of a scaffold: the intrinsic dimension, $D_\textrm{rep}$ and the compression ratio, $\rho$.
Specifically, $D_\textrm{rep}$ defines a reduced-dimensional space that stratifies crystal structures according to their intrinsic structural complexity.
%
%$D_\textrm{rep}$, which is always smaller than the 3$N$+6 degrees of freedom (DOFs) of a periodic crystal containing $N$ atoms.
%e for learning complex physical relationships among crystal structures.
%
Since $D_\textrm{rep}$ is always smaller than the 3$N$+6 degrees of freedom (DOFs) of a periodic crystal containing $N$ atoms, $\rho=D_\textrm{rep}/(3N+6)$ is defined to quantify the dimensionality reduction relative to the original crystal representation.
%in a reduced $D_\textrm{rep}$-dimensional space.}
% can be compressed into a reduced space with a dimension of $D_\textrm{rep}$.
%
%\rev{Accordingly, the complete Wyckoff space is naturally stratified into $D_\textrm{rep}$-dimensional subspaces, each representing a class of structure}.
%
By calculating the scaffold parameters ($D_\textrm{rep}$ and $\rho$) for all 154,875 crystal structures in the Materials Project (MP) database~\cite{jain2013materialsprojec}, we identify 32,550 trainable structures spanning 35 distinct space groups.
Further analysis of these structures demonstrates that the proposed scaffold parameters effectively capture intrinsic complexity of diverse crystal structures in the MP database (Supplementary Section 1.3).

%Based on $D_\textrm{rep}$, we further define a compression ratio, $\rho=D_\textrm{rep}/(3N+6)$, to quantify the reduction achieved compared to the original crystal structures.
%
%Distribution analyses of $D_\textrm{rep}$ and $\rho$ across the all 154,875 crystal structures in the Materials Project database (MPD)~\cite{jain2013materialsprojec} demonstrate that these descriptors effectively capture the intrinsic complexity of crystal structures \rev{(Supplementary Section 1.3).}
% %
% \rev{We thus adopt this two parameters to reorganize the entire MP database into a trainable database for {\model}.}

Although scaffolds encode key structural information, continuous fibers also play critical roles in crystal generation, as they preserve the symmetry changes associated with special Wyckoff positions within a given $D_\textrm{rep}$. 
These special positions typically correspond to high-symmetry Wyckoff coordinates that reduce the DOFs of a periodic crystal.
However, traditional generative models learn these special features on a fixed-dimensional space along with other Wyckoff information~\cite{kazeev2025wyckoff, chang2026space, levy2025symmcd}, thus overlooking the key scaffold features associate with these special positions.
%resulting in many potentially stable structure being overlooked.
%
% These special positions are often overlooked or treated as noisy data in conventional generative models because they introduce discontinuities into the smooth crystal manifold. 
%
To address this limitation, we treat these special positions as shared boundaries between continuous fibers with distinct $D_\mathrm{rep}$, thereby providing a unified description of fibers across different scaffolds and enabling structural transitions throughout the entire stratified Wyckoff space (Supplementary Section 2.1).
By processing the fibers of all 32,550 trainable structures, we construct a new crystal database with a unified Wyckoff representation (Supplementary Section 2.2).

% thereby capturing smooth structural transitions across different dimensions.
%which capturing the cross-dimensional symmetric structural changes}.
%
% To address this, we process these special positions as the boundaries of continuous fibers to connect \rev{different dimensional}
% % their neighboring
% fibers.  
%
%This topology-inspired construction provides a unified description of fibers across different scaffold and thus enables transitions throughout the stratified Wyckoff space \rev{(See Supplementary Section 2).}
%
%In this work, although terms of scaffold and fibers are introduced here for convenience, they can be mathematically defined in a more rigorous way, which will be presented in future work.

\subsection{Hierarchical UFO-MGen}
\begin{figure}[h]
\centering
\includegraphics[width=1\textwidth]{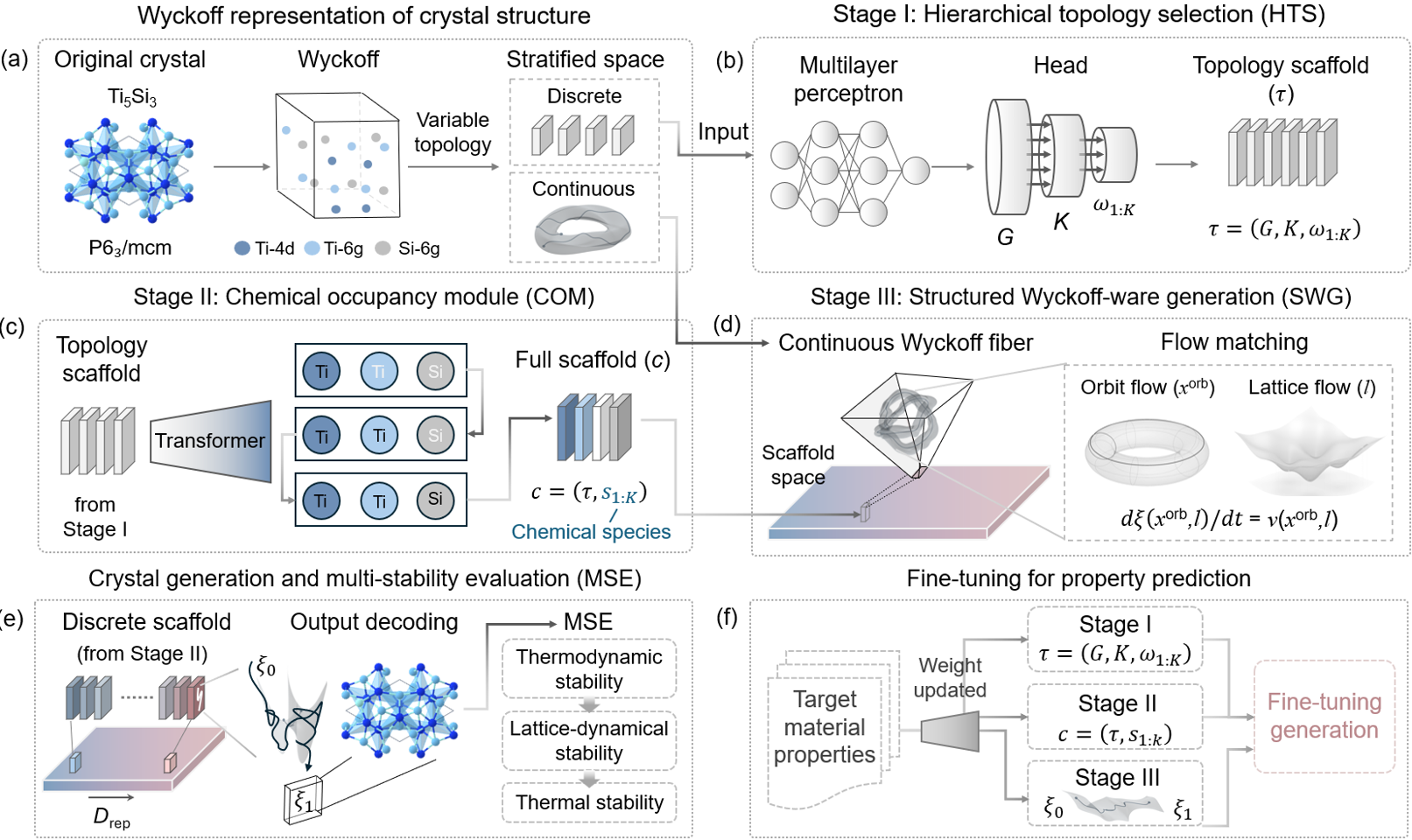}
\caption{\textbf
{Workflow of {\model} for generating accurate crystal structures and predicting their associated properties.}
\textbf{(a)} Wyckoff representation of crystal structure in latent space.
\textbf{(b)} Stage I: hierarchical topology section (HTS) for classifying crystal structures based on discrete information: space group ($G$), number of Wyckoff orbits ($K$), and Wyckoff letter sequence ($\omega_{1:K}$) using multilayer perceptron (MLP) network.
The topology scaffold ($\tau$) is a fused representation of $G, K$ and $\omega_{1:K}$.
\textbf{(c)} Stage II: chemical occupancy module (COM) for assigning chemical species ($s_{1:K}$) to each $\tau$ to form complete scaffold $c = (\tau,s_{1:K})$ using chemical constrains.
%
%Compression of classified data into the input for velocity field for {\model} model. 
%
\textbf{(d)} Stage III: structured Wyckoff-aware generation (SWG) for embedding the discrete scaffold and continuous fibers into hierarchical flow matching model to predict the velocity.
\textbf{(e)} Crystal structure generation by decoding the output and multi-stability evaluation (MSE).
\textbf{(f)} Fine-tuning of the {\model} model for property-constrained generation and inverse materials design.
}\label{fig:Fig1}
\end{figure}

%: a hierarchical flow model for Wyckoff fibers}
%
Building on the stratified feature of Wyckoff space, {\model} is designed as a hierarchical generative framework that separately models the discrete and continuous Wyckoff representation for crystal structure generation, as shown in Fig.~\ref{fig:Fig1}. 
The {\model} architecture includes the three sequential stages: Stage I, hierarchical topology selection (HTS); Stage II, chemical occupancy module (COM); and Stage III, structured Wyckoff-aware generation (SWG). 
A detailed description of {\model} is provided in Supplementary Section 3.

Stage I of {\model} begins by sampling a crystal scaffold that determines their Wyckoff topology (e.g., $G$, $K$, and $\omega_{1:K}$), and define a topology scaffold $\tau$ = ($G$, $K$, $\omega_{1:K}$) in Fig.~\ref{fig:Fig1}(b) (Supplementary Section 3.1). 
Next, Stage II determines the chemical occupancy of the predefined $\tau$ by assigning chemical species ($s_{1:K}$) to the Wyckoff orbit in $\tau$ under a set of chemical constrains (Supplementary Section 3.2).
This assignment constructs the full scaffold $c$ = ($G$, $K$, $\omega_{1:K}$, $s_{1:K}$), as illustrated in Fig.~\ref{fig:Fig1}(c).
Stage III uses flow matching~\cite{lipman2022flow} to generate the continuous Wyckoff information (\textit{i.e.}, $\ell$ and $x^{orb}$) within a predetermined scaffold ($c$) (Fig.~\ref{fig:Fig1}(d), Supplementary Section 3.3). 
Finally, the complete crystal structure is reconstructed by inserting these continuous variables into the corresponding Wyckoff templates and applying the associated space-group symmetry operations (Fig.~\ref{fig:Fig1}(e)). 

%\rev{Using these hiarchical architecture, {\model} is trained using the entire 154,875 crystal structures on the MPD~\cite{jain2013materialsprojec}. 
%
Using the unified database of all 32,550 trainable structures, {\model} is trained to generate previously unreported crystal structures. 
Detailed training objectives and sampling processes for each stage are provided in Supplementary Sections 3.4-3.6. 
The generated crystals are then subjected to rigorous stability evaluations to assess the performance of {\model}.

\subsection{Multi-stability evaluation}
Traditional evaluations of crystal stability primarily rely on the formation energy ($\Delta E_f$) or energy above the convex hull ($\Delta E_\textrm{hull}$) at zero K~\cite{ma2017PRB}.
However, $\Delta E_f$ and $\Delta E_{hull}$ only measure thermodynamic stability and do not capture other important physical stability, such as lattice-dynamic stability or finite-temperature thermal stability~\cite{oyeniran2026first, gu2021acs}.
%elevated temperatures. 
%
Therefore, a more comprehensive and rigorous stability evaluation is required to examine the stability of generated crystal structures.

\begin{figure}[h]
\centering
\includegraphics[width=1\textwidth]{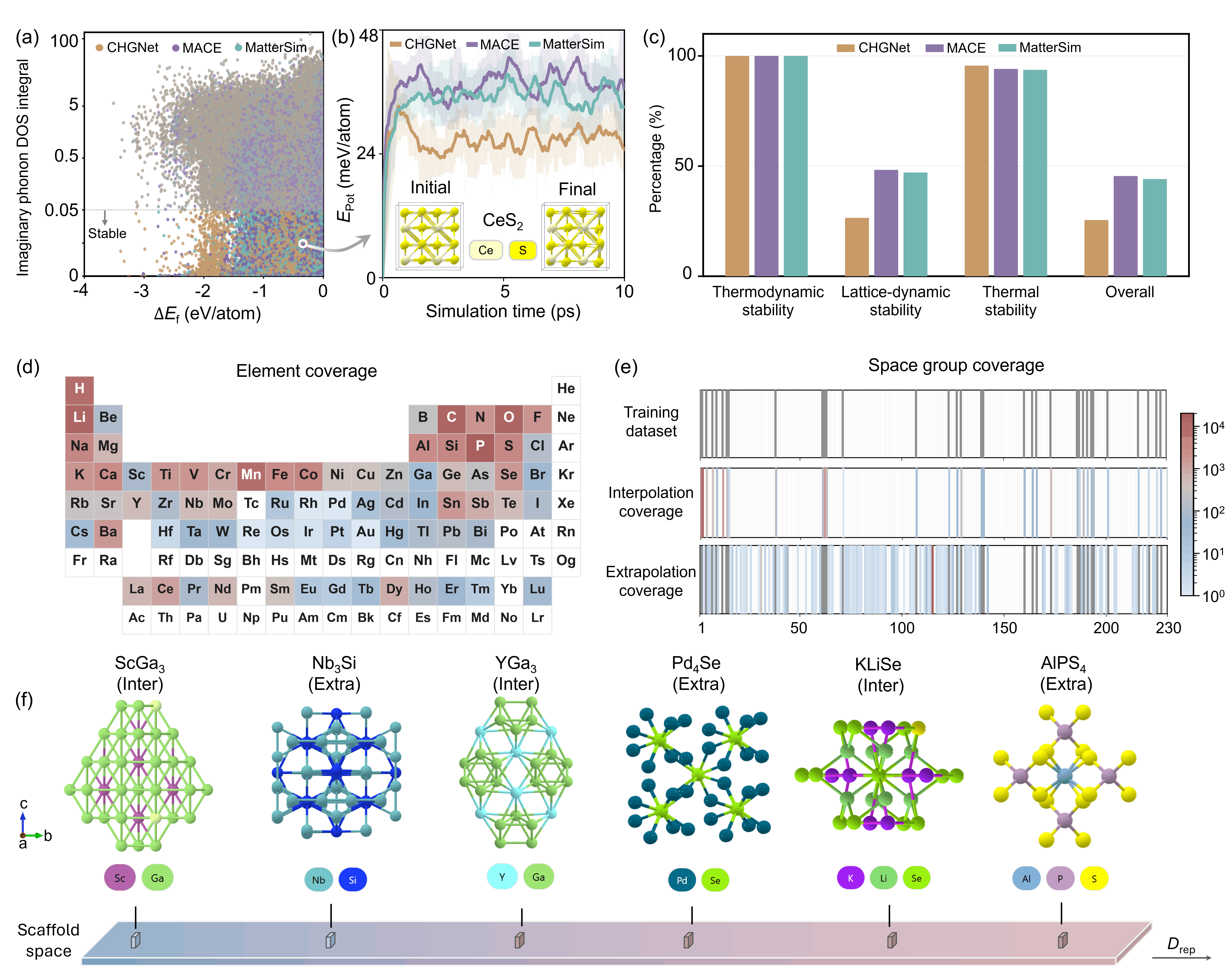}
\caption{\textbf{
Mutli-stability evaluation (MSE) and extrapolation capability of {\model}.}
\textbf{(a)} Distribution of generated crystal structure according to their thermodynamic and lattice-dynamical stabilities, screened using three universal machine learning interatomic potentials (MLIPs).
Thermodynamically stable structures are identified by negative formation energies ($\Delta E_\textrm{f}$ $<$ 0), and lattice-dynamically stable structures are identified by the absence of imaginary phonon frequencies, quantified by an integral of imaginary phonon density of states (DOS) below 0.05 phonon modes.
%density of states (DOS)  and formation energies ($\Delta E_\textrm{f}$).
%
\textbf{(b)} Thermal stability assessment of all stable structures that satisfy the first two stability criteria using MLIP-based molecular dynamics (MD) simulations.
Structures that maintain their initial configurations throughout the 10 ps NVT simulations at 300 K are classified as thermally stable structure, with CeS$_2$ shown as a representative example. 
\textbf{(c)} Success rates for each MSE criteria and the overall MSE success rate obtained using the three MLIPs.
\textbf{(d)} Distribution of chemical elements represented in the fully stable crystals predicted by CHGNet.
\textbf{(e)} Distribution of space group represented in the unified database, and in the interpolation and extrapolation predictions generated by {\model}.
\textbf{(f)} Representative generated crystals with different complexity dimension $D_\textrm{rep}$ values, all of which are validated by DFT calculations in Supplementary Section 4.1.
%
%Performance comparison between {\model} and other generative models based on \textbf{(e)} MSE criteria, \textbf{(f)} SUN, and \textbf{(g)} extrapolation ability. %99 
}
\label{fig:Fig2}
\end{figure}

In this work, we propose a multi-stability evaluation (MSE) framework based on three physically rigorous criteria: thermodynamic, lattice-dynamical, and thermal stability (Fig.~\ref{fig:Fig1}(e)). 
%{\model} unconditionally generated 50,000 crystal structures}
%
The thermodynamic stability is evaluated using the conventional criteria of $\Delta E_f$, where structures with  $\Delta E_f < 0$ are considered thermodynamically stable (Fig.~\ref{fig:Fig2}(a)).
The lattice-dynamical stability is assessed from the phonon spectrum and density of states (DOS), where the structures without significant imaginary frequencies are considered lattice-dynamically stable (Fig.~\ref{fig:Fig2}(a)). 
In this work, the structures with an integral of negative DOS below 0.05 phonon mode are considered as lattice-dynamically stable. 
Finally, thermal stability is further examined using molecular dynamics (MD) simulations at 300 K, where structures that retain their original structures over 10 ps canonical ensemble (NVT) simulations are considered thermally stable, see example of CeS$_2$ in Fig.~\ref{fig:Fig2}(b).
To ensure robust and consistent evaluations, MSE screening is performed using three universal machine-learning interatomic potentials (uMLIPs), including CHGNet~\cite{deng2023chgnet}, MACE~\cite{batatia2022mace}, and MatterSim~\cite{yang2024mattersim}.

Next, we use {\model} to unconditionally generate 50,000 crystal structures and then evaluate their stability using the MSE framework. 
The MSE screening results show that all structures satisfy the thermodynamic stability criterion ($\Delta E_f < 0$), while 13,326, 24,165, and 23,541 structures are identified as lattice-dynamically stable by CHGNet, MACE, and MatterSim, respectively (Fig.~\ref{fig:Fig2}(a)).
Subsequent MD simulations at 300 K identify 12,744, 22,738, and 22,064 thermally stable structures using these three uMLIPs, corresponding to overall MSE rates of 25.5\%, 45.5\%, and 44.1\%. 
Chemical coverage analysis of these MSE-screened stable crystals reveal that they span nearly all elements across the periodic table, as shown in Fig.~\ref{fig:Fig2}(c). 
Furthermore, their space-group distribution covers 34 of the 35 space groups represented in the training dataset (Fig.~\ref{fig:Fig2}(d)). 
These results demonstrate that {\model} not only generate crystals with rigorous physical stability but also produces chemically and structurally diverse structures. 

\subsection{Extrapolation capability and validation}
The large number of 195 space groups absent from our unified dataset provides a unique opportunity to evaluate the extrapolation capability of {\model}.
%for predicting the crystal structures with unprecedented space groups from the training data. 
%
Since Stage I of {\model} fixes the topology scaffold, we bypass this stage and use the well-trained model to generate 33,439 crystal structures with previously unseen crystal scaffolds. 
Following MSE screening, 5,921 generated crystals are fully stable, and subsequent structural analysis shows that they span 107 of the 195 space groups absent from the training set (bottom panel in Fig.~\ref{fig:Fig2}(e)).
%
%This is more than three times of the 35 available space groups captured in the training data.
%
%which is threes times of the 35 space groups covered in the training set
%
To the best of our knowledge, such extrapolation capability has not been demonstrated by existing crystal generative models.
These results suggest that extrapolation performance should be regarded as a critical benchmark for evaluating the ``true intelligence'' of crystal generative models.

To validate the accuracy and physical stability of the generated crystals, first-principles density functional theory (DFT) calculations~\cite{kresse1996efficiency, kresse1993ab} were performed to independently assess the MSE criteria of these structures from both interpolation and extrapolation groups of {\model}. 
Figure~\ref{fig:Fig2}(f) summarizes six representative crystals, three from interpolation and three from extrapolation, with different $D_\mathrm{rep}$ values (\textit{i.e.}, scaffold complexity). 
DFT structural relaxations show that the optimized lattice parameters of these generated crystals are in excellent agreement with those predicted by {\model} (Supplementary Section 4.1). 
Moreover, DFT-calculated $\Delta E_f$, phonon spectra, and \textit{ab initio} MD-simulated energy profiles are all consistent with the MSE screening using uMLIPs (Supplementary Sections 4.1-4.3).
Additional DFT validations of other generated crystals also show excellent agreement with {\model} predictions (Supplementary Section 4.4).
The strong agreement with quantum-accuracy DFT calculations demonstrates the high accuracy and reliability of the {\model} predictions and further validates its ability to generate physically stable crystal structures.
%is capable of generating not only stable but also chemically and structurally diverse crystal structures.
%The candidates show a clear separation in the space defined by formation energy and the imaginary phonon DOS integral. Unstable structures are concentrated in regions with high formation energy or large imaginary-frequency contribution, whereas the candidates that pass the screening are located in the low-formation-energy and low-imaginary-DOS region. Here, formation energy provides an initial energetic filter for thermodynamic feasibility, while the imaginary phonon contribution of the relaxed structure reflects local dynamical stability. Combining these two quantities can distinguish structures that only look like crystals from candidates with potential physical feasibility more directly than geometric validity or distributional similarity alone.

\subsection{Performance comparison}
\begin{figure*}[t]
\centering
\includegraphics[width=\textwidth]{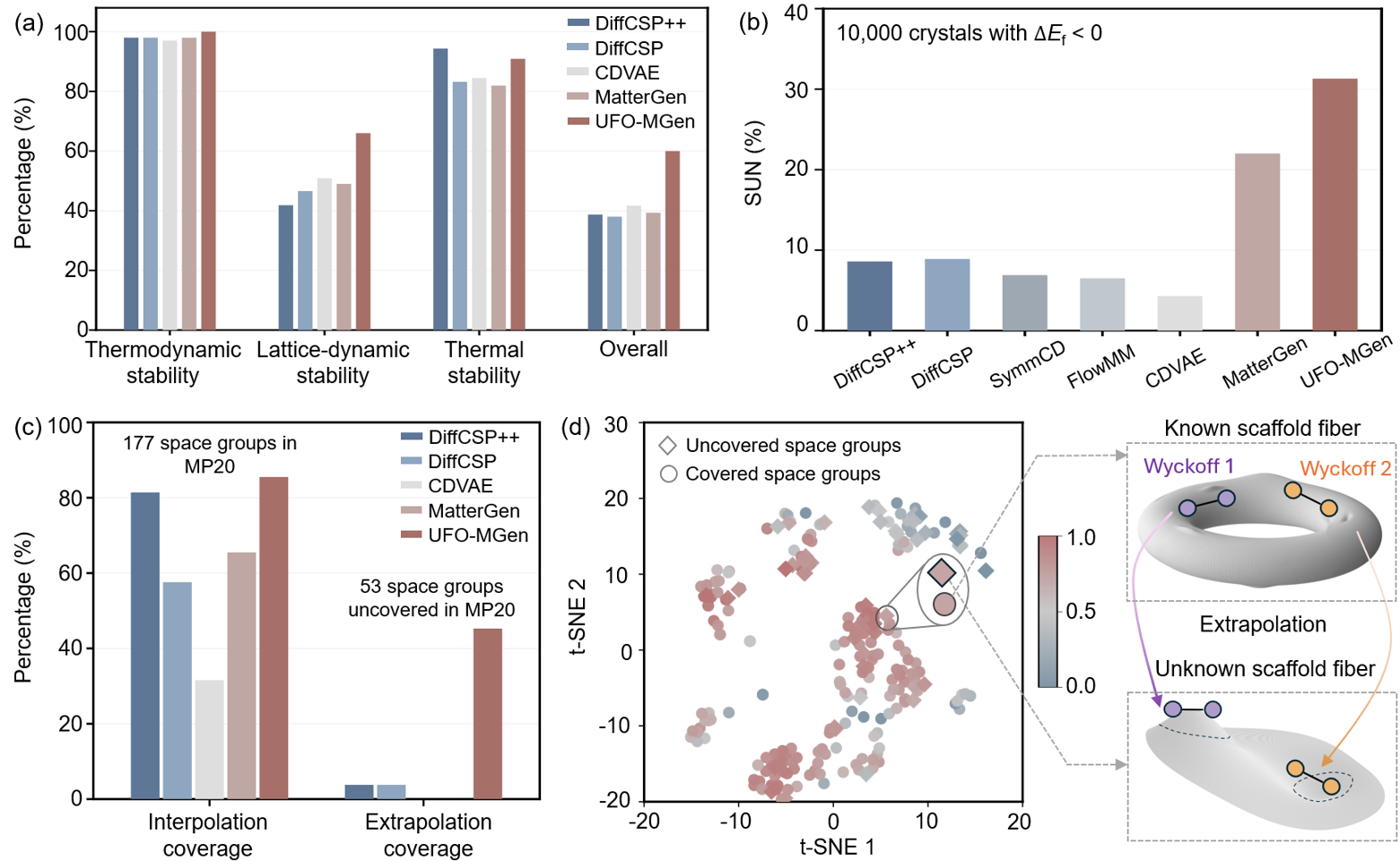}
\caption{\textbf{Performance comparison between {\model} and other generative models.}
\textbf{(a)} Comparison of the MSE success rates achieved by {\model} and other generative models.
\textbf{(b)} Comparison of the SUN rates achieved by {\model} and other generative models.
\textbf{(c)} Comparison of the interpolation and extrapolation capabilities among different generative models. 
\textbf{(d)} Similarity analysis of the scaffold features of 10,000 generated crystals visualized using t-SNE. 
The red color represents the high similarity and blue color means low similarity.
The right schematic illustrates the extrapolation mechanism of {\model}, in which crystals are generated by transferring similar Wyckoff features across neighboring scaffold fibers associated with different space groups.
}
\label{fig:Fig3}
\end{figure*}

Using the more rigorous MSE framework, we benchmark {\model} against several state-of-the-art crystal generative models, including MatterGen~\cite{zeni2025generative}, DiffCSP~\cite{jiao2023crystal, jiao2024diffcsppp}, FlowMM~\cite{miller2024flowmm}, SymmCD~\cite{levy2025symmcd}, and CDVAE~\cite{xie2021crystal}.
To ensure a fair comparison, we retrain {\model} using MP20 dataset that was used to train prior generative models.
We then generate 300 crystal structures from each model and evaluate their stability using the same MSE framework.
As shown in Fig.~\ref{fig:Fig3}(a), {\model} achieves the highest MSE success rate for every individual stability criteria, as well as the the highest overall success rate, demonstrating its superior performance in generating physically stable crystals. 
%demonstrating its superior ability to generate physical viable and accurate crystal structures.
%
In addition to MSE, we further evaluate {\model} using the widely adopted SUN metric~\cite{zeni2025generative} based on 10,000 generated structures that satisfy the first MSE criteria ($\Delta E_f<0$). 
Fig.~\ref{fig:Fig3}(b) shows that {\model} again exhibits the highest SUN score of 31.2\% among all models, which is also the highest reported SUN score to date.
More comprehensive benchmarks using other public metrics are presented in Supplementary Section 5, where {\model} consistently outperforms existing generative models.

Because {\model} has demonstrated superior extrapolation capability, it is also interesting to compare this capability against existing generative models.
Different from our processed database with unified Wyckoff representation, the original MP20 database contains 177 space groups, of which 53 are absent from this dataset.
Next, we regenerate 10,000 crystal structures for each generative model and quantify only the coverage of space groups that are present (interpolation) and absent (extrapolation) in the training data.
Fig.~\ref{fig:Fig3}(c) indicates that {\model} achieves not only the highest interpolation coverage rate, but also the highest extrapolation coverage rate of 45.3\% by generating 24 unseen space groups.  
In contrast, existing generative models exhibit little or no extrapolation capability (Fig.~\ref{fig:Fig3}(c)).
This comparison further demonstrates the superior performance of {\model} in intelligent crystal generation.
%Therefore, {\model} outperforms all existing generative models (Fig.~\ref{fig:Fig3}(c)), which superior performance of {\model} for both extrapolation and interpolation.
%

To better understand the extrapolation capability of {\model}, we performed a t-SNE analysis~\cite{maaten2008tsne} of the scaffold features based on all 10,000 generated crystals.
Crystal clusters belonging to known and unknown space groups exhibit large overlap in the latent space (Fig.~\ref{fig:Fig3}(d)), indicating that they share similar scaffold features, such as Wyckoff positions, crystal systems, and lattice parameters.
Owing to this large overlap, structural information can be easily transferred between distinct scaffolds through their continuous fibers, as schematically illustrated in Fig.~\ref{fig:Fig3}(d), which possibly explains the extrapolation capability of {\model}. 
%
%For instance, features in the known scaffold fibers can transmit to unknown scaffold fibers that correspond to a different space groups, thereby explaining the extrapolation capability of {\model}. 
%
%
%Such cross-fiber knowledge thus providing a plausible mechanistic explanation of the remarkable extrapolation ability of {\model}.
%
\rev{
%The extrapolation capability also highlights the importance of applying VT theory to crystal database to process the discontinuous and singular boundaries.
}

\subsection{Key Mechanisms of {\model}}
The superior performance of {\model} can be ascribed to two unique mechanisms:($i$) the unified Wyckoff representation and ($ii$) the use of a flow-matching model.
To quantitatively evaluate the importance of these mechanisms, we perform ablation studies using three additional comparative models: ($i$) a flow-matching model trained on the original MP database, ($ii$) a diffusion model trained on the original MP database, and ($iii$) a diffusion model trained on our processed database with unified Wyckoff representation. 
%and  replacing the flow-matching model with a traditional diffusion model on ($ii$) the native MP database and ($iii$) the VT-processed database.  
%
%replace the flow-matching model in {\model} by diffusion model and test the performance for crystal generation, both with and without Wyckoff representation.
%
%We then compare flow-based generation with diffusion-style iterative denoising in the same Wyckoff representation. 
%
%Both methods start from a base distribution and generate Wyckoff-space variables. Their sampling dynamics are different. Diffusion updates the sample through many local denoising steps. Flow matching learns a vector field that transports the sample from the base distribution to the target distribution.
%

\begin{figure*}[t]
\centering
\includegraphics[width=1\textwidth]{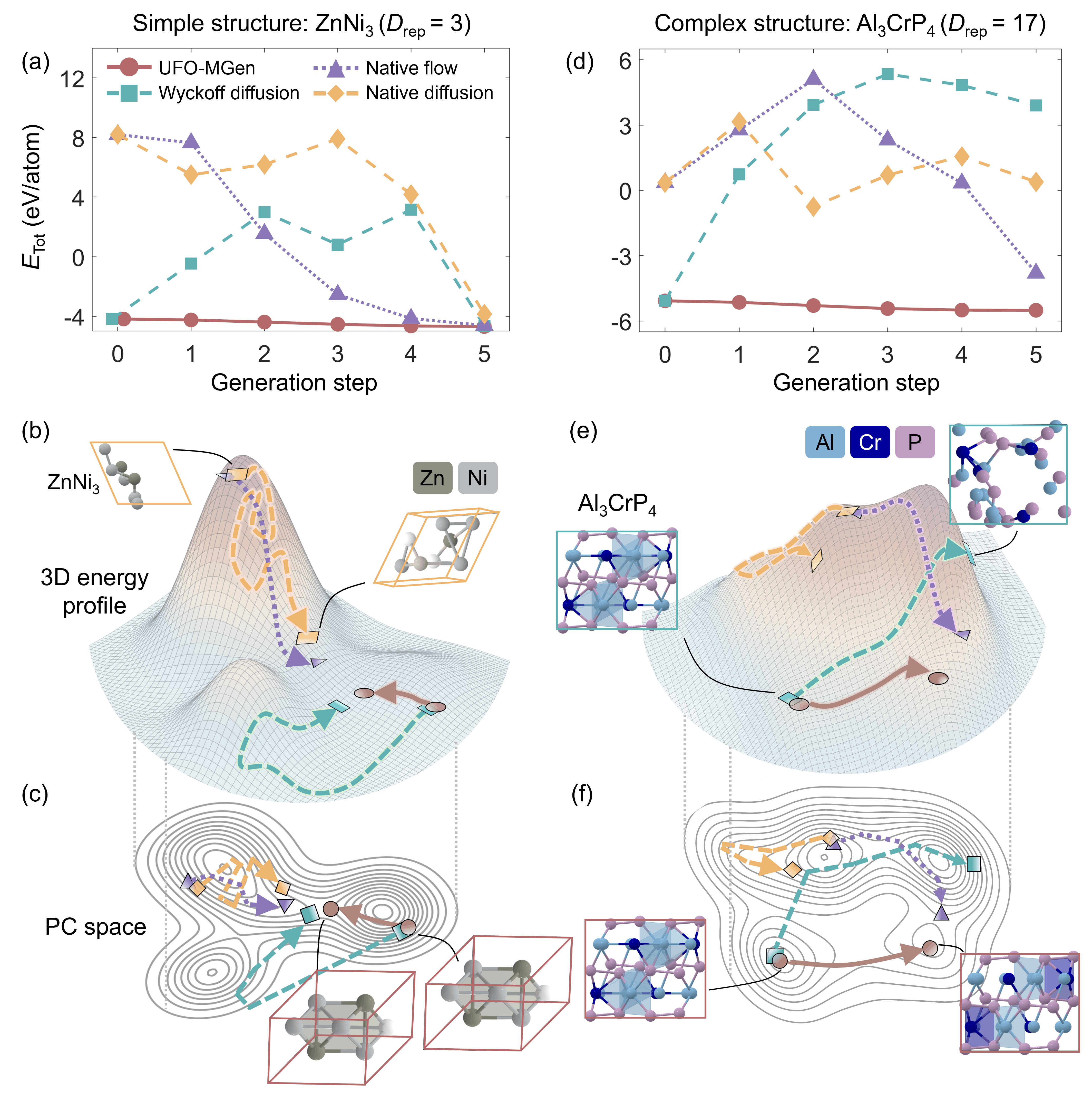}
\caption{\textbf{Ablation studies on the performance of {\model} for the importance of unified Wyckoff representation and flow-matching}.
\textbf{(a)} Evolution of the total energy during the generation of a representative ZnNi$_3$ system with a relatively simple crystal structure.
Four generative methods are compared: {\model}, a diffusion model based on Wyckoff space, a flow-matching model in native crystal space, and a diffusion model in native crystal space. 
%
%based on native crystal space.  
%
The corresponding schematics of three-dimensional (3D) energy trajectories and two-dimensional (2D) principal component (PC) projections are shown on the right side of panel.
\textbf{(b)} Evolution of the total energy during the generation of a representative Al$_3$CrP$_4$ system with a more complex crystal structure using the same four generative methods.   
The corresponding 3D energy profile and 2D PC projections are shown on the right side of the panel.
}
\label{fig:Fig4}
\end{figure*}

Figure~\ref{fig:Fig4} compares the generation trajectories and associated energy profiles of two representative material systems for the three ablation models and the full {\model}. 
For the simple ZnNi$_3$ crystal with $D_\textrm{rep} = 3$, the energy profile in Fig.~\ref{fig:Fig4}(a) shows that the full {\model} generates a physically reliable crystal structure from the first generation step, followed by a smooth and nearly flat trajectory through the remaining steps.
In contrast, when trained on the original MP database, the energy profile first increases and then decreases, indicating a less stable generation process.
Although the diffusion models eventually converge to a crystal structure similar to {\model}, their intermediate structures remain highly unrealistic during the early stages of generation.
%
%
%In contrast, both flow-matching and diffusion models without Wyckoff representation are less efficient in generating the initial structures. 
%
The schematic 3D energy landscapes and the corresponding principal component (PC) trajectories in Fig.~\ref{fig:Fig4}(b-c) clearly illustrate how {\model} rapidly identifies the physically accurate crystals and converges to the target structure, whereas other three ablation models exhibit lower performance.

Furthermore, the advantages of {\model} becomes even pronounced when generating the more complex Al$_3$CrP$_4$ crystal with $D_\textrm{rep} = 17$ (Fig.~\ref{fig:Fig4}(d)), as it directly generates a physically reliable structure from the initial step and rapidly converges to a low-energy state.
The schematic 3D energy profile and corresponding PC projection shown in Fig.~\ref{fig:Fig4}(e-f) illustrate the efficiency of {\model} in identifying the initial structure and faster converging to the target structure. 
%
%distribution toward the target region along a short and direct path. 
%
In contrast, the three ablation models fail to generate initial structures and do not converge to the correct final configurations. 
These ablation studies demonstrate that the combination of the unified Wyckoff representation and flow-matching framework provides an effective strategy for accurate and efficient crystal structure generation.
Additional ablation tests are provided in Supplementary Section 6, all of which demonstrate the advantages of the hierarchical architecture of {\model}.

\subsection{Inverse materials design}
Another key capability of start-of-the-art generative models is property-constrain generation for inverse materials design~\cite{zeni2025generative}. 
To this end, we introduce a fine-tuning module to {\model} and then evaluate its performance to generate crystal structures with target properties.
Since mechanical properties are important factors in the design of next-generation materials for extreme environments~\cite{prameela2023NRM,wyatt2024NRM}, we focus on three mechanical properties: bulk modulus, shear modulus, and Young's modulus.
To quantify the effective of fine-tuning, we use the base model to generate 540 crystals and adopt fine-tuned {\model} to generate 370 crystals with targeted high mechanical properties. 
Further details of the fine-tuning architectures and procedures are provided in Supplementary Sections 7.1-7.2.

Figure~\ref{fig:Fig5}(a) compares the distributions of crystals generated by the base {\model} and its fine-tuned version \textrm{(\textit{i.e.}, UFO-Mech)}, using the mechanical extreme score (Supplementary Section 7.2). 
A pronounced shift toward higher scores is observed for the fine-tuned model, indicating an enhanced capability to generate target crystal structures with superior mechanical properties. 
%suggesting that fine tuning is more efficient to generate crystal structures toward superior mechanical performance.   
%
This improvement is further supported by the shear-bulk modulus diagram in Fig.~\ref{fig:Fig5}(b), where crystals generated by UFO-Mech form a distinct red cluster with substantially higher bulk and shear moduli.
Compared with the existing materials known for their exceptional high mechanical properties, such as cubic boron nitride (c-BN)~\cite{zhang2011cbn}, tungsten carbide (WC)~\cite{brown1966wc}, and titanium diboride (TiB$_2$)~\cite{ledbetter2009tib2}, predicted crystals located in the upper-right region reaches values comparable to those of the benchmark materials.
Extensive DFT calculations further validate the accuracy of UFO-Mech in predicting the mechanical properties of generated crystals (Supplementary Section 7.3). 
Together, these results demonstrate the strong potential of the {\model} for the inverse design of mechanically robust materials. 
%with the potential to easilty extend to any other material properties.  

More interestingly, we find that scaffold features play an important role in influencing mechanical properties of materials.
For instance, we extract one of the most common scaffold feature, $c=(Imm2, 4, 4d, 4d, 4d, 4d)$, from crystals within the red cluster in Fig.~\ref{fig:Fig5}(b) and impose this scaffold on 50 different materials to computer their mechanical properties. 
As shown in Fig.~\ref{fig:Fig5}(c), crystals adopting this scaffold exhibit enhanced bulk, shear, and Young's moduli, with average improvements ranging from 45\% to 60\% relative to their original structures. 
A representative example is BC$_3$, which exhibits exceptionally high mechanical properties (Fig.~\ref{fig:Fig5}(b)).
Notably, its scaffold is very similar to that of WC, one of the strongest materials, indicating that the exceptional mechanical properties of B$_3$C may originate from these specific scaffold characteristics (Fig.~\ref{fig:Fig5}(d)). 
%
%\rev{$c=(Imm2, 4, B(4d), C(4d), C(4d), C(4d))$}, which shares similar scaffold characteristics with WC (Fig.~\ref{fig:Fig5}(d)). 
%

We also test three additional scaffold features selected from the red cluster in Fig.~\ref{fig:Fig5}(b) and find that they can significantly enhance mechanical properties across different materials (Supplementary Section 7.4).
These results further demonstrate the critical roles of crystal scaffolds in governing materials properties and inspire future studies into scaffold-property relationships.
Finally, although the present fine-tuning model is designed to target mechanical properties, it can be easily extended to other material properties.
This capability highlights the broad applicability of {\model} as a general framework for property-constrained inverse materials design.

%%%%%%%%%% conclusion %%%%%%%
%In conclusion, this work XXXXX

\bmhead{Supplementary information}

Supplementary information is available.

\bmhead{Acknowledgments}
J.Z. and C.H acknowledge the support of the National Science Foundation under Grant No. 2556184.
This research used resources of the National Energy Research Scientific Computing Center (NERSC), a DOE Office of Science User Facility supported by the Office of Science of the U.S. Department of Energy under Contract No. DE-AC02-05CH11231 using NERSC awards ERCAP0031213 and ERCAP0035988.
This work was also supported by a user project at the CNMS, a US DOE Office of Science User Facility, operated at Oak Ridge National Laboratory. 

\bmhead{Contribution}
J.Z. developed {\model} model, performed the ablation studies, and fine-tuned the model for inverse materials design.
%
%constructed all GB structures and performed atomistic simulations to calculate the GB properties;
%
O.C. and N.O. performed DFT calculations to validate the predictions of {\model}.
C.H. supervised this work. 
All authors contributed to the writing of the manuscript and agreed for publication. 

\bmhead{Data availability}
The {\model} model and datasets developed in this work are available through our Github repository: https://github.com/huhuhhhh/UFO-MGen.

\section*{Declarations}

The authors declare no conflict of interests.

%%%%%%%%%%%%%%% reference %%%%%%%%%%%

%\bibliographystyle{unsrt}
\bibliography{jobname}
\clearpage

%%%%%%%%%%%%%% Figure 1 for workflow %%%%%%%%%%%%%%

%%%%%%%%%%Figure 2 for stability evaluation%%%%%%%%%%%%%%
%%%%%%%%%%%%%%%%%%%%%%%%%%%%%%%%%%%%%%%%%%%%%%%%%%%%%%%%%%%

%%%%%%%%%%%%Figure 3 for performance comparison %%%%%%%%%%%%%%%%%

%%%%%%%%Figure 4  ufo vs diffusion model comparison %%%%%%%%%%%%%%%%%%

%%%%%%%%%%%%Figure 5 for fine tuning %%%%%%%%%%%%%%%%%%
\begin{figure*}[t]
\centering
\includegraphics[width=\textwidth]{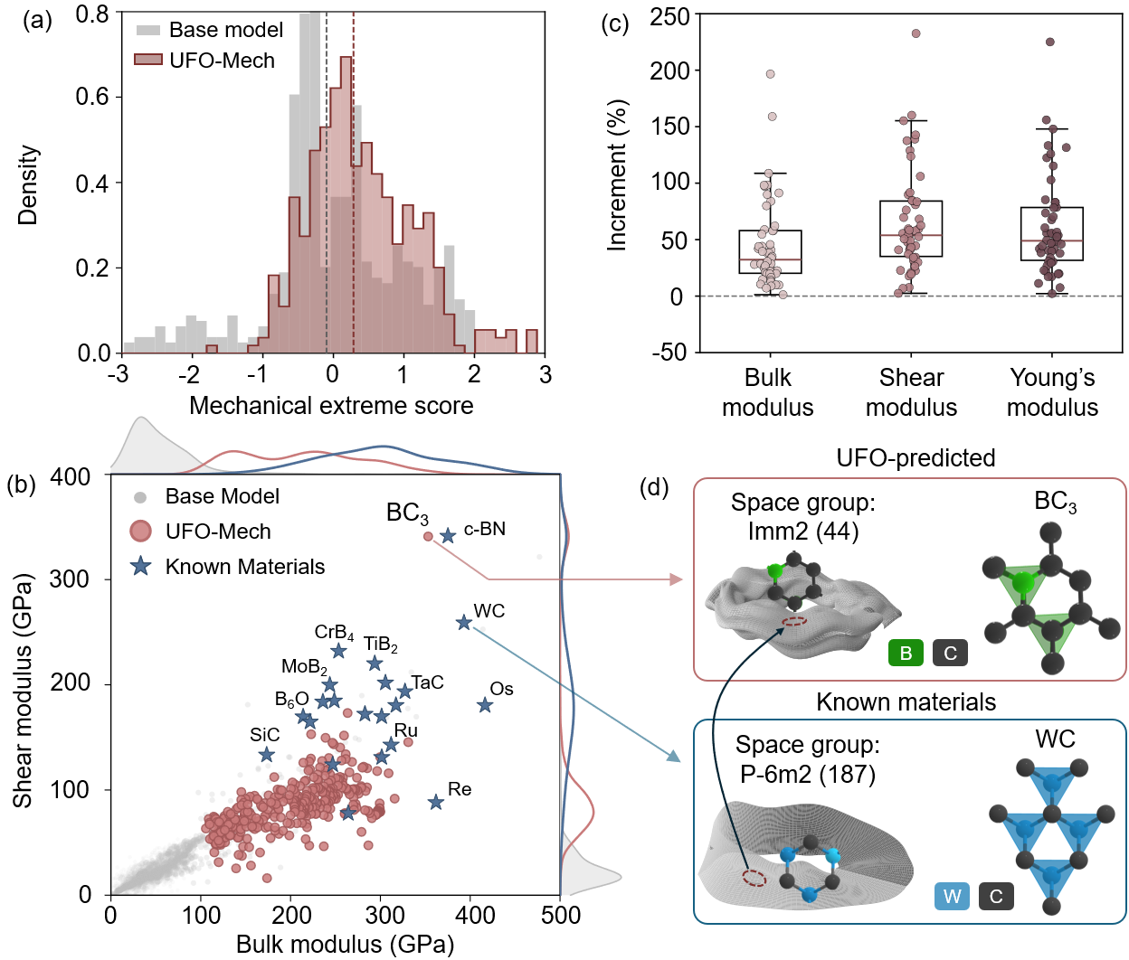}
\caption{\textbf{Fine-tuning of {\model} for property-constrained generations.}
\textbf{(a)} Distribution of crystal structures generated by the base {\model} model and the fine-tuned model (UFO-Mech), evaluated using the mechanical extreme score defined based on the bulk modulus, shear modulus, and Young's modulus.
\textbf{(b)} Distribution of crystal structures generated by both base and fine-tuned models in the shear-bulk modulus diagrams, compared with representative known materials exhibiting exceptional mechanical properties.
To avoid inconsistencies between experimental measurements and computational predictions, the mechanical properties of the known materials are recalculated using the MatterSim for consistent comparison.  
\textbf{(c)} Percentage improvements in the mechanical properties of 50 selected materials after imposing one of the most common scaffold features identified from the red cluster in panel \textbf{(b)}.
\textbf{(d)} Schematic comparison of the scaffolds of two representative materials with high mechanical properties: BC$_3$, predicted by UFO-Mech, and WC, an existing materials known for its high mechanical performance.  
}
\label{fig:Fig5}
\end{figure*}

\backmatter


\end{document}

\begin{center}
    {\LARGE \textbf{Supplementary Information}}\\[0.5em]
    {\large \textbf{Topology-Stratified Materials Discovery with A Flow-based Generative Model}}\\[0.5em]
    {\normalsize Jingyi Zhou$^1$, Oyshee Chowdhury$^1$, Noah Oyeniran$^1$, Chongze Hu$^{1,2}$}\\[0.5em]
    {\normalsize 1. Department of Aerospace Engineering and Mechanics, The University of Alabama, Tuscaloosa, Alabama 35487, United States}
    \\
    {\normalsize 2. Alabama Materials Institute, The University of Alabama, Tuscaloosa, Alabama 35487, United States}
\end{center}

\vspace{1em}

\tableofcontents
\clearpage

% ============================================================
% SECTION S1: Representation
% ============================================================
\section{Wyckoff Representation of Crystal Structures}
The difficulty in crystal generation lies not merely in the joint modeling of atomic species, lattice parameters, and fractional coordinates, but in the fact that the generation of real crystals does not occur within a smooth, natural coordinate space~\cite{hahn2005,adams2023representing,miller2024flowmm}. 
Traditional representations depict a crystal containing ${N}$ atoms as a concatenation of lattice parameters and the coordinates of all atoms; however, this approach suffers from two fundamental issues: (1) Due to symmetry, unit-cell choices, origin choices, and atom-index permutations, a single physical crystal can correspond to a multitude of distinct coordinate descriptions~\cite{mcgill2013geometry}; and (2) the orbit space corresponding to special positions is not a smooth subset of ordinary Euclidean space, but rather exhibits hierarchical and singular structures induced by site-symmetry enhancement and Wyckoff-position specialization~\cite{wondratschek2011international}. 
Consequently, directly learning the generative distribution within this ${3N+6}$-dimensional degrees of freedom (DOFs) space mixes symmetry equivalences, singular boundaries, and true DOF~\cite{xie2021crystal,luo2023towards,jiao2023crystal}.

This paper addresses these challenges by mapping the naive crystal representation into a Wyckoff representation based on the International Tables for Crystallography and standard crystallographic orbit conventions~\cite{aroyo2006bilbao,togo2018spglib}.
The advantages of this approach are twofold: it treats the independent orbits---defined under the crystal's space group---as the primary objects of interest, retaining only the elemental species, Wyckoff letters, and free coordinates associated with each orbit~\cite{hahn2005,aroyo2006bilbao,togo2018spglib}. This effectively compresses the dimensionality of the continuous representation from ${3N+6}$ to a lower dimension, which is dominated by the number of independent orbits. Furthermore, by leveraging space group operations to naturally preserve crystal symmetry, this method allows for the reconstruction of the complete crystal structure while shifting the metric of ``complexity'' away from a simple count of atoms and toward the degrees of freedom that truly reflect the inherent difficulty of the generative process~\cite{cao2024crystalformer,kelvinius2025wyckoffdiff}.

\subsection{Original presentation of Wyckoff points}
\label{subsec:s1-naive-to-wyckoff}
A periodic crystal can be defined as:
\begin{equation}
\mathcal{C} = \left(\mathbf{M},\{(s_i,x_i)\}_{i=1}^{N}\right),
\end{equation}
where $ \mathbf{M} \in \mathbb{R}^{3\times 3}$ is the lattice matrix, $s_i$ is the atomic species of atom $i$, and $x_i \in \mathbb{T}^3=[0,1)^3$ is its fractional coordinate~\cite{hahn2005}. 
If the lattice matrix $\mathbf{M}$ is replaced by lattice parameters $\ell=(a,b,c,\alpha,\beta,\gamma)$, the system has $3N+6$ continuous degrees of freedom. However, for a given structure, the action of the space group introduces significant redundancy through many equivalent coordinate sets; furthermore, the arrangement of equivalent atoms within an orbit introduces additional internal arrangement redundancy~\cite{aroyo2006bilbao}.

\begin{figure}[h]
    \centering
    \includegraphics[width=0.9\textwidth]{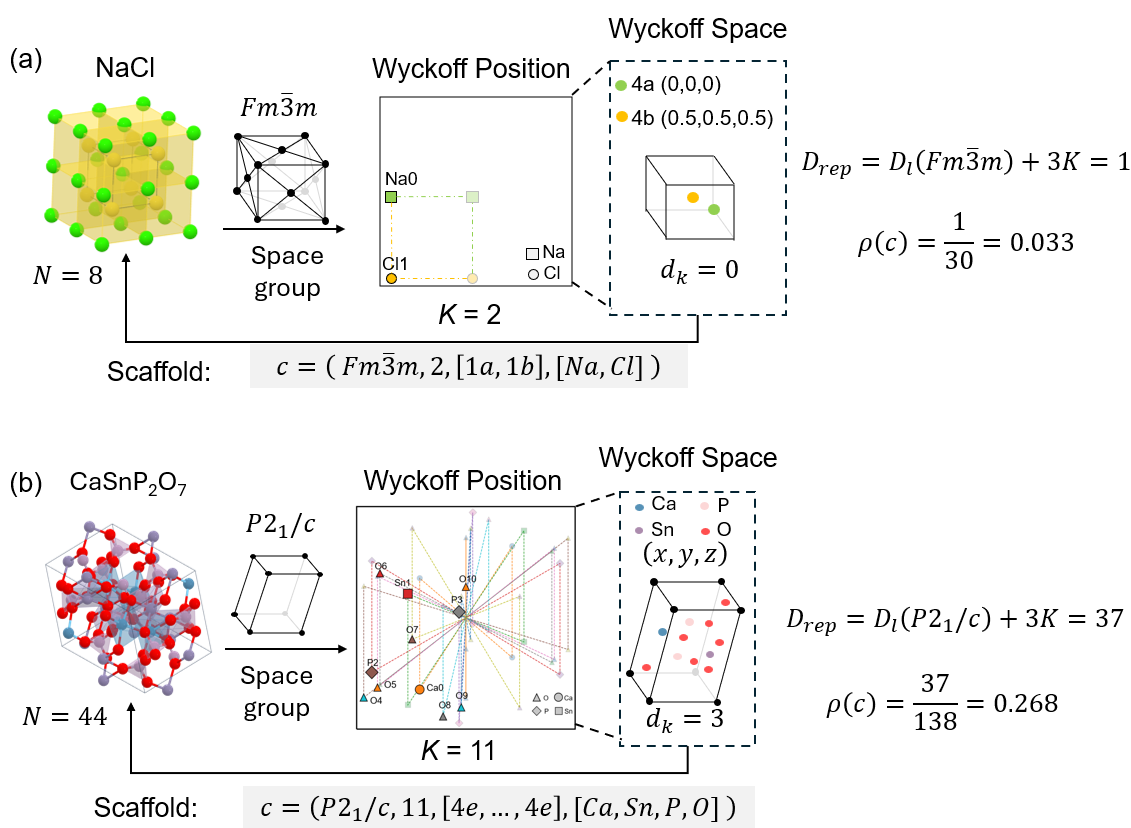}
    \caption{Workflow of transforming a three-dimensional crystal structure into the Wyckoff-space encoded representation:
    (a) Wyckoff representation of cubic NaCl with space group $Fm\bar3m$. The high symmetry of NaCl crystal has 2 representative points ($K=2$), 0 degrees of freedom ($d_k=0$), and a lower $D_\textrm{rep}$ value of 1.
    (b) Wyckoff representation of monoclinic CaSnP$_2$O$_7$ with space group $P2_1/c$. The more complex CaSnP$_2$O$_7$ crystal has 11 representative points ($K=11$), 3 degrees of freedom ($d_k=3$), and a higher $D_\textrm{rep}$ value of 37.
    }
    \label{fig:S1.1}
\end{figure}

To eliminate this redundancy, we employ the concept of symmetry orbits under a space group $G$. For a point $x$, its orbit under the space group $G$ is
\begin{equation}
\mathrm{Orb}_G(x)=\{g(x)\bmod \Lambda:\ g\in G\},
\end{equation}
and its stabilizer is given by:
\begin{equation}
\mathrm{Stab}_G(x)=\{g\in G:\ g(x)\equiv x \pmod{\Lambda}\}.
\end{equation}
A Wyckoff position is characterized by its site multiplicity ($m_k$), which is determined by:
\begin{equation}
m_k=\frac{|G_0|}{|\mathrm{Stab}_G(x_k)|},
\end{equation}
where $k$ is Wyckoff orbit and $x_k$ is representative coordinate of the $k$-th Wyckoff orbit.
Its number of free parameters, $d_k$, can be defined as:
\begin{equation}
d_k = 3-\dim\!\big(\mathrm{Fix}(\mathrm{Stab}_G(x_k))\big),
\label{s1:dk}
\end{equation}
where $d_k \in \{0, 1, 2, 3\}$ can be explained by the number of independent continuous parameters that must be explicitly specified for that Wyckoff orbit $k$~\cite{wondratschek2011international}. 
General Wyckoff positions typically exhibit higher $d_k$ values, whereas highly symmetric special positions often lead to $d_k = 0$.
For instance, Fig.~\ref{fig:S1.1}(a) shows that the NaCl structure has $d_k=0$ for both Na and Cl because both atoms occupy fixed high-symmetry Wyckoff positions.
In contrast, Fig.~\ref{fig:S1.1}(b) shows a more complex material system, CaSnP$_2$O$_7$ with $P2_1/c$ space group. 
In this structure, the symmetry-independent atoms occupy general Wyckoff positions whose coordinate templates contain three free fractional parameters (i.e., $x, y$ and $z$, see Fig.~\ref{fig:S1.1}(b)). 
As a result, the corresponding Wyckoff orbits have $d_k=3$, leading to a higher-dimensional Wyckoff representation than the highly symmetric NaCl crystal.

Following above procedures, we re-encode the original atom crystal representation  into a Wyckoff representation:
\begin{equation}
\phi(\mathcal C)=\Big(G,\,\ell,\,\{(w_k,s_k,x_k^{\mathrm{free}})\}_{k=1}^{K}\Big),
\label{eqn:eq6}
\end{equation}
where $G$ is the space group, $\ell$ denotes the symmetry-constrained lattice parameters, $K$ is the number of symmetry-independent Wyckoff orbits~\cite{togo2018spglib}, $w_k$ is the Wyckoff letter, $s_k$ is the species assigned to orbit $k$, and $x_k^{\mathrm{free}}\in\mathbb{T}^{d_k}$ are the free coordinates of that Wyckoff orbit. 

    \subsection{Crystal scaffolds and associated parameters}
\label{subsec:s1-scaffold}

The Wyckoff representation in Eq.~\ref{eqn:eq6} contains both discrete Wyckoff information (e.g., $G$, $K$, $\omega_{1:K}$, $s_{1:K}$, and others) and continuous parameters (e.g., $\ell$ and $x_k^\textrm{free}$). 
To separate the discrete and continuous information, we define a crystal scaffold ($c$) to only represent the discrete Wyckoff information only, and thus $c$ can be expressed as:
\begin{equation}
    c=(G,K,w_{1:K},s_{1:K})
\label{eq:scaffold}
\end{equation}
which specifies the space group ($G$), the number of Wyckoff orbits ($K$), the Wyckoff-letter sequence ($\omega_{1:K}$), and chemical species ($s_{1:K}$) to each Wyckoff orbit ($k$).
%
%Once a scaffold is fixed, the remaining continuous variables are the lattice parameters $\ell$ and the orbit-level free coordinates ${x_k^{\mathrm{free}}}_{k=1}^{K}$.
%and these continuous variables will be defined as fibers.
%\rev{in Section \ref{subsec:s2-fibers})}.

Using this scaffold representation, we introduce a parameter, called scaffold intrinsic dimension ($D_\textrm{rep}$), to describe the space and complexity of a certain scaffold $c$: 
\begin{equation}
D_{\mathrm{rep}}(c)=D_{\ell}(G)+\sum_{k=1}^{K} d_k,
\label{s1:Drep}
\end{equation}
where $D_{\ell}(G)$ is the number of independent lattice degrees of freedom allowed by the crystal system of $G$.
According to the ITA lattice constraints~\cite{hahn2005}, $D_{\ell}$ has following relations: $D_{\ell}=1$ for cubic systems, $D_{\ell}=2$ for tetragonal, hexagonal, and trigonal systems, $D_{\ell}=3$ for orthorhombic systems, $D_{\ell}=4$ for monoclinic systems, and $D_{\ell}=6$ for triclinic systems.
Since $D_\textrm{rep}$ directly quantifies the number DOFs of a certain crystal structure in Wyckoff space, this parameter can be considered as a core measure of the complexity of a crystal scaffold.

%a core measure of representational complexity of crystal structure in this work: it 
%that a generative model truly needs to learn, while preserving symmetry information. 
%

Based on the value of $D_\textrm{rep}$, we can introduce another scaffold parameter, called compression ratio ($\rho$) as:
\begin{equation}
\rho(c)=\frac{D_{\mathrm{rep}}(c)}{3N+6},
\end{equation}
This quantity describes the extend to which crystal symmetry reduces the dimensionality of the naive coordination representation.
Accordingly, the Wyckoff representation reframes the concept of ``complexity''—shifting it from ``how many atoms are present'' to ``how many independent orbits exist, how many free parameters are associated with each orbit, and how many degrees of freedom remain within the lattice itself.''

\subsection{Database analysis using scaffold parameters}
\label{subsec:s1-efficiency}
\begin{figure}[h]
    \centering
    \includegraphics[width=0.9\textwidth]{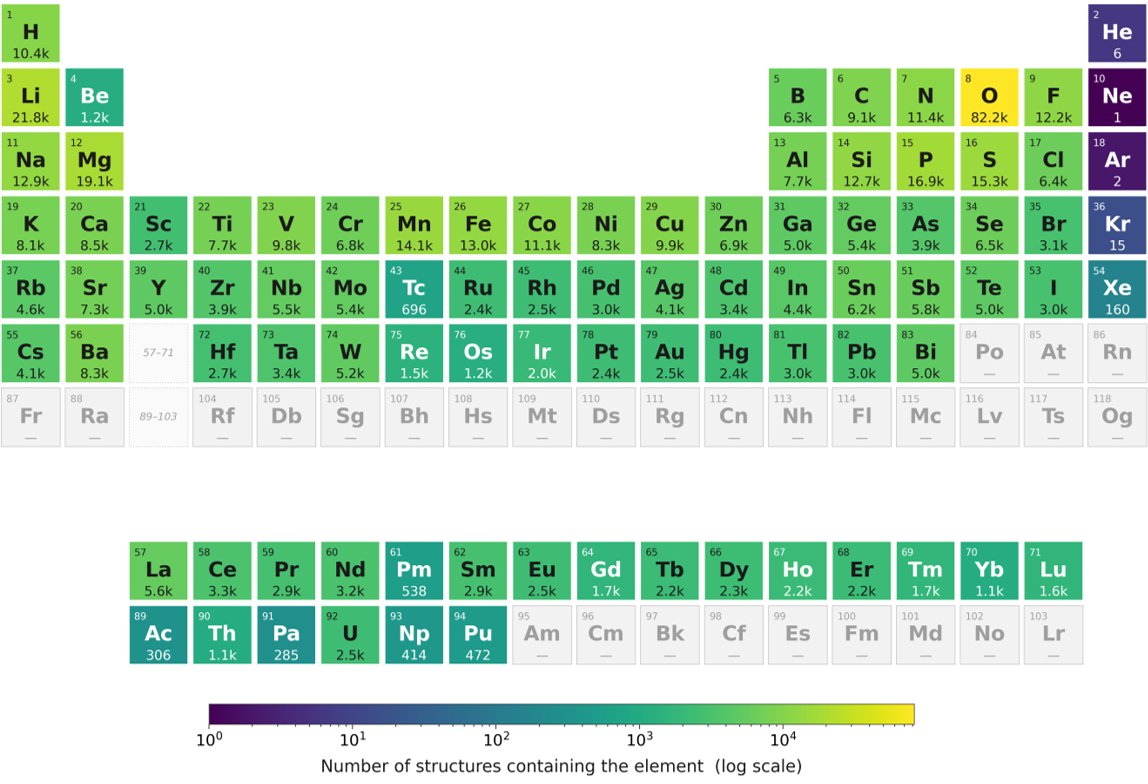}
    \caption{
    Element coverage in the Materials Project dataset containing 154,875 crystal structures. 
    The dataset covers 89 elements, but the distribution is strongly imbalanced, with common inorganic elements appearing much more frequently than noble gases, radioactive elements, and several rare elements. 
    Colors are shown on a logarithmic scale: dark blue indicates elements appearing in only a few structures, green indicates intermediate-frequency elements, and yellow indicates highly frequent elements appearing in tens of thousands of structures. Gray cells indicate elements absent from the dataset.
    }
    \label{fig:S1.2}
\end{figure}

%
%\subsection{Full database analysis}
To evaluate the efficiency of the proposed Wyckoff parameters ($D_\textrm{rep}$ and $\rho$) for representing the crystal structures, we calculated both quantities for all structures in the Materials Project database (v2026.03.15)~\cite{jain2013materialsproject}.
This database contains 154,875 crystal structures with unit cells ranging from 1 to 444 atoms and their compositions spanning 89 chemical elements. 
It is worth noting that although the MP database covers most of the chemical elements in periodic table, the element occurrence is highly uneven, as shown in Fig.~\ref{fig:S1.2}. 
For instance, common inorganic constituents such as O, Li, Na, K, Ca, Fe, Si, F and Cl appear in many structures, whereas noble gases, radioactive elements and several rare elements are weakly represented or absent. 

After classifying these data points based on their $D_{rep}$ and $\rho$ values, the total MP dataset yields 18,074 distinct scaffolds across 35 different space groups.
We further select the trainable data points based on three criteria:
First, the corresponding structures can be successfully standardized and encoded using the Wyckoff representation. 
Second, the scaffold contains no failed, duplicated, or unstable encodings.  
Third, each scaffold has enough training examples. In the expanded setting, we use trainable quantity $n_\textrm{train} \geq$ 100 as the minimum support threshold. 
These criteria yield 32,550 crystal structures with 119 trainable scaffold features.
%
%These scaffolds cover a wide range of $D_L(G)$, $K$, $\sum_k d_k$, and $D_{\mathrm{rep}}$, demonstrating that our proposed scaffold parameters are efficient and compact to classify diverse crystal structures. 

% Figure~\ref{fig:S1.3} presents the distribution of all crystal structures in the $D_\textrm{rep}$--$\rho$ space. 
% %
% After coloring them with different crystal systems, a clear stratified distribution appears: high-symmetry crystals, including cubic, hexagonal, tetragonal, and trigonal structures, primarily locate in the low-$D_\textrm{rep}$ and low-$\rho$ regions.
% %
% In contrast, low-symmetry crystals, such as triclinic and monoclinic structures, primarily locate the high-$D_\textrm{rep}$ and high-$\rho$  region.
% %
% %Furthermore, the distribution of all crystal structures shows clear stratification based on their crystal systems, where high-symmetry crystals locate in low $D_\textrm{rep}$ and $\rho$ regions, but low-symmetry structures locate in high $D_\textrm{rep}$ and $\rho$ regions.
% %
% Such a stratified distribution demonstrates that $D_\textrm{rep}$ and $\rho$ parameters can effectively capture the complexity of crystal structures in Wyckoff space, making them effective descriptors for representing crystal structures in latent space.

\begin{figure}[htbp]
    \centering
    \includegraphics[width=1\textwidth]{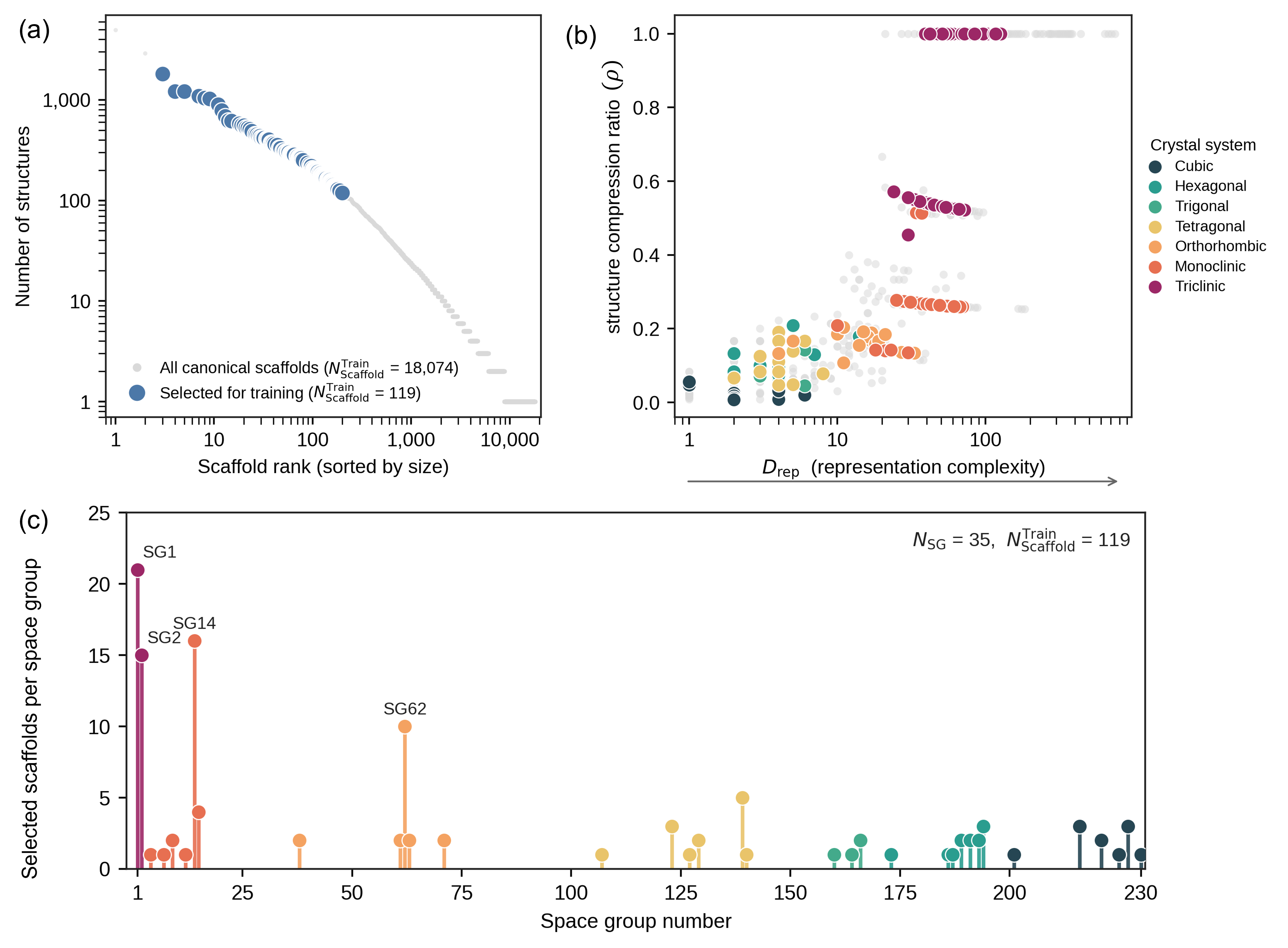}
    \caption{
    Trainable scaffold pool and long-tailed scaffolds distribution.
    (a) Scaffold-support distribution after Wyckoff encoding and canonicalization. The expanded corpus contains 18,074 canonical scaffolds, but most scaffolds are weakly supported, meaning that they contain only a small number of structures. The 119 scaffolds selected for training lie in the high-support region and satisfy the minimum support threshold \(n_{\mathrm{train}}\geq100\).
    (b) Coverage of the trainable scaffolds in the \(D_{\mathrm{rep}}-\rho\) complexity space. Grey points denote all canonical scaffolds, while colored points denote the selected trainable scaffolds. Colors indicate crystal systems. The selected scaffolds span both low- and high-complexity regions and cover all seven crystal systems.
    (c) Distribution of the selected scaffolds over space groups. The trainable scaffold pool covers 35 space groups, with 119 selected scaffolds in total. The vertical bars show the number of selected scaffolds associated with each space group, and colors again indicate crystal systems.
    }
    \label{fig:S1.3}
\end{figure}

Based on these trainable scaffolds, we further analyze their structural characteristics and their coverage of crystal structures and chemical elements.
Figure~\ref{fig:S1.3}(a) shows the scaffold-size distribution, where a small number of scaffolds contain hundreds to thousands of structures, whereas most scaffolds are represented by only a few configurations.
The 119 scaffolds used for training are selected from the left region in Fig.~\ref{fig:S1.3}(a), as they contain sufficient examples to reliably train the {\model}.
In contrast, the remaining scaffolds contain insufficient examples for stable scaffold-specific flow training and are therefore retained for dataset statistics and coverage analyses but are not used as primary routes for continuous flow training.

The crystal structure distribution of the trainable scaffold is further analyzed in Fig.~\ref{fig:S1.3}(b), where the scaffolds are mapped onto the \(D_{\mathrm{rep}}-\rho\) space and colored based on their crystal systems. 
Grey points denote all canonical scaffolds, while colored points denote the scaffolds selected for training. 
The selected set covers all seven crystal systems and spans a broad range of representation complexity and Wyckoff density.
Thus, the trainable scaffold pool is not limited to the simplest cases; it also includes high-complexity and weakly compressed regions, especially in monoclinic and triclinic systems.
Such a diverse distribution demonstrates that $D_\textrm{rep}$ and $\rho$ parameters can effectively capture the complexity of crystal structures in Wyckoff space, making them effective descriptors for representing crystal structures in latent space.

Finally, we analyze the space-group distribution of the trainable scaffolds.
Figure~\ref{fig:S1.3}(c) summarizes the distribution of the 119 trainable scaffolds across space groups.
Although the training dataset covers only 35 space groups, these space groups are broadly distributed across the full crystallographic range from No.~1 to No.~230.
This broad distribution ensures that Stage~III is trained across diverse symmetry regimes rather than being restricted to a narrow subset of common space groups.
%
%he distribution remains non-uniform: several space groups, such as No. 1, No.2, No. 14, and No.62, contribute multiple trainable scaffolds, whereas many other covered space groups contribute only one or a few scaffolds. 
%
%This reflects the same long-tailed support pattern at the space-group level. Importantly, the selected pool is not restricted to a single crystal family; it includes scaffolds from triclinic, monoclinic, orthorhombic, tetragonal, trigonal, hexagonal, and cubic systems. 
%
%This diversity ensures that Stage~III is trained on multiple symmetry regimes rather than on a narrow subset of common space groups.
%

\clearpage

% ============================================================
% SECTION S2: Wyckoff Stratified Space and Variable Topology Theory
% ============================================================
\section{Stratified Wyckoff Space and Unification}
\label{sec:s2-wyckoff-stratified-space}

This section explains why crystal generation in Wyckoff space is not naturally defined on a single smooth manifold. 
The key point is that high-symmetry atomic positions create lower-dimensional geometric limits, while different scaffolds induce different continuous fibers. 
Together, these three facts motivate the formulation of {\model} as a multi-stage frameworks to handle the non-smooth manifold of crystal structure generation in Wyckoff space.
%: ($i$) a select discrete topology scaffold ($ii$) a occupancy for discrete chemical elements ($iii$) Continuous variable: Fiber.}

\subsection{Wyckoff-stratified parameter space}
\label{subsec:s2-whitney-background}

Conventional generative modeling is typically formulated on a fixed-dimensional smooth space, such as an Euclidean space or a single Riemannian manifold~\cite{kingma2013autoencoding,mathieu2020riemannian}. 
However, this assumption is not valid for crystal generation. 
This is because even when the space group and the scaffold are fixed, the Wyckoff crystal parameter space is not globally smooth. 
As continuous Wyckoff coordinates approach special positions, the local site symmetry increases and the number of free degrees of freedom decreases, leading to changes in the dimensionality of the representation space. 
As a result, the Wyckoff representation space is more appropriate to define as a union of multiple smooth manifolds with different dimensions, rather than a single smooth manifold.
Accordingly, the effective Wyckoff representation for a fixed scaffold ($c$) can be expressed as:

% This viewpoint appears naturally in the \rev{Wyckoff representation of crystal structure}. 
% %
% \rev{For a fixed scaffold $c$ Eq.~\eqref{eq:scaffold}}, let $\mathcal{W}_\textrm{raw}(c)={L}_G \times \prod_k \mathbb{T}^{d_k}$ denote the raw \rev{continuous structural paramater space}, where ${L}_G$ is \rev{the lattice parameters constrained by symmetry asscociated with space group $G$}, and \rev{$X_{\omega_k}\subseteq \mathbb{T}^{d_k}$} is the \rev{free fractional coordinate space of the $k$-th Wyckoff orbit}. 

%
%\rev{The effective Wyckoff representation requires explicitly modified for special boundary regions ($\mathcal{W}_{\textrm{spec}}$) and physically invalid regions ($\mathcal{W}_{\textrm{coll}}$).}
%
\begin{equation}
\mathcal{W}(c)=\mathcal{W}_\textrm{raw}(c)\setminus\big(\mathcal{W}_{\mathrm{spec}}(c)\cup \overline{\mathcal{W}_{\mathrm{coll}}(c)}\big),
\label{eq:s2-valid-space}
\end{equation}
where $\mathcal{W}_\textrm{raw}$ is the raw Wyckoff space, $\mathcal{W}_{\mathrm{spec}}$ denotes the special subset of $\mathcal{W}_\textrm{raw}$, and $\mathcal{W}_{\mathrm{coll}}$ denotes the the collision subset of $\mathcal{W}_\textrm{raw}$.
More specifically, the $\mathcal{W}_{\mathrm{spec}}$ corresponds to cases where free parameters at special positions with high symmetry (e.g., $x=0$ or $x=y$).  
The $\mathcal{W}_{\mathrm{coll}}$ corresponds to cases where orbit distances are too short and hence can consider atoms collide.
%
% A key message based on Eq.~\eqref{eq:s2-valid-space} is that, since the boundaries of $\mathcal{W}(c)$ are not arbitrary and correspond to different symmetry specialization and physical validity constraints, they naturally create lower-dimensional geometric pieces. Therefore, the effective Wychoff space is not a smooth and unified \rev{space}, but rather a stratified space composed of several \rev{subspace}.

\begin{figure}[h]
    \centering
    \includegraphics[width=0.6\textwidth]{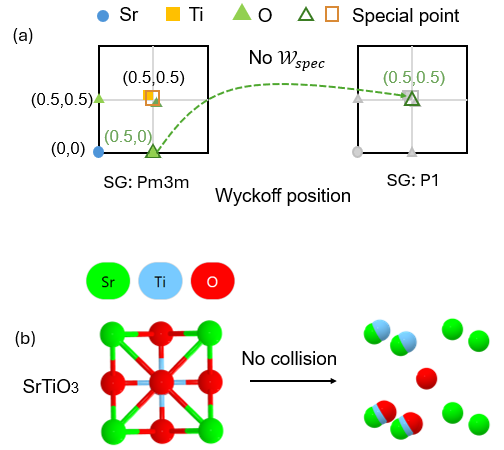}
    \caption{Representation of discontinuous features in Wyckoff space that needs a stratified framework to treat discontinuous and continuous features.
    (a) Special position treatment.
    (b) Collision treatment.
    }
    \label{fig:S2.1}
\end{figure}

The continuous nature of Wyckoff space has important implications for generative modeling of crystal structures, as special Wyckoff positions must be treated explicitly.
For instance, Fig.~\ref{fig:S2.1}(a) shows that when the entire Wyckoff space is treated as an ordinary Euclidean space~\cite{adams2023representing}, symmetry-induced singularities may be incorrectly interpreted as noise, thereby obscuring the physical significance of special Wyckoff positions such as ($x$ = 0.5 and $y$ = 0.5).
Figure~\ref{fig:S2.1}(b) further illustrates the physical constraint that atoms cannot overlap or approach each other too closely, thereby excluding physically invalid configurations from the accessible Wyckoff space.

To address this issue, we adopt a stratified framework for continuous generation, which transforms the continuous Wyckoff parameters from a base distribution to the data distribution within a fixed scaffold ($c$). 
The resulting dynamics are defined only on the smooth and physically valid region of each scaffold parameter space, while the excluded lower-dimensional subsets are interpreted as geometric limits corresponding to symmetry-specialized Wyckoff strata or physically invalid collision configurations.

\clearpage

\subsection{Unification of Wyckoff representation}
\label{subsec:s2-variable-topology}
Conventional generative models, such as diffusion and flow matching~\cite{miller2024flowmm,jiao2023crystal,ho2020denoising,lipman2023flow}, generally assume that crystal structures can be represented on a single smooth manifold. 
However, this assumption does not hold true for the Wyckoff representation, where different scaffolds have different number of continuous DOFs and are connected through discontinuous symmetry transitions.
For instance, statistical analysis of the distribution of $D_{\mathrm{rep}}$ of all crystal structures in MP data suggest that about 89\% of sampled scaffold pairs have different representation dimensions. 
This clearly demonstrates that a single fixed-dimensional intrinsic manifold is not a good model for the crystal generation process, which also demonstrates the importance of adopting Wyckoff unification to process the crystal data~\cite{jiao2023crystal}.

\begin{figure}[h]
    \centering
    \includegraphics[width=1\textwidth]{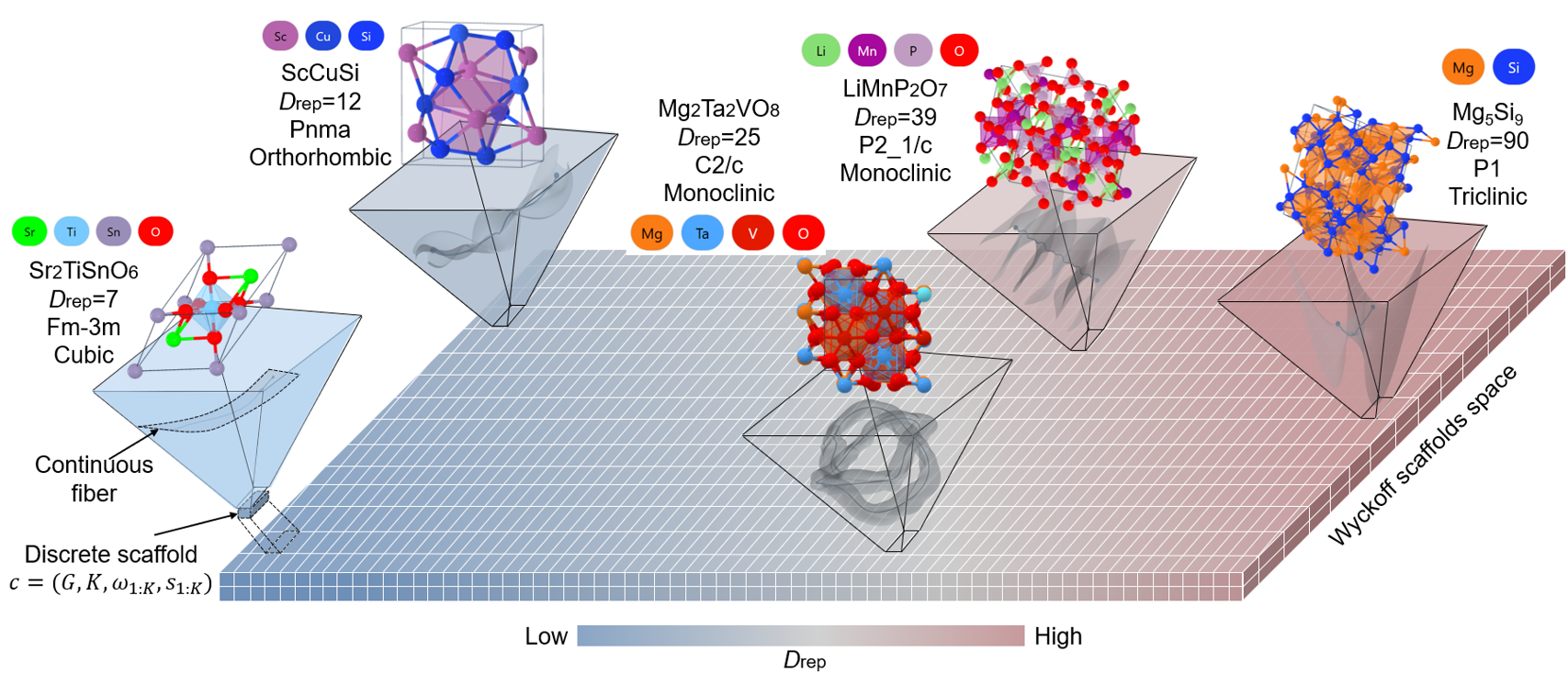}
    \caption{Unification of Wyckoff space.
    For each scaffold, the special positions in its continuous fibers will be processed as the boundary of the fibers and connect with other fibers more smoothly.
    Such an unification can be extended to the entire Wyckoff space characterized by the $D_\textrm{rep}$ values.
    %Shows the selected scaffolds of crystals of varying complexity and the fibers in different dimensions within them.
    }
    \label{fig:S2.2}
\end{figure}

In this work, we introduce the concept of ``unified Wyckoff representation'' that provides a coherent description of the stratified Wyckoff space.
Specifically, each scaffold ($c$) defines a local continuous manifold, referred as a fiber ($\mathcal{F}$), that contains all continuous crystal structure information, such as lattice variables ($\ell$) and free Wyckoff coordinates ($x^{orb}$).
%
%These continuous information derived quantities, such as bond length and angles, are smooth functional of these continuous parameters.
%
%ne origin of the stratified nature of the Wyckoff representation is that changes in site symmetry reduce the number of independent Wyckoff parameters. 
%
Since fibers corresponding to different scaffolds generally have different dimensions, they need to be connected through lower-dimensional boundary regions representing special Wyckoff positions.
Rather than treating these special configurations as disconnected states, the topological fiber bundle theory~\cite{husemoller1994fibre} provides a natural framework for interpreting them as boundary connections between neighboring fibers of different dimensions.
%interprets them as boundary connections between neighboring fibers.
%
A rigorous mathematical formulation and proof of this concept will be presented in a separate study.
Briefly speaking, this process unifies the description of both continuous structural variations within a scaffold and discrete transitions between different scaffolds, enabling crystal structures with different crystallographic representations to be modeled within a single framework, see Fig.~\ref{fig:S2.2}.

Beyond crystallographic constraints, physical constraints such as atomic overlap, geometric validity, and energy stability further modify the topology of each fiber by excluding physically inaccessible regions.
We refer to these topology changes as physically variable topology. 
Within each valid fiber, local structural variations are described using a Riemannian metric that measures how infinitesimal changes in lattice parameters and Wyckoff coordinates affect the crystal geometry. 
Together, the global unification framework captures the connectivity among different crystallographic manifolds, while the local Riemannian geometry characterizes structural variations within each manifold, providing a unified mathematical description of crystal generation in Wyckoff space.

%\subsection{Importance of unified Wyckoff representation}
%\label{subsec:s2-empirical-prevalence}
%
The critical role of the unified Wyckoff representation have been systematically evaluated through ablation studies, as shown in main-text Fig. 4.
For isntance, the {\model} significantly outperforms its counterparts trained on the MP databases without the unified Wyckoff representation. 
%

%Although a padded tensor can be useful for implementation~\cite{hoogeboom2022equivariant}}, but the active geometry still remains scaffold-specific, further indicating the critical need to use VT theory.

\clearpage

% ============================================================
% SECTION S3: {\model} Model
% ============================================================

\section{{{\model} Architecture and Training Objective}}
\begin{figure}[h]
    \centering
    \includegraphics[width=1\textwidth]{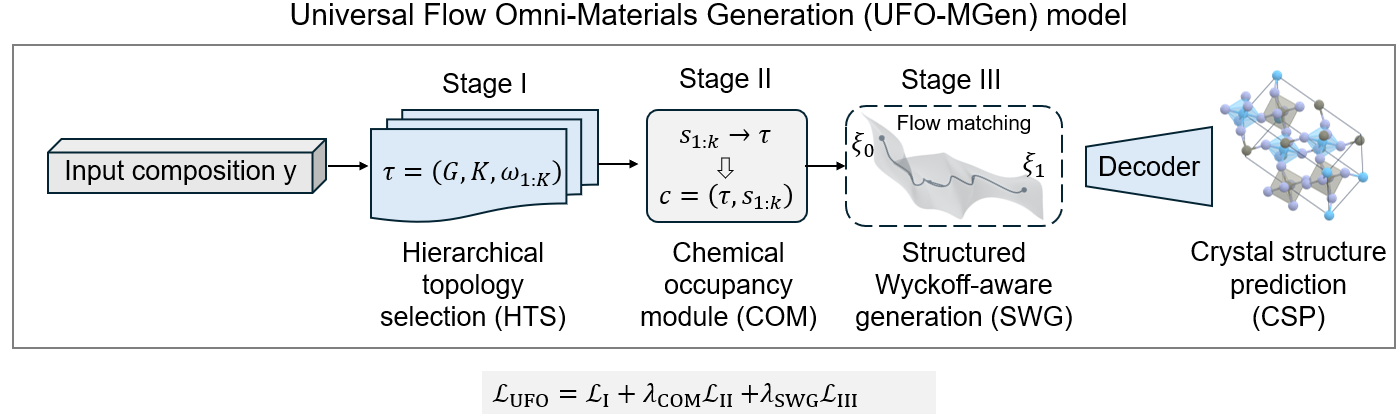}
    \caption{Workflow of {\model} with three key stages:
    Stage I: hierarchical topology selection (HTS) for topology-fixed scaffold ($\tau$) prediction; 
    Stage II: chemical occupancy module (COM) for assigning chemical species to $\tau$ to generate complete scaffold ($c$);
    and Stage III: Structured Wyckoff-aware generation (SWG) for using flow matching model to generate the continues lattice parameters and Wyckoff coordination that was finally used for crystal structure predictions (CSP).
    }
    \label{fig:S3.1}
\end{figure}

\label{sec:s3-{\model}-model}
\textbf{Architecture of {\model}}:
Owing to the stratified Wyckoff space, we design the {\model} as a hierarchical crystal generation process consisting of three stages: ($i$) Stage I: hierarchical topology selection (HTS), ($ii$) Stage II: chemical occupancy module (COM), and ($iii$) Stage III: Structured Wyckoff-aware generation (SWG), as illustrated in Fig.~\ref{fig:S3.1}.
We briefly discuss the three stages below.

Stage I is motivated by the stratified features of Wyckoff space, in which the crystallographic topology is determined before chemical species are assigned to each Wyckoff orbit.
Specifically, discrete Wyckoff parameters such as space group ($G$), number of Wyckoff orbits ($K$), Wyckoff letter ($\omega_{1:K}$), define a topology-only scaffold, while the chemical species ($s_{1:K}$) are introduced later in Stage II.
%
%symmetry scaffold before chemical species are assigned. 
%
Since the full scaffold defined in Eq.~\eqref{eq:scaffold} contains both crystallographic topology and orbit-level chemistry, we separate its topology-only part ($\tau$) as:
\begin{equation}
    \tau=(G,K,w_{1:K}).
\label{eq:s3-tau}
\end{equation}
%
%where $\tau$ specifies the space group, the number of symmetry-independent Wyckoff orbits, and the Wyckoff-letter sequence, but does not include the species assignment $s_{1:K}$.
%Thus, Stage I predicts the topology-only scaffold $\tau$, while the chemical assignment is handled separately in Stage II.
%
%The species assignment is denoted by $s_{1:K}=(s_1,\ldots,s_K)$, where $s_k$ is the element placed on the $k$-th Wyckoff orbit. 
%
%This can be assigned to $\tau$ to form a complete scaffold as: $c = (\tau,s_{1:K})$.
%
%\begin{equation}
%    c = (G,K,w_{1:K},s_{1:K}) = (\tau,s_{1:K}).
%    \label{eq:s3_c}
%\end{equation}
%
%As discussed in Section~\ref{subsec:s2-variable-topology}, the full Wyckoff crystal space is a disjoint union of scaffold-conditioned top strata \rev{(Eq.~\eqref{eq:s2-total-space})}. 
%
The motivation to define $\tau$ is that, once the scaffold $c$ is determined, the continuous variables lie in the corresponding smooth fiber $\mathcal W_0(c)$, which contains the valid $\ell$ and $x_k^\textrm{free}$ for that scaffold. 
This geometric decomposition motivates the design of hirarchical design of {\model}: the model first selects a discrete scaffold $c$, thereby selecting the target fiber, and then learns a continuous flow only within $\mathcal W_0(c)$. 
In this way, {\model} avoids learning a single continuous flow over a mixture of fibers with different dimensions and topological structures.

Stage II applies a chemical constrains based on the species occupancy on each orbits to improve the accuracy.
The species assignment follows several physical and chemical constrains, such as charge balance, neural balance, and others to ensure the reasonable chemical occupation at each specific Wyckoff orbit site.

Stage III integrates lattice variables together with the Wyckoff free coordinates to define the continuous crystal geometry.
Flow-matching is incorporated in Stage III to generate continuous fibers, which are then combined with the complete scaffolds obtained from Stage II to reconstruct the corresponding 3D crystal structures. 
\\\\
\textbf{Probability propagation in {\model}}:
Based on the architecture of {\model} discussed above, the overall probability $p(\tau, s_{1:k},\xi\mid y)$ can be defined using the three-stage factorization:
\begin{equation}
    p(\tau,s_{1:K},\xi\mid y)
    =
    \pi_\alpha(\tau\mid y)\,
    p_\phi(s_{1:K}\mid \tau,y)\,
    p_\theta(\xi\mid \tau,s_{1:K},y).
    \label{eq:s3_main_factorization}
\end{equation}
Here $y$ is the the input for {\model} as shown in Fig.~\ref{fig:S3.1}, which contains basic information of a crystal, such as chemical species ($s_{1:K}$), total number of atoms ($N$), and others.
Stage I learns $\pi_\alpha(\tau\mid y)$, the route-aware topology selector. 
Stage II learns $p_\phi(s_{1:K}\mid \tau,y)$, the chemical occupancy module (COM). 
Stage III learns $p_\theta(\xi\mid \tau,s_{1:K},y)$, the structured Wyckoff-aware generator, and $\xi$ is the coordination representation that combines $\ell$ and $x^{orb}$:
%
% Since we used $\eta\in\mathcal W_0(c)$ to denote an abstract continuous point in the scaffold-conditioned fiber. 
% %
% However,  the neural generator cannot handle with this abstract point, and thus we must represent $\eta$ in explicit coordinates. 
% %
% For convenience, we introduce $\xi$ as the coordinate representation of $\eta$ under the scaffold-specific chart $\chi_c$ as:
%
% \begin{equation}
%     \xi=\chi_c(\eta).
% \end{equation}
% %
% After Stage I determines the topological scaffold $\tau$ and Stage II assigns the orbit-level species $s_{1:K}$, the complete scaffold $c=(\tau,s_{1:K})$ is fixed. 
% %
% Next, Stage III uses this full scaffold to generate $\xi\in\mathcal W_0(c)$.
% %after the complete scaffold $c$ has been fixed by Stage I and II.
% %
% Indeed, the generation process in Stage III is continuous and takes place within $\mathcal W_0(c)$, but the model operates on its coordinate representation:
%
\begin{equation}
\xi=(\ell,x^{\mathrm{orb}})
\label{eq:s3_xi}
\end{equation}
%
%where $\ell$ is the normalized lattice representation and $x^{\mathrm{orb}}$ contains the orbit-level free fractional coordinates.
%\rev{where $\ell$ is a six-dimensional normalized lattice-parameter vector and $x^{\mathrm{orb}}$ contains the orbit-level free fractional coordinates.} 
%
%In this {\model} framework, $\ell$ is a six-dimensional normalized lattice-parameter vector. 
\\
\textbf{Loss functions of {\model}}:
Based on the Fig.~\ref{fig:S3.1}, the overall loss of the entire framework can be simply expressed as:
\begin{equation}
    \mathcal L_{\mathrm{UFO}}
    =
    \mathcal L_{\mathrm I}
    +
    \lambda_{\mathrm{COM}}\mathcal L_{\mathrm{II}}
    +
    \lambda_{\mathrm{CFM}}\mathcal L_{\mathrm{III}}.
    \label{eq:s3_total_loss}
\end{equation}
where $\mathcal L_{\mathrm I}$, $\lambda_{\mathrm{COM}}\mathcal L_{\mathrm{II}}$, and $\lambda_{\mathrm{CFM}}\mathcal L_{\mathrm{III}}$ are the loss of Stage I  (Eq.~\ref{eq:s3_stage1_loss}), II (Eq.~\ref{eq:s3_sam_loss}), and III (Eq.~\ref{eq:s3_stage3_loss_full}), respectively.
In the following sections, we described the detailed architecture of each stage and the corresponding loss functions used to train models.

%%%%%%%%% section 3.1 for stage I %%%%%%%%%%%%%%%%%%%%
\subsection{Stage I: Hierarchical topology selection (HTS)}
\label{subsec:s3-stage1}

\subsubsection{Architecture of Stage I}
\label{subsubsec:s3-stage1-arch}
The main target of Stage I is to predict the topology scaffold, $\tau$ (Eq.~\ref{eq:s3-tau}), of a crystal in the Wyckoff parameter space from the input composition $y$, where $y$ is a 125-dimensional feature vector constructed from the target chemical formula:
\begin{equation}
    y=
    [
    f_{\mathrm{elem}},
    \log(1+n_{\mathrm{atoms}}),
    b_{\mathrm{atoms}}
    ]
    \label{eq:s3_y}
\end{equation}
where $f_{\mathrm{elem}}\in\mathbb R^{120}$ is the element-fraction vector over the fixed element vocabulary, $\log(1+n_{\mathrm{atoms}})$ is a scalar that represents the total number of atoms, and $b_{\mathrm{atoms}}\in\mathbb R^4$ is a one-hot atom-count bucket feature.
\begin{figure}[h]
    \centering
    \includegraphics[width=0.99\textwidth]{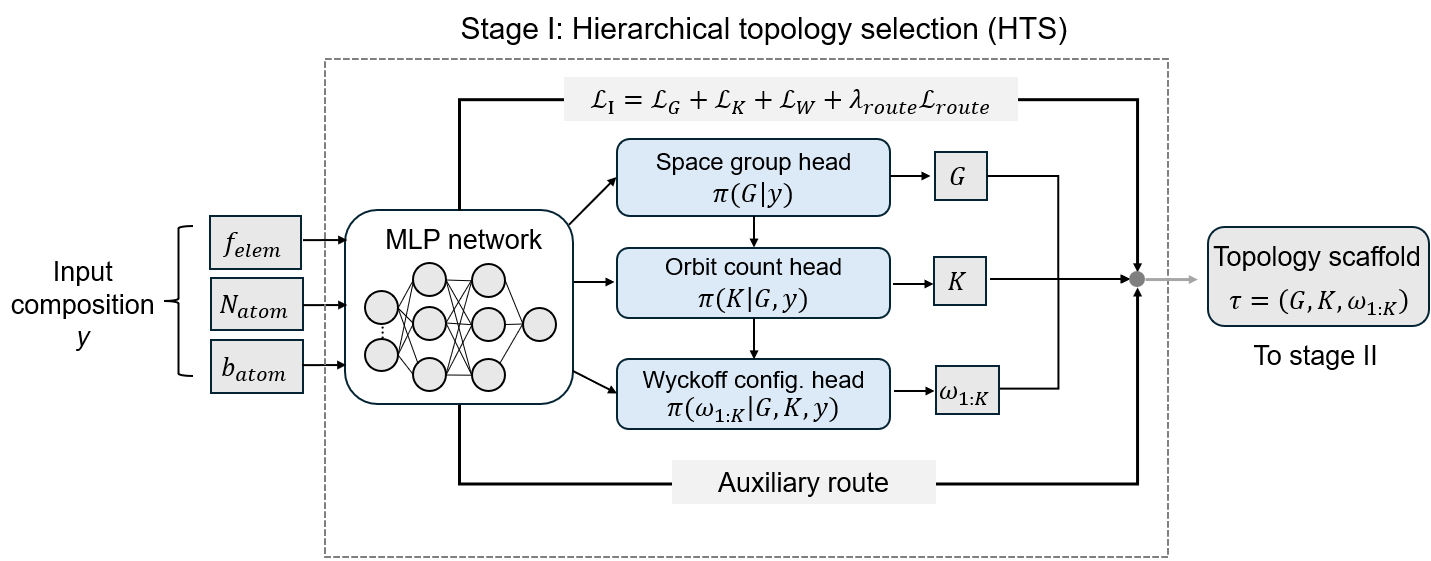}
    \caption{Workflow of {\model} Stage I for hierarchical topological selection (HTS).
    }
    \label{fig:S3.2}
\end{figure}
In addition to predicting $\tau$, Stage I is designed to use a two-layers MLP with hidden dimension of 256 to learn the basic structural relationships among crystal scaffolds underlying the unified Wyckoff parameter space.
By learning this unified space, {\model} distinguishes crystals with different $D_\mathrm{rep}$ values and learns their corresponding topological features, providing a shared feature space for the hierarchical prediction of the topology scaffold $\tau$. 
Therefore, it is more efficient than learning within the traditional atom-by-atom learning space.

The topology labels used for training are obtained from dataset unification process, as discussed in Section~\ref{sec:s2-wyckoff-stratified-space}.
During this step, topology scaffold features belonging to the special $\mathcal W_{\mathrm{spec}}$ are examined according to hierarchical ordering $G\rightarrow K\rightarrow w_{1:K}$, with following hierarchical factorization:
\begin{equation}
    \pi_\alpha(\tau\mid y)
    =
    \pi_\alpha(G\mid y)\,
    \pi_\alpha(K\mid G,y)\,
    \pi_\alpha(w_{1:K}\mid G,K,y),
    \label{eq:s3_stage1_factorization}
\end{equation}
where $\pi_{\alpha}$ is a conditional distribution. 
If the re-encoding criteria are satisfied, a boundary configuration associated with a high-dimensional topology scaffold is re-encoded as the corresponding low-dimensional topology scaffold.
Otherwise, the structure is labeled as \texttt{boundary-ambiguous}.
The resulting canonicalized topology scaffolds serve as the training targets for Stage I.

%
% Accordingly, Stage I models the conditional distribution over the canonicalized strata through the following hierarchical factorization:
% %
% \begin{equation}
%     \pi_\alpha(\tau\mid y)
%     =
%     \pi_\alpha(G\mid y)\,
%     \pi_\alpha(K\mid G,y)\,
%     \pi_\alpha(w_{1:K}\mid G,K,y),
%     \label{eq:s3_stage1_factorization}
% \end{equation}
% %
% where the ordered combination $G\rightarrow K\rightarrow w_{1:K}$ uniquely identifies a Wyckoff stratum in the stratified Wyckoff parameter space and implicitly determines its effective dimension Drep.

The MLP learns the mapping from the input composition to the topology features used to predict \(G\), \(K\), and \(\omega_{1:K}\). The auxiliary route helps separate different topology scaffolds in the feature space and improves scaffold predicted ability.
%
% Based on Eq.~(25), Stage I predicts and assigns one of three labels: \texttt{Normal, Special, boundary-ambiguous}, to the topology scaffold by a boundary head $p$:
% \begin{equation}
%     p(\texttt{Normal, Special, boundary-ambiguous} \mid y, \tau)
% \end{equation}
% %

\subsubsection{Loss function of Stage I}

Based on the architecture of Stage I shown in Fig.~\ref{fig:S3.2}, the overall loss function of stage I ($\mathcal L_{\mathrm I}$) can be determined as:
\begin{equation}
    \mathcal L_{\mathrm I}
    =
    \mathcal L_G
    +
    \mathcal L_K
    +
    \mathcal L_w
    +
    \lambda_{\mathrm{state}}
    \mathcal L_{\mathrm{state}}
    +
    \lambda_{\mathrm{spec}}
    \mathcal L_{\mathrm{spec}}
    +
    \lambda_{\mathrm{amb}}
    \mathcal L_{\mathrm{amb}}
    \label{eq:s3_stage1_loss}
\end{equation}
where $\mathcal{L_\mathrm{state}}$ is the model explicitly distinguishes between the three states, $\mathcal L_{\mathrm{spec}}$ is used for mapping to low-dimensional topology scaffolds, $\mathcal L_{\mathrm{amb}}$ handles boundary structures.

\subsection{Stage II: Chemical occupancy module (COM)}
\label{subsec:s3-stage2-sam}

\subsubsection{Architecture of Stage II}
\label{subsubsec:s3-stage2-arch}
Stage II is a Chemical Occupancy Module (COM) that assigns chemical species $s_{1:K}$ to the Wyckoff orbits of the specific topology scaffold $\tau$.
This assignment is designed to satisfy the target composition, the Wyckoff multiplicities ($m_k$), and basic chemical priors.
%
%Given the topology-only scaffold $\tau=(G,K,w_{1:K})$ predicted by Stage I, COM assigns a species sequence $s_{1:K}$ to the $K$ symmetry-independent Wyckoff orbits. 
%
Once COM is complete, the full scaffold, $c$, as defined in Eq.~\eqref{eq:scaffold} is obtained.
The completed scaffold $c$ is then passed to Stage III, where it identifies the corresponding Wyckoff fiber $\mathcal{F}_c$ and serves as the basis for generating the geometric features of the crystal.

\begin{figure}[h]
    \centering
    \includegraphics[width=0.9\textwidth]{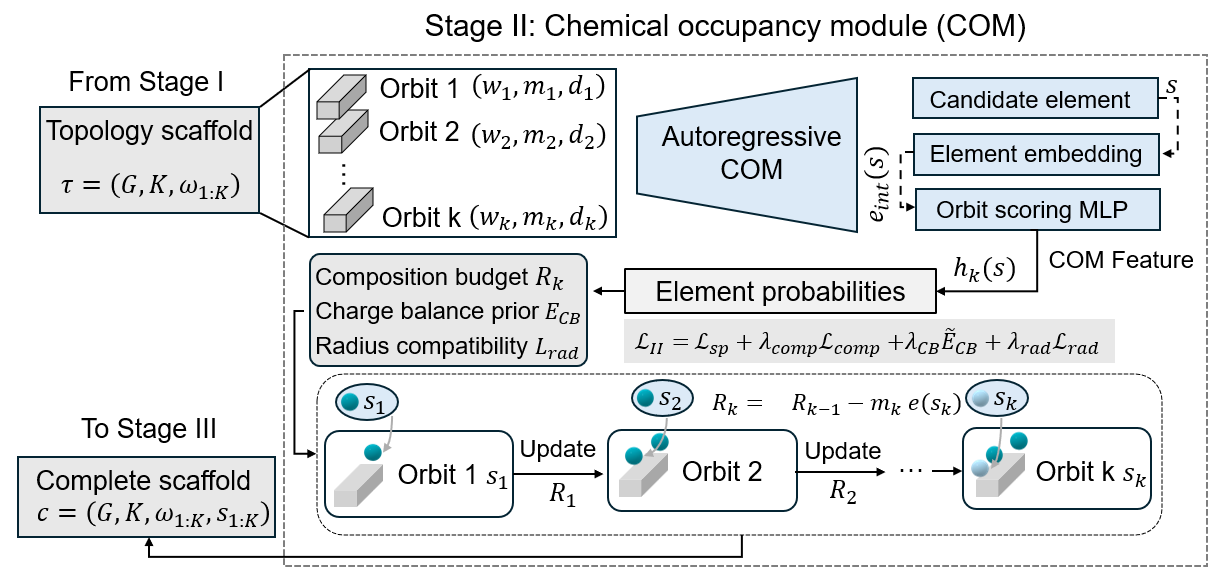}
    \caption{Workflow of {\model} Stage II for chemical occupancy module (COM). 
    This stage is designed to assign the chemical species at each Wyckoff orbit on the top of the scaffold.
    }
    \label{fig:S3.3}
\end{figure}

Fig.~\ref{fig:S3.3} illustrates the detailed architecture of Stage II for COM.
Conditioned on the topology-only scaffold \(\tau=(G,K,w_{1:K})\) predicted by Stage I,
COM assigns $s_{1:K}$ to the Wyckoff orbits in an autoregressive manner, thereby constructing the full scaffold \(c=(\tau,s_{1:K})\).
Accordingly, the conditional probability distribution learned by COM ($p_{\phi}$) is factorized as:
\begin{equation}
    p_\phi(s_{1:K}\mid \tau,y)
    =
    \prod_{k=1}^{K}
    p_\phi(s_k\mid s_{1:k-1},\tau,y,R_{k-1}),
    \label{eq:s3_sam_factorization}
\end{equation}
where $R_{k-1}$ is the residual composition budget after assigning the first $k-1$ orbits. 
If Wyckoff orbit $k$ has multiplicity $m_k$, assigning species $s_k$ contributes $m_k$ atoms of that element to the unit-cell composition.

To determine the chemical occupancy of each Wyckoff orbit, COM integrates three groups of information, including ($i$) global composition feature $y$, ($ii$) topology features, such as $G$, $K$, $w_{1:K}$, $m_k$, and orbit degrees of freedom, and ($iii$) autoregressive history and residual budget $R_{k-1}$.
For each candidate chemical species ($s$), it can be represented by a hybrid element embedding: 
\begin{equation}
    e_{\mathrm{int}}(s)
    =
    [
    e_{\mathrm{learn}}(s)
    \parallel
    f_{\mathrm{phys}}(s)
    ],
    \label{eq:s3_element_embedding}
\end{equation}
where $e_{\mathrm{learn}}(s)$ is a a trainable element embedding learned during Stage-II training, $f_{\mathrm{phys}}(s)$ contains fixed chemical descriptors such as periodic-table group, electronegativity, common oxidation states, ionic or covalent radius, and valence information. 
These descriptors are used as statistical chemical priors that guide the occupancy assignment, rather than imposing explicit quantum-mechanical constraints.

\subsubsection{Loss functions of Stage II}
\label{subsubsec:s3-stage2-loss}
Based on the COM architecture shown in Fig.~\ref{fig:S3.3}, the overall loss function of COM is defined as:
\begin{equation}
    \mathcal L_{\mathrm{II}}
    =
    %-\log p_\phi(s_{1:K}\mid \tau,y)
    \mathcal L_{\mathrm{sp}}
    +
    \lambda_{\mathrm{comp}}\mathcal L_{\mathrm{comp}}
    +
    \lambda_{\mathrm{CB}}\widetilde E_{\mathrm{CB}}
    +
    \lambda_{\mathrm{rad}}
    %\sum_{k=1}^{K}
    %E_{\mathrm{rad}}(s_k,w_k).
    \mathcal L_{\mathrm{rad}}
    \label{eq:s3_sam_loss}
\end{equation}
where the first term $\mathcal L_{\mathrm{sp}}$ is the species-assignment negative log-likelihood, which can be determined by $\mathcal L_{\mathrm{sp}} = -\log p_\phi(s_{1:K}\mid \tau,y)$;
the second term $\mathcal L_{\mathrm{comp}}$ is penalizes violations;
the third term $\lambda_{\mathrm{CB}}\widetilde E_{\mathrm{CB}}$ is a relaxed charge-balance penalty;
and last term $\lambda_{\mathrm{rad}}\mathcal L_{\mathrm{rad}}$ is a weak radius-compatibility penalty, which is determined by $\mathcal L_{\mathrm{rad}} = \sum_{k=1}^{K}E_{\mathrm{rad}}(s_k,w_k)$.

\subsubsection{Chemical constrains}
\label{subsubsec:s3-stage2-constrains}

The charge-balance prior is based on:
\begin{equation}
    \sum_{k=1}^{K}m_k q_k=0,
    \label{eq:s3_charge_balance}
\end{equation}
where $q_k$ is an oxidation state assigned to species $s_k$. This should not be treated as a universal hard rule. It is useful for many ionic compounds, but it may not apply cleanly to metals, intermetallics, or strongly covalent systems. We therefore use the relaxed form:
\begin{equation}
    \widetilde E_{\mathrm{CB}}
    =
    \min_{\tilde q_1,\ldots,\tilde q_K}
    \left(
    \sum_{k=1}^{K}m_k\tilde q_k
    \right)^2
    +
    \lambda_{\mathrm{ox}}
    \sum_{k=1}^{K}
    \min_{q\in \mathrm{OxStates}(s_k)}
    (\tilde q_k-q)^2.
    \label{eq:s3_charge_penalty}
\end{equation}
Here $\tilde q_k$ is a continuous relaxed oxidation state and $\mathrm{OxStates}(s_k)$ is the candidate oxidation-state set for element $s_k$. The weight $\lambda_{\mathrm{CB}}$ can be reduced or set to zero for systems where oxidation-state models are not appropriate.

% The radius term $E_{\mathrm{rad}}(s_k,w_k)$ is also introduced as a weak statistical prior. Since the geometry has not yet been generated at Stage II, this term should not be interpreted as a bond-length constraint. Rather, it encourages the model to learn whether elements of certain sizes are statistically more compatible with particular Wyckoff-site contexts.

\subsection{Stage III: Structured Wyckoff-aware generation (SWG)}
\label{subsec:s3-stage3-vtcfm}

\subsubsection{Architecture of Stage III}
Stage III is responsible for generating the continuous Wyckoff parameters $\xi$, defined in Eq.~(\ref{eq:s3_main_factorization}), based on the complete scaffold $c$.
Once the discrete scaffold has been determined by Stage I and II, the corresponding Wyckoff fiber $\mathcal{F}_c$ is uniquely specified, and the remaining task is to generate the continuous parameter distribution associated with each scaffold lies on a smooth manifold, while different scaffolds correspond to different fibers with distinct topologies.
To model these heterogeneous continuous distribution efficiently, Stage III employs a share neural network whose latent space is conditioned by the discrete scaffold.

The continuous parameters $\xi$ are generated using conditional flow matching ~\cite{lipman2023flow,tong2023improving}, a generative framework that learns a continuous-time transport from a simple base distribution to the target data distribution by regressing a velocity field along interpolation paths.
In this framework, the flow model learns a time-dependent velocity field $v_\theta(\xi_t,t,c,y)$ that transports samples from the base distribution to the data distribution of continuous Wyckoff parameters associated with the fixed scaffold. 
The corresponding continuous dynamics are given by:
\begin{equation}
    \frac{d\xi_t}{dt}
=
v_\theta(\xi_t,t,c,y),
\qquad t\in[0,1],
\end{equation}
%
%where $c=(\tau,s_{1:K})$ is the full scaffold selected by Stages I and II, and \rev{$y$ is the input of {\model}}.
%
The scaffold $c$ conditions the velocity field by specifying the target Wyckoff fiber, while $\xi_t$ evolves only through the continuous lattice and orbit-coordinate variables.
As a result, Stage III is only used to generate continuous features of the crystal ($\ell$ and $x^{orb}$), while keep all the discrete features of a crystal fixed. 
%
%Instead, it is used only to generate the continuous geometry, including the normalized lattice variables (\(\ell\)) and the orbit-level free fractional coordinates (\(x^{\mathrm{orb}}\)), as discussed in \rev{Eq.~\eqref{eq:s3_xi}}. 
%
%In the current {\model} framework, \(\ell\) is a six-dimensional normalized lattice-parameter vector and \(x^{\mathrm{orb}}\) are represented in a padded form, because different scaffolds have different numbers of orbits and different numbers of active free coordinates.
%
The important modes included in Stage III are discussed below.

\begin{figure}[h]
    \centering
    \includegraphics[width=0.99\textwidth]{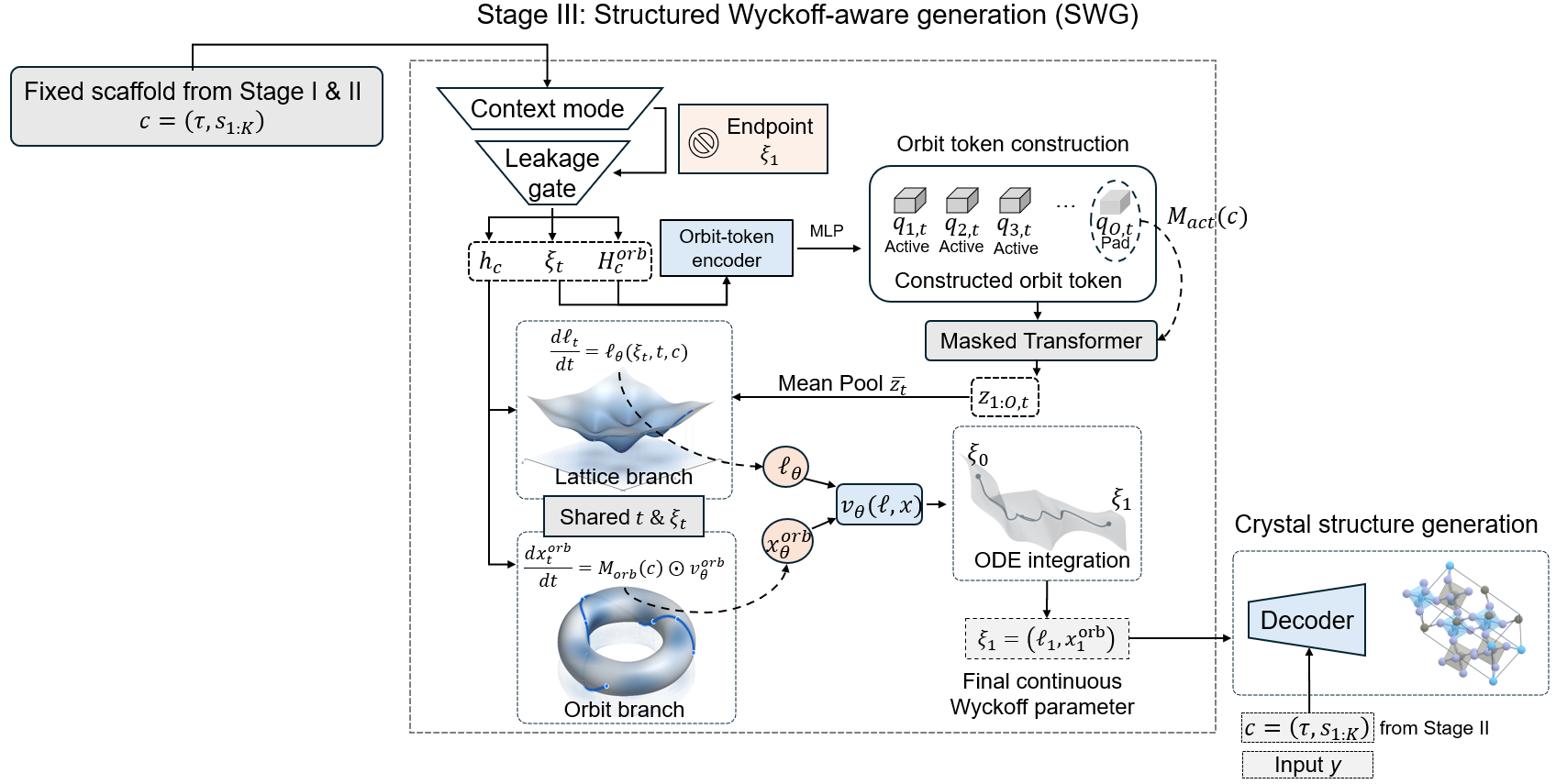}
    \caption{Workflow of {\model} Stage III for structured Wyckoff-aware generation (SWG) using flow matching generative model.
    }
    \label{fig:S3.4}
\end{figure}

Because different scaffolds contain different numbers of Wyckoff orbits and different numbers of free coordinates per orbit, the padded state representation contains many entries that do not correspond to real continuous variables. 
To enable a single neural network to process all scaffolds, Stage III introduces scaffold-specific masks that identify the valid degrees of freedom while ignoring padded entries. As a result, the same fixed-size tensor can represent crystals with different intrinsic dimensions.

Stage III employs two complementary masks. 
The orbit activity mask, ($M_{\mathrm{act}}(c)\in{0,1}^{O}$), identifies which padded orbit slots correspond to real Wyckoff orbits in scaffold (c). This mask is applied in the Transformer so that attention is computed only among active orbits, while padded or inactive orbits are completely ignored. 
The orbit coordinate mask, ($M_{\mathrm{orb}}(c)\in{0,1}^{O\times d_{\max}}$), specifies which coordinate dimensions are valid free parameters for each active orbit. During velocity prediction, this mask forces all invalid or padded coordinates to have zero velocity and excludes them from the training loss.

For example, a fixed Wyckoff position has no free coordinates and therefore an all-zero coordinate mask, whereas an orbit with one or three free coordinates activates only the corresponding entries. Consequently, different Wyckoff positions can be represented using the same padded tensor without introducing invalid degrees of freedom. Together, the activity and coordinate masks allow a shared Stage III architecture to operate efficiently across all scaffolds while ensuring that only physically meaningful continuous variables participate in prediction and optimization.

\subsubsection{Relation to the Wyckoff fiber}
\label{subsubsec:s3-fiber-relation}

For a fixed scaffold \(c\), the active part of \(\xi_t\) is a point in the padded coordinate chart of scaffold-specific fiber \(\mathcal W_0(c)\) at certain time $t$.
Throughout the ODE trajectory, Stage III remains confined to this fiber and does not move samples between different fibers in the total Wyckoff Space, \(\mathcal W(c)\). 
Therefore, the vector field only evolves on the continuous coordinates that are valid for the selected scaffold.

This design differs from atom-wise crystal generators that predict all atomic coordinates in the full unit cell~\cite{xie2021crystal,jiao2023crystal}. 
In contrast, Stage III in our {\model} framework predicts the independent orbit-level free coordinates ($x^{orb}$) and then reconstructs the full crystal using Wyckoff coordinate templates and space-group operations. 
As a result, crystal symmetry is imposed by the representation during generation, rather than recovered after unconstrained atom-wise generation.

\subsubsection{Decoding for crystal structure prediction}
\label{subsubsec:s3-output-decoding}

The output of Stage III is the final continuous parameter \(\xi_1=(\ell_1,x_1^{\mathrm{orb}})\), see Fig.~\ref{fig:S3.4}. 
Together with the scaffold \(c\) from Eq.~\eqref{eq:scaffold}, this parameter is decoded into a crystal structure. The lattice variables are mapped back to physical lattice parameters. The orbit-level free coordinates are inserted into the corresponding Wyckoff coordinate templates. Space-group operations then generate the full set of atoms in the unit cell.

Because the ODE is integrated with a finite number of steps, small numerical deviations can occur. Lattice projection and symmetry refinement are applied as numerical cleanup. These steps are not additional generative stages; they only enforce numerical consistency of the decoded structure.

\subsubsection{Loss functions of Stage III}
\label{subsubsec:s3-stage3-loss}
Stage III is trained with a flow-matching velocity regression loss, following the structured vector field in Fig.~\ref{fig:S3.4}. For a sampled flow time (t), the target velocity is decomposed into lattice and orbit-coordinate components:
\begin{equation}
    u=(u^\ell,u^{\mathrm{orb}})
\end{equation}
corresponding to the same decomposition $\xi=(\ell,x^{\mathrm{orb}})$. The overall Stage-III loss is
\begin{equation}
    \mathcal L_{\mathrm{III}}
    =
    \mathcal L_\ell
    +
    \mathcal L_{\mathrm{orb}}
    \label{eq:s3_stage3_loss_full}
\end{equation}
The reported runs use equal weight for the lattice and orbit terms. Thus no additional coefficient is placed between $\mathcal L_\ell$ and $\mathcal L_{\mathrm{orb}}$.
The lattice loss is
\begin{equation}
    \mathcal L_\ell
    =
    \left\|
    \dot\ell_\theta(\xi_t,t,c,y)
    -
    u^\ell
    \right\|_2^2.
    \label{eq:s3_lattice_loss}
\end{equation}
The orbit loss is masked:
\begin{equation}
    \mathcal L_{\mathrm{orb}}
    =
    \frac{
    \left\|
    M_{\mathrm{orb}}(c)\odot
    \left(
    \dot x_\theta^{\mathrm{orb}}(\xi_t,t,c,y)
    -
    u^{\mathrm{orb}}
    \right)
    \right\|_2^2
    }{
    \|M_{\mathrm{orb}}(c)\|_1+\epsilon
    }.
    \label{eq:s3_orbit_loss}
\end{equation}
Here $\epsilon=10^{-6}$. This normalization keeps the loss scale comparable across scaffolds with different numbers of active orbit degrees of freedom.

\subsection{Training objective and loss decomposition}
\label{subsec:s3-training}

Since {\model} adopts a hierarchical architecture (Fig.~\ref{fig:S3.1}), the three stages can be trained independently or jointly fine-tuned after pretraining.
As a result, the overall objective in Equation~\eqref{eq:s3_total_loss} of {$\mathcal{L}_\textrm{UFO}$} does not require every training run to backpropagate through all three stages at once.

For Stage I, the objective is to learn the topology selector $\pi_\alpha(\tau\mid y)$. It is trained with supervised classification losses for the space group, orbit count, and Wyckoff sequence, together with an auxiliary route-role loss.
For Stage II, the objective is to learn the chemical occupancy distribution $p_\phi(s_{1}\mid \tau,y)$. COM is trained with the species-assignment loss and auxiliary penalties enforcing composition consistency and weak chemical priors.
For Stage III, a training sample is $(c,\xi_1)$, where $\xi_1=(\ell_1,x_1^{\mathrm{orb}})$. We draw $t\sim U(0,1)$ and a base sample $\xi_0=(\ell_0,x_0^{\mathrm{orb}})$.

The lattice bridge uses a Gaussian base and Euclidean interpolation in normalized lattice space:
\begin{equation}
    \ell_0\sim \mathcal N(0,I),
    \qquad
    \ell_t=(1-t)\ell_0+t\ell_1,
    \qquad
    u^\ell=\ell_1-\ell_0.
    \label{eq:s3_lattice_bridge}
\end{equation}
This interpolation is an engineering choice in normalized lattice-parameter space. It is not claimed to be a geodesic on the full lattice manifold.

The orbit bridge uses a uniform base on the torus:
\begin{equation}
    x_0^{\mathrm{orb}}\sim U[0,1),
    \label{eq:s3_orbit_base}
\end{equation}
with wrapped displacement
\begin{equation}
    \Delta_{\mathrm{wrap}}(x_1,x_0)
    =
    \mathrm{wrap}_{[-1/2,1/2)}(x_1-x_0).
    \label{eq:s3_wrapped_delta}
\end{equation}
The interpolated orbit state and target velocity are
\begin{equation}
    x_t^{\mathrm{orb}}
    =
    \mathrm{wrap}_{[0,1)}
    \left(
    x_0^{\mathrm{orb}}
    +
    t\Delta_{\mathrm{wrap}}(x_1,x_0)
    \right),
    \qquad
    u^{\mathrm{orb}}=
    \Delta_{\mathrm{wrap}}(x_1,x_0).
    \label{eq:s3_orbit_bridge}
\end{equation}
This bridge is defined for free fractional coordinates. The discontinuity at the wrap boundary is a measure-zero issue and is not observed as a practical training problem.

\subsection{Sampling procedure}
\label{subsec:s3-sampling}

Given a composition-side condition $y$, {\model} samples in three stages.
First, Stage I samples or ranks topology-fixed scaffolds $\tau$ as:
\begin{equation}
    \tau\sim\pi_\alpha(\tau\mid y).
    \label{eq:s3_sample_tau}
\end{equation}
The auxiliary route role can be used for ranking or filtering, but it is not part of the scaffold.
Second, Stage II samples species assignments using Eq.~\eqref{eq:s3_sam_factorization}, which gives:
\begin{equation}
    s_k
    \sim
    p_\phi(s_k\mid s_{1:k-1},\tau,y,R_{k-1}).
    \label{eq:s3_sample_species}
\end{equation}
After assigning $s_k$, the residual budget is updated as:
\begin{equation}
    R_k
    =
    R_{k-1}
    -
    m_k e(s_k),
    \label{eq:s3_residual_update}
\end{equation}
where $e(s_k)$ is the one-hot count vector for element $s_k$. A hard composition mask can be used to prevent choices that make impossible.

After Stage II, the full scaffold $c=(\tau,s_{1:K})$ is fixed. Stage III initializes
\begin{equation}
    \ell_0\sim\mathcal N(0,I),
    \qquad
    x_0^{\mathrm{orb}}\sim U[0,1).
    \label{eq:s3_sampling_base}
\end{equation}
The ODE is then integrated as
\begin{equation}
    \frac{d\ell_t}{dt}
    =
    \dot\ell_\theta(\xi_t,t,c),
    \label{eq:s3_ode_lattice}
\end{equation}
\begin{equation}
    \frac{dx_t^{\mathrm{orb}}}{dt}
    =
    M_{\mathrm{orb}}(c)\odot
    \dot x_\theta^{\mathrm{orb}}(\xi_t,t,c).
    \label{eq:s3_ode_orbit}
\end{equation}
The current implementation uses 48-step explicit Euler integration. With $\Delta t=1/48$, the update is
\begin{equation}
    \ell_{t+\Delta t}
    =
    \ell_t+\Delta t\,\dot\ell_t,
    \label{eq:s3_euler_lattice}
\end{equation}
\begin{equation}
    x_{t+\Delta t}^{\mathrm{orb}}
    =
    \mathrm{wrap}_{[0,1)}
    \left(
    x_t^{\mathrm{orb}}
    +
    \Delta t\,
    M_{\mathrm{orb}}(c)\odot
    \dot x_t^{\mathrm{orb}}
    \right).
    \label{eq:s3_euler_orbit}
\end{equation}
The modulo-one wrapping in Eq.~\eqref{eq:s3_euler_orbit} is applied after every step. Inactive dimensions are kept fixed by the mask.

The final state $\xi_1=(\ell_1,x_1^{\mathrm{orb}})$ is decoded together with scaffold $c$ to form a crystal structure. Lattice projection and symmetry refinement may be applied as numerical cleanup. They are not additional generative stages.

% \subsection{Regime-aware base distribution for support-scarce fibers}
% \label{subsec:s3-regime-aware-base}

% The standard Stage~III base distribution uses a Gaussian lattice base and a uniform torus base for orbit coordinates. This works well when the scaffold has enough training support. For support-scarce high-dimensional fibers, we also use a regime-aware base distribution. It stays inside the same scaffold fiber but starts samples closer to high-density regions of the training support.

% For a fixed scaffold $c$, the regime-aware base is
% \begin{equation}
% p_0^c(\xi)
% =
% (1-\alpha_{\mathrm{mix}})p_{\mathrm{unif}}^c(\xi)
% +
% \alpha_{\mathrm{mix}}
% \frac{1}{k}
% \sum_{j=1}^{k}
% \mathcal N_{\mathbb T}
% (\xi;\mu_j^c,\sigma_{\mathrm{loc}}^2 I),
% \label{eq:s3-regime-aware-base}
% \end{equation}
% where $p_{\mathrm{unif}}^c$ is the default base on the active coordinates of $\mathcal W_0(c)$, $\mathcal N_{\mathbb T}$ denotes a wrapped local Gaussian on active torus coordinates combined with the lattice base, and $\mu_j^c$ are scaffold-specific anchor centers chosen from the training support. The current high-dimensional preset uses $k=5$, $\alpha_{\mathrm{mix}}=0.85$, and $\sigma_{\mathrm{loc}}=0.06$.

% This change does not break the correct-fiber guarantee. The mixture is defined only on the active coordinates of the selected scaffold $c$. It changes where sampling starts inside $\mathcal W_0(c)$, but it does not allow motion into another scaffold fiber.

\clearpage

\subsection{Training and sampling pseudocode}
\label{subsec:s3-pseudocode}

\begin{algorithm}[!htbp]
\caption{UFO-MGen training procedure}
\label{alg:s3_training}
\footnotesize
\begin{algorithmic}[1]
\Require training set $\mathcal D_{\mathrm{train}}$
\Ensure trained parameters $\alpha,\phi,\theta$

\State Encode each crystal as $(y,\tau,s_{1:K},c,\xi_1)$, where $c=(\tau,s_{1:K})$ and $\xi_1=(\ell_1,x_1^{\mathrm{orb}})$.

\For{mini-batches $\mathcal B\subset\mathcal D_{\mathrm{train}}$}
    \State Predict $(\hat G,\hat K,\hat\omega_{1:\hat K},\hat r)\gets f_\alpha(y)$.
    \State Compute $\mathcal L_{\mathrm I}$ using Eq.~\eqref{eq:s3_stage1_loss}.
    \State Update $\alpha$ using AdamW on $\mathcal L_{\mathrm I}$.
\EndFor

\For{mini-batches $\mathcal B\subset\mathcal D_{\mathrm{train}}$}
    \ForAll{$(\tau,y,s_{1:K})\in\mathcal B$}
        \State Initialize $R_0\gets R(y)$.
        \For{$k=1,\ldots,K$}
            \State Evaluate $p_\phi(s_k\mid s_{1:k-1},\tau,y,R_{k-1})$ with teacher forcing.
            \State Update $R_k\gets R_{k-1}-m_k e(s_k)$.
        \EndFor
    \EndFor
    \State Compute $\mathcal L_{\mathrm{II}}$ using Eq.~\eqref{eq:s3_sam_loss}.
    \State Update $\phi$ using AdamW on $\mathcal L_{\mathrm{II}}$.
\EndFor

\For{mini-batches $\mathcal B=\{(c_i,\xi_{i,1})\}_{i=1}^{B}$}
    \State Sample $t_i\sim U(0,1)$ and $\xi_{i,0}\sim p_0^{c_i}$.
    \State Build $\xi_{i,t}$ and target velocities using Eqs.~\eqref{eq:s3_lattice_bridge}--\eqref{eq:s3_orbit_bridge}.
    \State Evaluate $\hat v_i\gets v_\theta(\xi_{i,t},t_i,c_i)$.
    \State Compute $\mathcal L_{\mathrm{III}}$ using Eq.~\eqref{eq:s3_stage3_loss_full}.
    \State Update $\theta$ using AdamW on $\mathcal L_{\mathrm{III}}$.
\EndFor

\State \Return $\alpha,\phi,\theta$.
\end{algorithmic}
\normalsize
\end{algorithm}

\begin{algorithm}[!htbp]
\caption{UFO-MGen sampling procedure}
\label{alg:s3_sampling}
\footnotesize
\begin{algorithmic}[1]
\Require composition-side condition $y$, trained parameters $\alpha,\phi,\theta$
% number of Euler steps $T=48$
\Ensure generated crystal structure $\hat{\mathcal C}$

\State Sample or rank $\tau=(G,K,w_{1:K})\gets \mathrm{Select}(\pi_\alpha(\tau\mid y))$.
\State Initialize residual composition budget $R_0\gets R(y)$.

\For{$k=1,\ldots,K$}
    \State Sample $s_k\sim p_\phi(s_k\mid s_{1:k-1},\tau,y,R_{k-1})$.
    \State Update $R_k\gets R_{k-1}-m_k e(s_k)$.
\EndFor

\State Form the complete scaffold $c\gets(\tau,s_{1:K})$.
\State Sample $\ell_0\sim\mathcal N(0,I)$ and $x_0^{\mathrm{orb}}\sim U[0,1)$.
\State Set $\xi_0\gets(\ell_0,x_0^{\mathrm{orb}})$ and $\Delta t\gets 1/T$.

\For{$i=0,\ldots,T-1$}
    \State Set $t_i\gets i/T$.
    \State Evaluate $(\dot\ell_i,\dot x_i^{\mathrm{orb}})\gets v_\theta(\xi_{t_i},t_i,c)$.
    \State Update $\ell_{t_{i+1}}\gets \ell_{t_i}+\Delta t\,\dot\ell_i$.
    \State Update $x_{t_{i+1}}^{\mathrm{orb}}\gets \mathrm{wrap}_{[0,1)}(x_{t_i}^{\mathrm{orb}}+\Delta t\,M_{\mathrm{orb}}(c)\odot\dot x_i^{\mathrm{orb}})$.
    \State Set $\xi_{t_{i+1}}\gets(\ell_{t_{i+1}},x_{t_{i+1}}^{\mathrm{orb}})$.
\EndFor

\State Decode $\hat{\mathcal C}\gets\mathrm{Decode}(c,\xi_1)$.
\State Apply lattice projection and symmetry refinement as numerical cleanup.
\State \Return $\hat{\mathcal C}$.
\end{algorithmic}
\normalsize
\end{algorithm}

\FloatBarrier

\section{DFT Validation of {\model} Prediction}
\label{sec:s4-validation}
First-principles density functional theory (DFT) calculations are performed to validate the predictive performance of {\model}.
All DFT calculations are performed using Vienna Ab initio Simulation Package (VASP)~\cite{kresse1996efficiency, kresse1993ab}.
The projector augmented-wave (PAW) method was employed together with the Perdew--Burke--Ernzerhof (PBE) exchange--correlation functional within the generalized gradient approximation (GGA)~\cite{blochl1994projector,grimme2006semiempirical}. 
A plane-wave energy cutoff of 500~eV is employed for all DFT calculations, together with a dense $\Gamma$-centered $k$-point mesh to ensure numerical convergence. 
%s adopted to ensure the convergence of DFT calculations.
%
%
Full structural optimization are performed by relaxing the lattice parameters and atomic positions until the total energy and atomic force satisfy the specific convergence criteria of $1 \times 10^{-5}$ eV and $0.1$ eV~\AA$^{-1}$, respectively. 
Using the DFT-optimized structures, physical stability calculations are subsequently carried out for the generated crystal structures from both the extrapolation and interpolation groups following the multi-stability evaluation (MSE) framework discussed in main-text Section 2.3.
%are performed for both extrapolation and interpolation group.
%

%
% Even though the novel crystal structures generated by {\model} can satisfy the target space group and Wyckoff scaffold requirements, they may not be the most stable structures due to kinetic trapping at their local minimum. 
% %
% In addition, these generated structures may be lattice-dynamically unstable or undergo phase transformation at finite temperatures. 
% %
% To rigorously assess the stability of the generated materials, we perform a comprehensive stability evaluation using stringent criteria, including thermodynamic, lattice-dynamic, mechanical, and thermal stability.
% %
% The detailed validation tests will be discussed as follows.

\subsection{DFT structural optimization and thermodynamic stability}
\label{subsec:s8-relaxation}
We first validate the generated candidates from both the interpolation and extrapolation groups presented in the main text by comparing their lattice parameters and formation of energy ($\Delta E_f$).
Table~\ref{tab:structure_validation} summaries the DFT-optimized lattice parameters and $\Delta E_f$ with the corresponding {\model} predictions for the six representative material systems shown in main-text Fig. 2(f).

\begin{table}[htbp]
\centering
\caption{DFT-calculated formation energy ($\Delta E_f$) and optimized lattice parameters ($a, b, c, \alpha, \beta, \gamma$) of the six representative crystals shown in main-text Fig. 2(f), compared with the corresponding {\model} predictions.}
\label{tab:structure_validation}
\begin{tabular}{lccccccc}
\toprule
Materials & $\Delta E_f$ & $a$ (\AA) & $b$ (\AA) & $c$ (\AA) & $\alpha$ ($^\circ$) & $\beta$ ($^\circ$) & $\gamma$ ($^\circ$) \\
\midrule
$\textrm{ScGa}_3$ ({\model}) & -0.516 & 4.12 & 4.12 & 4.12 & 90.00 & 90.00 & 90.00 \\
ScGa$_3$ (DFT)      & -0.492 & 4.13 & 4.13 & 4.13 & 90.00 & 90.00 & 90.00 \\
YGa$_3$ ({\model})  & -0.559 & 6.26 & 6.26 & 4.62 & 90.00 & 90.00 & 120.00 \\
YGa$_3$ (DFT)       & -0.545 & 6.27 & 6.27 & 4.59 & 90.00 & 90.00 & 120.00 \\
KLiSe ({\model})  & -1.213 & 4.53 & 4.53 & 7.30 & 90.00 & 90.01 & 90.00 \\
KLiSe (DFT)       & -1.221 & 4.53 & 4.53 & 7.29 & 90.00 & 90.00 & 90.00 \\
Nb$_3$Si ({\model}) & -0.438 & 5.13 & 5.13 & 5.13 & 90.00 & 90.00 & 90.00 \\
Nb$_3$Si (DFT)      & -0.361 & 5.11 & 5.11 & 5.11 & 90.00 & 90.00 & 90.00 \\
Pd$_4$Se ({\model}) & -0.121 & 5.33 & 5.33 & 5.69 & 90.00 & 90.00 & 90.00 \\
Pd$_4$Se (DFT)      & -0.143 & 5.33 & 5.33 & 5.70 & 90.00 & 90.00 & 90.00 \\
AlPS$_4$ ({\model}) & -0.603 & 5.73 & 5.73 & 10.55 & 90.01 & 89.99 & 90.00 \\
AlPS$_4$ (DFT)      & -0.597 & 5.72 & 5.72 & 10.46 & 90.00 & 90.00 & 90.00 \\
\bottomrule
\end{tabular}
\end{table}

Based on Table~\ref{tab:structure_validation}, the DFT-calculated $\Delta E_f$ and optimized lattice parameters show excellent agreement with those of the original structures generated by {\model}, indicating that the generated crystals are not only thermodynamically stable, but also already close to their locally optimized configurations, requiring only minimal structural relaxation.
This agreement also demonstrates the high structural accuracy and effectiveness of the {\model} generation process.
%
%This consistency further demonstrates the reliability of uMLIPs for screening thermodynamically stable structures and supports their use in the MSE screening framework described in \rev{Section XXX of the main text}.

\subsection{DFT validation of lattice-dynamic stability}
\label{subsec:s4-phonon}
To validate the second MSE criteria, namely the lattice-dynamical stability of the generated crystals, DFT-based phonon calculations were carried out using the finite-difference method as implemented in the Phonopy package~\cite{togo2023jpcm, togo2015scr, togo2023phonopy}.
The energy cutoff, energy convergence, and force convergence criteria were set to the same values as those used in structural optimizations.
%
%The phonon dispersions were sampled along the reciprocal-space path $\Gamma$--X--M--$\Gamma$--Z--R--A--Z, corresponding to the fractional reciprocal coordinates $\Gamma$ $(0,0,0)$, X $(1/2,0,0)$, M $(1/2,1/2,0)$, Z $(0,0,1/2)$, R $(1/2,0,1/2)$, and A $(1/2,1/2,1/2)$.
%
The high-symmetry points and corresponding reciprocal-space paths for the phonon dispersions were automatically determined for each crystal structure using SeeK-path~\cite{Hinuma2017}.
Atomic displacements in $2\times2\times2$ supercells were used for the phonon calculations.
The calculated phonon spectra and corresponding density of state (DOS) are used to assess the lattice-dynamic stability of the generated crystals.
The structures with ``clean'' phonon spectrum or only minor imaginary models are considered lattice-dynamically stable. 
Here, we define minor imaginary modes as those for which the integrated phonon DOS in the negative region is less than 0.05 phonon modes.
%as the criteria for the tiny imaginary modes.
%) are considered as stable configurations.

\begin{figure}[h]
    \centering
    \includegraphics[width=0.7\textwidth]{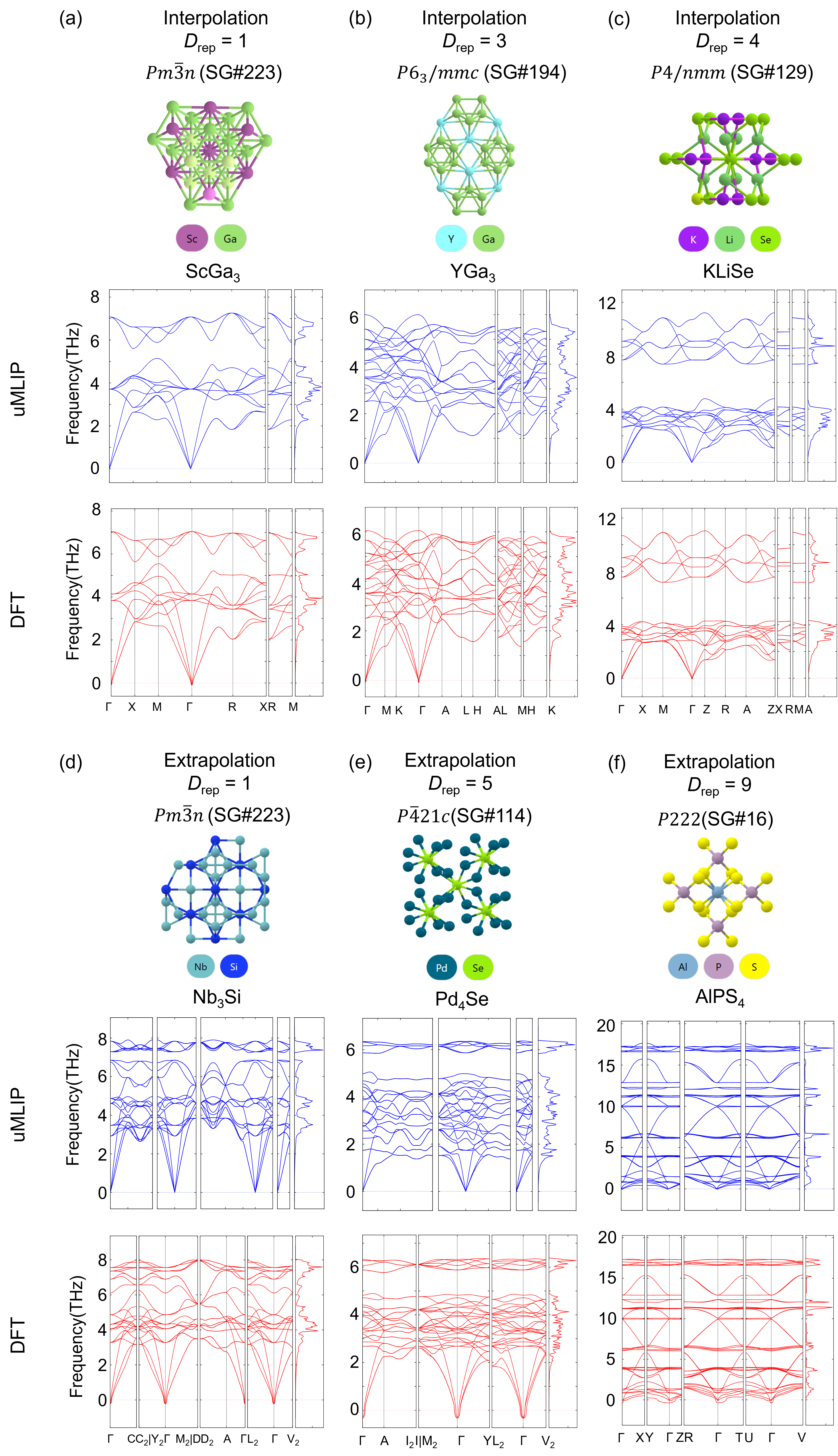}
    \caption{DFT validation of phonon dispersion relations for the six representative crystal structures from {\model}-predicted interpolation and extrapolation groups.
    }
    \label{fig:S4.1}
\end{figure}

Figure~\ref{fig:S4.1} presents the phonon dispersion relations of six representative crystal structures generated by {\model}, selected from both the interpolation and extrapolation groups.
The DFT-calculated phonon dispersions show excellent agreement with the corresponding uMLIP predictions, with no imaginary phonon frequencies observed for these structures.
These results confirm the lattice-dynamic stability of the {\model}-generated crystals and also validate the reliability of the uMLIP for predicting the phonon properties of the generated materials.

\subsection{AIMD validation of thermal stability}
\label{subsec:s4-aimd}
\begin{figure}[h]
    \centering
    \includegraphics[width=0.82\textwidth]{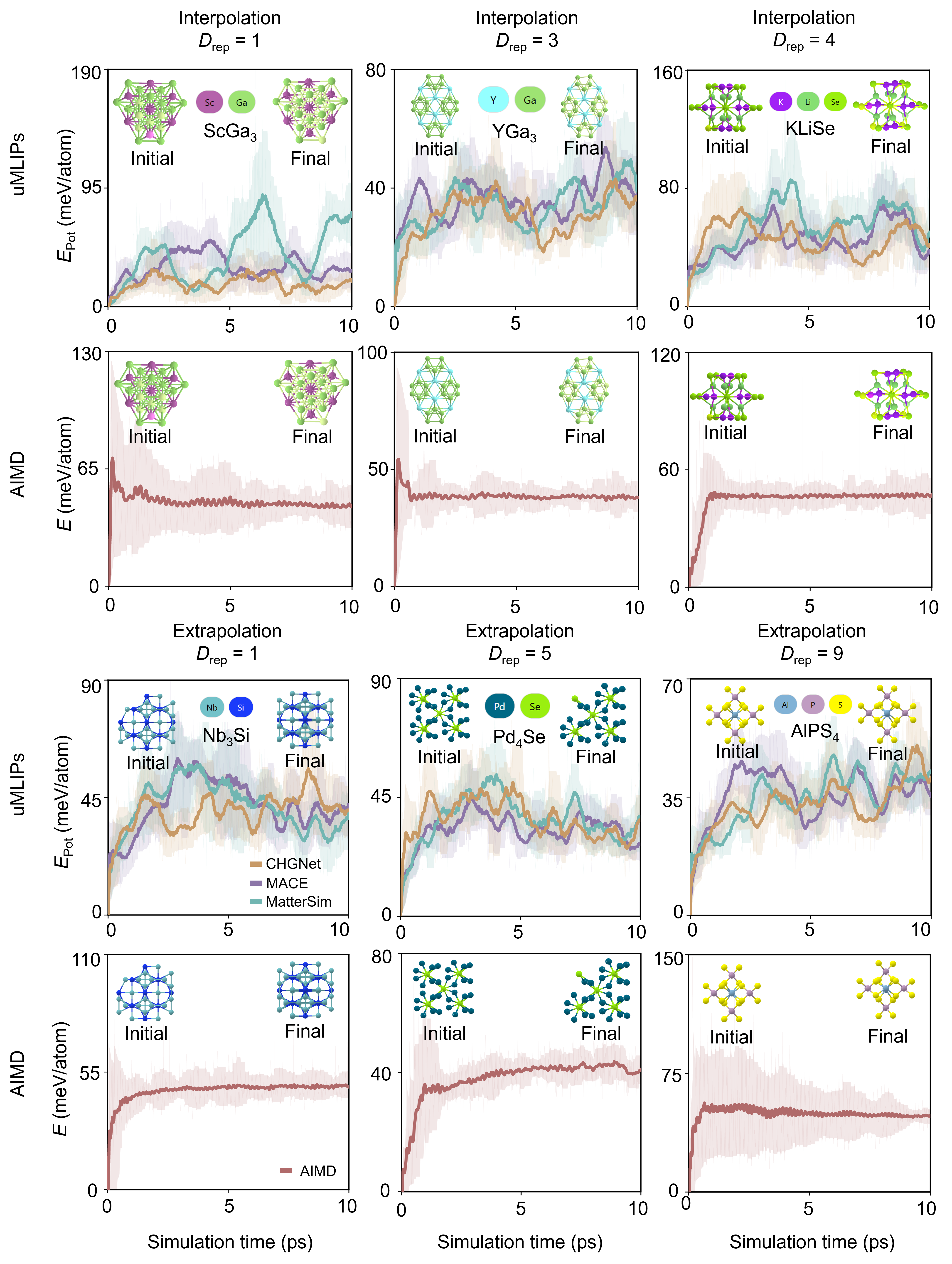}
    \caption{AIMD validation of thermal stability of six representative crystal structures from {\model}-predicted interpolation and extrapolation groups, respectively.
    }
    \label{fig:S4.2}
\end{figure}

For the final MSE criteria, we further \textit{ab initio} molecular dynamics (AIMD) simulations to validate the thermal stability of all six representative crystals at 300~K.
All AIMD simulations were performed under the NVT ensemble for 10 ps, consistent with the uMLIP-based MD simulations.
Fig.~\ref{fig:S4.2} shows that the energy profiles remain stable throughout the AIMD simulations, with no structural instabilities or phase transitions observed for any of the representative crystals.
Moreover, MD simulations using three uMLIPs for the same 10~ps duration yield consistent results, further supporting the thermal stability of these structures.
Together, the AIMD and uMLIP-based MD results demonstrate that the {\model}-generated crystals remain structurally stable at finite temperature.

\clearpage

\subsection{DFT validation of additional generated crystals}
\label{subsec:s4-others}
In addition to the six representative generated structures shown in main-text Fig. 2(f), we performed extensive validation on additional crystal structures for further validate the predictive capability of {\model} and associated uMLIP prediction accuracy.
%for demonstrating the high accuracy and efficiency of our MSE screening framework.
%
%Candidates that pass single-point energetic screening are first subjected to a short structural relaxation of up to 50 steps MLIP models. 
%
% This step aims to check whether the generated structures can be locally relaxed without any collapses or large structural changes. 
% %
% Successful candidates are then subjected to longer structural relaxations of 500 to 800 steps to further reduce the total energy and residual stress in the cell. 
% %
% \rev{[one sentence to discuss what is the energy and force convergence criteria used for MLIP]}. 
% %
% For the fully optimized structures, the formation energy calculation ($\Delta E_f$) is performed. Only the generated crystals with negative $\Delta E_f$ are considered as thermodynamically stable structures.
For instance, Table~\ref{tab:othercrystals} compares the DFT-optimized lattice parameters of additional 21 crystal structures with the corresponding {\model} predictions.
The perfect agreement between the DFT calculations and {\model} predictions further demonstrates the high structural accuracy of the generated crystal structures.

\begin{table}[htbp]
\centering
\caption{DFT-calculated optimized lattice parameters ($a, b, c, \alpha, \beta, \gamma$) of additional 21 crystals, compared with the corresponding {\model} predictions.}
\label{tab:othercrystals}
\begin{tabular}{lccccccc}
\toprule
Formula  & $a$ (\AA) & $b$ (\AA) & $c$ (\AA) & $\alpha$ ($^\circ$) & $\beta$ ($^\circ$) & $\gamma$ ($^\circ$) \\
\midrule
$\textrm{CeSe}_2$ ({\model})    & 4.88 & 4.88 & 13.79 & 90.00 & 90.00 & 90.00 \\
$\textrm{CeSe}_2$ (DFT)         & 4.09 & 7.08 & 8.42  & 90.00 & 110.53 & 90.00 \\
TiAlV$_2$ ({\model})   & 4.44 & 4.44 & 6.06  & 90.00 & 90.00 & 90.00 \\
TiAlV$_2$ (DFT)        & 4.39 & 4.40 & 6.09  & 90.00 & 90.00 & 90.00 \\
GaAs ({\model})      & 4.07 & 4.07 & 5.76  & 89.99 & 90.00 & 90.00 \\
GaAs (DFT)           & 4.08 & 4.07 & 5.79  & 90.00 & 90.00 & 90.00 \\
Mo$_2$C ({\model})     & 4.76 & 6.04 & 5.21  & 90.00 & 90.00 & 90.00 \\
Mo$_2$C (DFT)          & 4.71 & 6.08 & 5.24  & 90.00 & 90.00 & 90.00 \\
Nb$_4$CoSi ({\model})  & 6.22 & 6.22 & 5.03  & 90.00 & 90.00 & 90.00 \\
Nb$_4$CoSi (DFT)       & 6.19 & 6.19 & 5.02  & 90.00 & 90.00 & 90.00 \\
KLiTe ({\model})     & 4.85 & 4.85 & 7.78  & 90.00 & 90.01 & 90.00 \\
KLiTe (DFT)          & 4.84 & 4.84 & 7.78  & 90.00 & 90.00 & 90.00 \\
Zr$_3$O ({\model})     & 5.65 & 5.65 & 5.21  & 90.00 & 90.00 & 120.01 \\
Zr$_3$O (DFT)          & 5.66 & 5.66 & 5.21  & 90.00 & 90.00 & 120.00 \\
LiScI$_3$ ({\model})   & 7.48 & 7.48 & 6.74  & 90.00 & 90.00 & 120.00 \\
LiScI$_3$ (DFT)        & 7.40 & 7.40 & 6.71  & 90.00 & 90.00 & 120.00 \\
NiAsSe ({\model})    & 7.41 & 5.86 & 4.81  & 90.00 & 90.00 & 90.00 \\
NiAsSe (DFT)         & 7.43 & 5.85 & 4.84  & 90.00 & 90.00 & 90.00 \\
V$_6$SiOs ({\model})   & 4.75 & 4.75 & 4.75  & 90.00 & 90.00 & 90.00 \\
V$_6$SiOs (DFT)        & 4.73 & 4.73 & 4.73  & 90.00 & 90.00 & 90.00 \\
CrN$_2$ ({\model})     & 4.75 & 4.75 & 4.75  & 90.00 & 90.00 & 90.00 \\
CrN$_2$ (DFT)          & 4.69 & 4.69 & 4.69  & 90.00 & 90.00 & 90.00 \\
BC$_3$ ({\model})        & 5.01 & 4.53 & 2.57 & 90.01 & 104.87 & 116.88 \\
BC$_3$ (DFT)             & 5.04 & 4.55 & 2.61 & 89.99 & 105.02 & 116.78 \\
MnV$_2$C$_3$ ({\model})  & 4.82 & 4.82 & 5.29 & 90.00 & 90.00 & 120.00 \\
MnV$_2$C$_3$ (DFT)       & 4.84 & 4.84 & 5.21 & 90.00 & 90.00 & 120.00 \\
MnMo$_2$C$_3$ ({\model}) & 4.96 & 4.96 & 5.57 & 90.00 & 90.00 & 120.00 \\
MnMo$_2$C$_3$ (DFT)      & 4.94 & 4.94 & 5.53 & 90.00 & 90.00 & 119.99 \\
VCo$_2$C$_3$ ({\model})  & 4.66 & 4.66 & 5.25 & 90.00 & 90.00 & 119.99 \\
VCo$_2$C$_3$ (DFT)       & 4.65 & 4.65 & 5.28 & 90.00 & 90.00 & 120.00 \\
V$_3$C$_3$N ({\model})   & 8.90 & 2.90 & 9.92 & 90.00 & 90.00 & 90.00 \\
V$_3$C$_3$N (DFT)        & 8.84 & 2.91 & 9.90 & 90.00 & 90.00 & 90.00 \\
V$_2$SiC$_2$ ({\model})  & 4.06 & 4.06 & 6.81 & 90.00 & 90.00 & 90.00 \\
V$_2$SiC$_2$ (DFT)       & 4.04 & 4.04 & 6.89 & 90.00 & 90.00 & 90.00 \\
Ti$_2$CrC$_2$N ({\model}) & 6.00 & 4.21 & 9.03 & 90.00 & 90.00 & 90.00 \\
Ti$_2$CrC$_2$N (DFT)      & 6.03 & 4.19 & 9.05 & 90.00 & 90.00 & 90.00 \\
VCrC$_2$ ({\model})      & 2.91 & 7.04 & 7.69 & 69.43 & 90.34 & 88.64 \\
VCrC$_2$ (DFT)           & 2.91 & 7.04 & 7.66 & 68.39 & 90.70 & 88.67 \\
TiNiC$_2$ ({\model})     & 3.29 & 3.29 & 8.11 & 90.00 & 90.00 & 90.00 \\
TiNiC$_2$ (DFT)          & 3.32 & 3.32 & 8.18 & 90.00 & 90.00 & 90.00 \\
Cr$_4$C$_3$ ({\model})   & 7.84 & 2.79 & 11.77 & 90.00 & 90.00 & 90.00 \\
Cr$_4$C$_3$ (DFT)        & 7.87 & 2.81 & 11.64 & 90.00 & 90.00 & 90.00 \\
\bottomrule
\end{tabular}
\end{table}

\begin{figure}[h]
    \centering
    \includegraphics[width=0.99\textwidth]{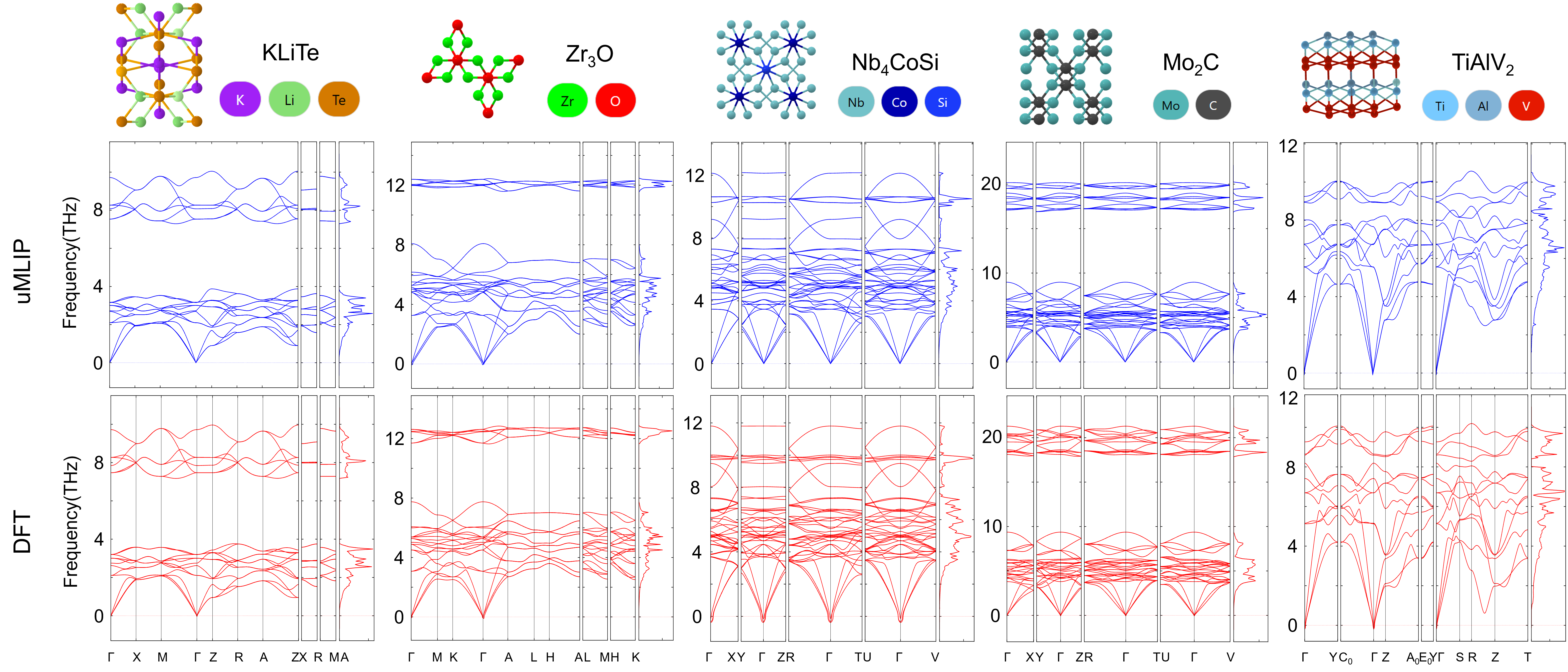}
    \caption{DFT validation of the phonon dispersion relations for an additional 5 crystal structures generated by {\model} model, compared with the corresponding uMLIP-calculated phonon spectra.
    }
    \label{fig:S4.3}
\end{figure}

% \subsection{Lattice-dynamical stability}
% \label{subsec:s8-phonon-dos}
%
Furthermore, extensive DFT-based phonon calculations were performed to validate the lattice-dynamic stability of the generated crystals.
%
%Among the \rev{above XXX structures, we selected XX } of them to calculate their phonon dispersion relationships. 
%
Fig.~\ref{fig:S4.3} shows that both DFT-calculated phonon spectra of 5 additional crystals exhibit no imaginary frequencies, consistent with the corresponding uMLIP-calculated spectra. 
These results confirm the lattice-dynamical stability of these generated crystals.
%predicted spectra predict clean phonon dispersion relations, suggesting the high lattice dynamic stability of these genreated crystals. 
%
In addition, the excellent agreement between the DFT and uMLIP results further demonstrate the high accuracy of uMLIPs in predicting lattice-dynamical stability, supporting the reliability and efficiency of the MSE framework proposed in this work.
%

%
% The generated structures that are thermodynamically stable ($\Delta E_f$ < 0) are next subjected to next level stability assessment, namely lattice-dynamical stability testing.
% %
% The lattice-dynamical stability is evaluated by calculating the phonon spectrum and density of states (DOS) of fully-optimized structures.
% %
% \rev{The phonon calculations are performed using finite difference method~\cite{togo2015scr}, and XXX supercell are adopted.}
% %
% If imaginary frequencies are observed in either phonon spectrum or DOS, the structures are lattice-dynamically unstable.
% %
% Only the generated structures with a clean phonon spectrum or a small/or most negligible imaginary frequencies are considered as lattice dynamically stable structures and will be chosen for further stability evaluation.

% \subsection{Thermal stability at elevated temperatures}
% \label{subsec:s8-chgnet-md}
%
% Although the prior sections evaluate the thermodynamic and lattice-dynamical stability of the crystal structures generated by {\model}, these evaluations are primarily performed at zero (0) K.
% %
% As such, some generated structures may trap at their local minima at low temperatures and may undergo structural transformations at elevated temperatures.
% %
% To assess their stability at elevated temperature, molecular dynamics (MD) simulations are performed in this section.
% %
% Because \textit{ab initio} MD simulations are very computationally expensive and time-consuming, all MD simulations are conducted using foundation MLIP models.
%

\begin{figure}[h]
    \centering
    \includegraphics[width=0.8\textwidth]{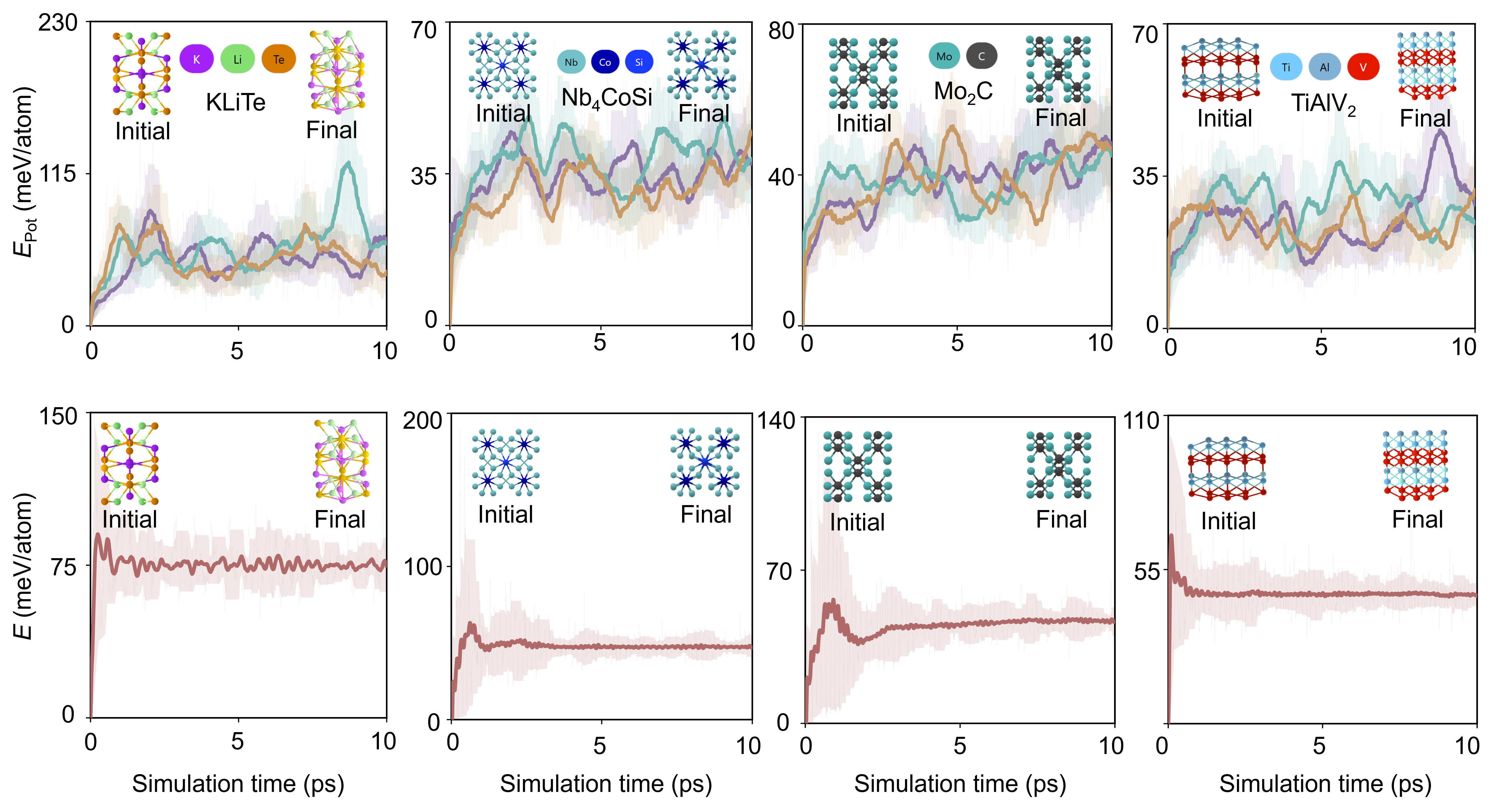}
    \caption{AIMD validation of the thermal stability for an additional 4 crystal structures generated by {\model} model, compared with the corresponding uMLIP-calculated energy profiles
    }
    \label{fig:S4.4}
\end{figure}

Finally, we performed AIMD simulations to further validate the thermal stability of other crystal structures. 
%The fully-optimized structures obtained at 0 K are used as the initial structure for the MD simulations.
%
As shown in Fig.~\ref{fig:S4.4}, the AIMD-simulated energy profiles remain stable thorough the 10 ps simulations, with no significant structural transitions or distortion observed.
This indicates that all examined crystals remain thermally stable at elevated temperatures, which is also consistent with uMLIP-based MD simulations. 
%during the MD simulations.
%
Consequently, these extensive DFT and AIMD validations demonstrate that the crystals generated by {\model} can successfully pass the rigorous MSE screening framework, highlighting the superior performance of {\model} in generating crystal structures.

\subsection{Comparison with Baseline Generative Models}
\label{subsubsec:s8-baseline-comparison}

Using the MSE criteria, we further evaluated the performance of baseline crystal generative models and compared their results with {\model}.
As described in the main text, {\model} was retrained on the MP20 dataset to ensure a fair comparison.
Each generative model was then used to generate 300 crystal structures, all of which were subjected to the same MSE screening framework based on the three stability criteria.
As shown in Table~\ref{tab:s8-baseline-comparison}, the success rates for each MSE criterion are summarized and compared across all models.
%
%\rev{A structure is considered stable when the structural relaxation is converged, 
%$\Delta E_f<0$, and the integrated imaginary phonon DOS satisfies 
%$I_{\mathrm{imag}}\leq 0.05$.}
%
\begin{table}[h]
\centering
\caption{Baseline comparison under the {\model}-MP20 downstream screening protocol. Screening-stage results are reported within the shortlisted candidates, while the final column reports stable candidates normalized per 10,000 generated structures.}
\label{tab:s8-baseline-comparison}
\begin{tabular}{lcccccc}
\toprule
Model & Shortlist & Thermodynamic & Lattice-dynamic & Thermal & overall MSE  \\
\midrule
CDVAE      & 300  & 291 (97.0\%)  & 148 (49.3\%) & 125 (84.5\%) & 41.7\% \\
DiffCSP    & 300  & 294 (98.0\%)  & 137 (45.7\%) & 114 (83.2\%) & 38.0\% \\
DiffCSP++  & 300  & 294 (98.0\%)  & 123 (41.0\%) & \textbf{116 (94.3\%)} & 38.7\% \\
MatterGen  & 300 & 294 (98.0\%)  & 144 (48.0\%) & 118 (81.9\%) & 39.3\% \\
{\model}     & 300  & \textbf{300 (100\%)} & \textbf{196 (65.3\%)} & 178 (90.9\%) & \textbf{59.3\%} \\
\bottomrule
\end{tabular}
\end{table}

{\model} consistently yields the largest number of stable candidates under each MSE criterion, achieving the highest overall success rate of 59.3\%.
In contrast, the other generative models exhibit substantially lower success rates, demonstrating the superior performance of {\model} in generating physically stable and valid crystal structures.

%this shared screening protocol. It produces 196 stable candidates from the final shortlist, corresponding to a stable ratio of 65.33\% and 233 stable candidates per 10,000 generated samples. This is higher than CDVAE, DiffCSP, DiffCSP++, and MatterGen under the same {\model}-MP20 downstream evaluation.

%%%%%%%% figure 4.3 %%%%%%%%%%%%%

%%%%%%%% figure 4.4 %%%%%%%%%%%%%

\clearpage
%

% ============================================================
% SECTION S5: Metrics
% ============================================================
\section{Evaluation Metrics}
\label{sec:s5-metrics}

\subsection{Public metrics}
In addition to validating the MSE criteria discussed in Section~\ref{sec:s4-validation}, we conducted another round of comprehensive evaluation of {\model} on the \textit{ab initio} crystal generation using established public metrics, including structural validity, compositional validity, coverage of the reference set, distributional agreement, stability--uniqueness--novelty (SUN), and geometric error~\cite{xie2021crystal,jiao2023crystal,zeni2023mattergen}. 
Unless stated otherwise, these metrics are computed with the public evaluator adopted by the corresponding benchmark.

For baseline methods, we report values from the original papers or from released evaluation scripts when a matched re-evaluation is not available. Metrics that are not reported, or not computed under the same protocol, are marked as ``/''. 
Higher values indicate better performance for validity, coverage, and SUN, whereas low values are preferred for distribution distances and root mean square derivation (RMSD).
The evaluation using these public metrics is conducted separately from the MSE validation and screening based on uMLIPs and DFT calculations.

\subsubsection{Validity, coverage and distribution metrics}
We evaluate generated crystals using validity, coverage, and distributional alignment metrics following the public benchmark protocols~\cite{xie2021crystal,jiao2023crystal}. All metric definitions and thresholds are kept consistent with the public evaluator, and we do not introduce additional validity or matching criteria in this work.

\textbf{Validity metrics.}
Structural validity is denoted as Val-Struct, measures the fraction of generated crystals that pass basic structural checks, including valid lattice parameters, positive cell volume, valid periodic coordinates, and the absence of severe atomic overlap. Compositional validity, denoted as Val-Comp, measures the fraction of generated crystals with chemically valid compositions under the benchmark rules, such as element validity, charge or valence sanity checks, and allowed stoichiometry.

\textbf{Coverage metrics.}
Coverage metrics evaluate whether the generated structures can recover the reference crystal distribution under the benchmark structure matcher and a distance threshold. We report both COV-R and COV-P. COV-R measures the fraction of reference structures covered by the generated set, while COV-P measures the fraction of generated structures that can be matched to the reference set.

\textbf{Distributional alignment metrics.}
We further compare generated and reference crystals using distributional distances computed over scalar or categorical properties of the full generated set, rather than over individual matched structures. Lower values indicate better agreement with the reference distribution. Specifically, we report the density distance $d_{\rho}$, the energy-related distance $d_E$, the elemental distance $d_{\mathrm{elem}}$, and the space-group distance $d_{\mathrm{sg}}$. The space-group distance is especially relevant for {\model}, since the model explicitly generates the symmetry scaffold before sampling continuous Wyckoff parameters.

\subsubsection{Stability, uniqueness and novelty (SUN) metric}
\label{subsec:s5-sun}

We report the SUN rate~\cite{zeni2023mattergen} as the fraction of generated structures that are
stable, unique and novel at the same time. This metric is more restrictive
than stability alone, since a structure must also be non-duplicated within
the generated set and absent from the reference database.

\textbf{Stability}: A generated structure is counted as stable if it satisfies the energetic-stability criterion used in the corresponding evaluation protocol. Because different crystal generative models may use different relaxation pipelines, energy models, and stability thresholds, we report SUN under the protocol aligned with MatterGen~\cite{zeni2023mattergen} for direct comparison. Specifically, energetic stability is evaluated after structural relaxation and is defined by the near-hull criterion:
\begin{equation}
\Delta E_{\mathrm{hull}} \leq 0.1~\mathrm{eV/atom},
\end{equation}
where $\Delta E_{\mathrm{hull}}$ is the energy above the convex hull. For baseline models, the reported SUN are taken from the published results rather than recomputed in our workflow. This choice avoids mixing different stability-evaluation pipelines and ensures that the comparison follows the same reported benchmark convention.
% Let $\mathcal{G}$ be the set of generated structures. We define three subsets:
% $\mathcal{G}_{\mathrm{S}}$ for stable structures,
% $\mathcal{G}_{\mathrm{U}}$ for unique structures, and
% $\mathcal{G}_{\mathrm{N}}$ for novel structures. The SUN rate is

% \begin{equation}
% \mathrm{S.U.N}
% =
% \frac{
% \left|
% \mathcal{G}_{\mathrm{S}}
% \cap
% \mathcal{G}_{\mathrm{U}}
% \cap
% \mathcal{G}_{\mathrm{N}}
% \right|
% }{
% |\mathcal{G}|
% }.
% \end{equation}

%A generated structure is counted as stable if it satisfies the energy criterion used in the evaluation protocol. In our current {\model} SUNevaluation, energetic stability is estimated after structural relaxation \rev{via several universal MLIPs including CHGNet, MACE [refs]?}, using following criteria:
%\begin{equation}
%\Delta E_{\mathrm{hull}} \leq 0.1~\mathrm{eV/atom}.
%\end{equation}
%where $\Delta E_{\mathrm{hull}}$ is the energy above hull.
%This threshold follows the common near-hull criterion used in recent crystal generation studies.
%
%When comparing against baselines, the energy model and relaxation protocol must be stated, since SUN values based on CHGNet, DFT, or another MLIPs are not directly interchangeable.

\textbf{Uniqueness}: A generated structure is counted as unique if it does not match any
earlier structure in the same generated set. Matching is performed with
the structure matcher used by the public evaluator. The standard
\texttt{pymatgen} \texttt{StructureMatcher}~\cite{ong2013pymatgen} is used, we use the default
tolerance setting.
%
%\begin{equation}
%\mathrm{ltol}=0.2,\qquad
%\mathrm{stol}=0.3,\qquad
%\mathrm{angle\_tol}=5^\circ ,
%\end{equation}
%Here, $\mathrm{ltol}$ is the fractional tolerance for matching lattice lengths, $\mathrm{stol}$ is the site-position tolerance for matching atomic coordinates after lattice normalization, and $\mathrm{angle_tol}$ is the tolerance for matching lattice angles. 

%
\textbf{Novelty}: A generated structure is counted as novel if it does not match any
structure in the training set or reference database under the same
structure matcher. In this work, novelty is evaluated after reducing the
structure to the same representation used by the benchmark matcher. 

The three conditions are applied jointly. Thus, a structure with low
energy is not counted as SUN if it is a duplicate of another generated
structure or if it matches a known training/reference structure. 
Because SUN depends on the energy estimator, relaxation settings, structure matcher and novelty database, we use it as a summary metric but report the underlying stable, unique and novel rates whenever possible.

\subsubsection{RMSD and relaxed RMSD}
\label{subsec:s5-rmsd}
Root mean square deviation (RMSD) measures the geometric difference between a generated crystal structure and a reference state by comparing their local atomic coordination. 
This metric can be defined as:
\begin{equation}
\mathrm{RMSD}
=
\sqrt{
\frac{1}{N}
\sum_{i=1}^{N}
\left\|
r_i - \hat{r}_{\pi(i)}
\right\|^2
},
\end{equation}
where $r_i$ is the Cartesian coordinate of atom $i$ in the reference structure, $\hat{r}_{\pi(i)}$ is the matched atom in the generated structure, and $\pi$ is the atom mapping from the structure matcher.
Due to ordered and periodic nature of crystal structures, RMSD must account for periodic boundary conditions (PBCs), local atom ordering, species matching, and possible cell choices. 
We therefore rely on the benchmark matcher rather than direct coordinate subtraction.

\subsubsection{Comparison with MP-20 baseline}
\label{subsec:s5-mp20-baseline}

Table~\ref{tab:s5_mp20_comparison} summarizes the public benchmark metrics of the baseline models and {\model} for the \textit{ab initio} crystal generations using MP20 database.
%public metrics for the \textit{ab initio} crystal generation task. 
All public metrics are evaluated based on 10,000 crystal structures generated by each model.
It clearly shows that the {\model} exhibits the best performance in both Validity and SUN rate among the 12 mainstream generative models.

% The current COV-P version has high SUN but weaker $d_E$ and $d_{\mathrm{elem}}$. This is not a contradiction. The selector shifts the pool toward structures that pass stability, uniqueness and novelty filters, and this can move the energy and element distributions away from the MP-20 training distribution. Therefore $d_E$ and SUN measure different behavior. In the final headline table, both rows should be reported, and the raw $n=10{,}000$ row should be regenerated with the final script.

\begin{table*}[htbp]
\centering
\caption{Comparison of crystal generation performance on MP-20 under public evaluation metrics. Values marked with $\dagger$ are literature-reported rows. Values marked with $\ddagger$ are {\model} rows from the current evaluation logs. A slash indicates that the metric is not yet available under the same protocol.}
\label{tab:s5_mp20_comparison}
\small
\setlength{\tabcolsep}{3.2pt}
\begin{adjustbox}{max width=\textwidth}
\begin{tabular}{lcccccccccc}
\toprule
Model 
& Val-Struct. 
& Val-Comp. 
& COV-R 
& COV-P 
& $d_{\rho} \downarrow$ 
& $d_E \downarrow$ 
& $d_{\mathrm{elem}} \downarrow$ 
& $d_{\mathrm{sg}} \downarrow$ 
& SUN 
& RMSD \\
\midrule

CDVAE~\cite{xie2021crystal}$^\dagger$
& 100.0& 86.70& 99.15& 99.49& 0.6875& 0.2778& 1.432& 0.69& 4.26~\cite{levy2025symmcd}& / \\

DiffCSP~\cite{jiao2023crystal}$^\dagger$
& 100.0& 83.25& 99.71& 99.76& 0.3502& 0.1247& 0.3398& /& 8.92~\cite{levy2025symmcd} 
& / \\

DiffCSP++~\cite{jiao2024diffcspplusplus}$^\dagger$
& 99.94& 85.12& 99.73& 99.59& 0.2351& 0.0574& 0.3749& /& 8.62~\cite{levy2025symmcd} 
& / \\

FlowMM~\cite{miller2024flowmm}$^\dagger$
& 96.85& 83.19& 99.49& 99.58& 0.239& 0.083& /& /& 6.49~\cite{levy2025symmcd} 
& / \\

SymmCD~\cite{levy2025symmcd}$^\dagger$
& 94.32& 85.85& 99.64& 98.87& 0.0901& 0.1166& 0.3990& 0.0899& 6.89~\cite{levy2025symmcd} 
& / \\

MatterGen~\cite{zeni2023mattergen}$^\dagger$
& $\mathbf{100.0}$ 
& $82.60$ 
& / 
& / 
& $0.2059$ 
& / 
& 0.2416& 0.4331& $22.0$ 
& $\mathbf{0.11}$ \\

SGEquiDiff~\cite{chang2025sgequidiff}$^\dagger$
& 99.81& 84.06& /
& /
& 0.6247& /
& 0.1988& 0.1769& 12.47& / \\

WyFormer~\cite{kazeev2025wyformer}$^\dagger$
& 99.56& 80.44& 98.67& 96.72& 0.74& 0.053& 0.097& 0.223& 6.9& / \\

OMatG~\cite{hoellmer2026omatg}$^\dagger$
& 99.64& 87.02& 99.39& 99.86& 0.0834& /
& 0.0784& /
& 18.58& 0.294\\

CrystalFlow~\cite{luo2024crystalflow}$^\dagger$
& 99.55& 81.96& 98.21& 99.84& 0.169& 0.259& /
& /
& 3.7& / \\

FlowLLM~\cite{sriram2024flowllm}$^\dagger$
& 99.94& 90.84& 96.95& 99.82& 1.14& /
& 0.15& /
& 4.92& 0.023\\

{\model} (MP20) $^\ddagger$
& $\mathbf{100.0}$ 
& $\mathbf{91.30}$ 
& $96.58$ 
& $99.45$ 
& $0.2049$ 
& $0.3214$
& $0.4754$ 
& $0.099$ 
& $\mathbf{31.3}$ 
& $0.29$ \\

% {\model} (Final) $^\ddagger$
% & $\mathbf{100.0}$ 
% & $\mathbf{91.30}$ 
% & $96.58$ 
% & $99.45$ 
% & $0.2049$ 
% & $0.3214$
% & $0.4754$ 
% & $0.099$ 
% & $\mathbf{31.3}$ 
% & $0.29$ \\

\bottomrule
\end{tabular}
\end{adjustbox}
\end{table*}

%\rev{\subsection{Multi-stability evaluated (MSE) metics}}

\subsection{Interpolation and extrapolation metrics}
Space groups are one of most fundamental descriptors of crystal structures, with a total of 230 distinct space groups. 
As discussed in main-text Section 2.4, examining whether generated crystals cover the space groups represented in the training data provides a measure of their ability to generate structures within the domain knowledge. 
%groups represented in the training data can reflect the ability to generate structures within the training distribution.
%
On the other hand, the ability to generate unprecedented crystal structures belonging to space groups absent from the training dataset represent an important measure of the models' capability of extrapolation. 
%is another key ability to predict innovative crystal structures. 
%
%rovided that the structure is valid. This is also a key metric for evaluating its exploratory and extrapolation capabilities.
%
As such, we introduce two metrics: interpolation rate and extrapolation rate, to evaluate the generative capability of models. 
%
%The former measures the proportion of valid generated crystals belonging to space groups present in the training set, while the latter measures the proportion belonging to space groups not found in the training set.

%
To evaluate these two capabilities, we used each generative model to generate 10,000 crystals and then calculate interpolation and extrapolation rates.
Since existing generative models were trained on the MP20 database, we also retrained {\model} using MP20 database and generated 10,000 crystals under the same setting for a fair comparison.
During retaining, we only adopted Stages II and III for crystal generation, as Stage I fixes the crystal scaffold and thus limits the exploration of space groups beyound those represented in the training dataset. 
Because the MP20 database covers 177 space groups, leaving 53 space groups absent from the training set, the interpolation and extrapolation rates can be easily determined by calculating the percentage of generated crystals belonging to space groups within and beyound the training set, respectively.
%space groups in each catogeory. 
%and thus we calculate the coverage of the space groups for both interpolation and extrapolation.
%

Table~\ref{SGcoverage} summarizes the interpolation and extrapolation rates of leading generative models and our {\model}.
%the space-group distribution of 10,000 crystals generated using models trained on the MP20 database. 
%
Once again, {\model} exhibits the highest coverage for both interpolation and extrapolation among all evaluated generative models.
These results further demonstrate the strong capability of {\model} to generate chemically and structurally diverse crystal structures within and beyond the space groups represented in the training domain.

% To maximize the {\model}'s extrapolation capabilities, {\model} based only on the generation of Stages II and III. The results show that, although some stability is sacrificed, the model expands the thermodynamic stability of the space group distribution of crystals beyond the 23 MP20 training datasets. Compared to models based on diffusion models, {\model} shows stronger scalability and exploratory potential.

% %
% To further test the extrapolation ability, we applied a stricter evaluation to {\model}, which was trained on the full MP database. 
% %
% Because of crystal complexity and the long-tail distribution of the data, only 35 space groups were included in the training set of {\model}. Therefore, we tested whether the same extrapolation module could overcome this data limit. 
% %
% As before, we used only Stages II and III for generation. For each of the other 195 space groups, we recorded how many crystals had to be generated before an emergent structure passed the MSE metrics. This stricter physics-based evaluation shows that the observed \texttt{emergence} is not caused by data noise or distribution shift. Instead, it indicates that a breakthrough result from the model's successful learning of the Wyckoff parameter space.

%
\begin{table}[htbp]
\centering
\caption{Model performance comparison on coverage and extrapolation rate.}
\label{SGcoverage}
\begin{tabular}{lcc}
\toprule
\textbf{Model} & \textbf{Interpolation rate (\%)} & \textbf{Extrapolation rate (\%)} \\
\midrule
DiffCSP++ & 81.4 & 3.8 \\
DiffCSP   & 57.6 & 3.8 \\
CDVAE     & 31.6 & 0.0 \\
MatterGen & 65.5 & 0.0 \\
UFO-MGen  & \textbf{85.5} & \textbf{45.3} \\
\bottomrule
\end{tabular}
\end{table}

\clearpage

% ============================================================
% SECTION 6: Ablations
% ============================================================

\section{Ablation Studies of {\model} Performance}
\label{sec:s6_ablation}
Ablation tests are commonly adopted to evaluate how individual components, layers, or features contribute to the overall performance of a machine-learning model. 
In main-text Fig. 4, we perform a series of ablation studies on revealing the key mechanisms underlying {\model}, where different types of database and diffusion-based models are adopted to systematically compare their crystal generation performance with {\model}.
In addition, we conduct another round of ablation study to understand which components of {\model} actually dominate its overall performance.
Rather than treating all ablation experiments as equally important, we categorize them according to the specific model components and capabilities they are designed to evaluate.
Since main performance of {\model} primarily comes from three components: Stage I, Stage II, and  Stage III, our ablation tests primarily focus on quantifying the contribution of each state to the overall model performance.

\subsection{Ablation test for Stage I: HTS}
\label{subsec:s8_stageI}
Stage I constrains the crystal topology scaffold prior to generation, thereby guiding the model toward more stable and physically plausible structures.
However, because Stage I is trained on datasets containing only 35 trainable space groups, it also restricts the generated crystals to these space groups and limits the model's ability to explore beyond the training domain.
Table~\ref{tab:ablation} summarizes the results obtained after removing Stage I.
Although the overall stability decreases, the model gains substantial extrapolation capability, enabling it to explore beyond the training domain and generate crystals across 52 previously unseen space groups that successfully pass the MSE screening framework.

\subsection{Ablation test for Stage II: COM}
\label{subsec:s8_stageII}
Stage II enables diverse chemical occupancies and elemental compositions, allowing {\model} to explore novel compositional spaces within physically reasonable crystal scaffolds.
When Stage II is removed, the generated distribution shifts toward more frequently occurring and inherently stable elemental assignments, making the generation process more deterministic and primarily focused on generating continuous structural features within familiar scaffolds.
Consequently, in terms of the physical stability of generated crystals, the ablated model exhibits a higher success rate than {\model}-Full.
However, further evaluation using the SUN metric (Table~\ref{tab:ablation}) reveals that, although most generated crystals are physically valid and stable, the ablated model tends to reproduce known structures rather than discover unique ones.
These results demonstrate that Stage II plays a critical role in expanding the chemical diversity and novelty of {\model}-generated crystals.

\subsection{Ablation test for Stage III: SWG}
\label{subsec:s8_stageIII}
Stage III determines whether a discrete scaffold design can be realized as a geometrically valid three-dimensional crystal structure.
Although the thermodynamic and lattice-dynamic stabilities of crystals generated without Stage III show only minor differences after structural relaxation, their reduced AIMD stability indicates that removing Stage III leads to less physically reasonable lattice parameters and atomic coordinates, resulting in atomic drift and eventual scaffold collapse.
Compared with {\model}-Full, {\model} without Stage III generates one stable structure that is classified as belonging to an unexplored space group (Table~\ref{tab:ablation}).
However, this structure does not represent true extrapolation, as it is assigned to a different space group only after minor structural changes during relaxation.
Therefore, the importance of Stage III is further reflected in its ability to generate geometrically valid structures that preserve their crystallographic symmetry during structural relaxation.

In conclusion, our ablation studies demonstrate that each component of {\model} plays a critical role in determining the overall model performance.
These results further highlight the importance of the hierarchical architecture of {\model} and demonstrate the effectiveness of its three-stage design.

%efficient architecture of {\model} and design strategy.
%
\begin{table}[htpb]
\centering
\caption{Ablation study of each stage in {\model}. 
The symbol ``$^\textbf{*}$'' denotes an extrapolated space group whose symmetry changes during structural optimization.}
\label{tab:ablation}
\resizebox{\linewidth}{!}{
\begin{tabular}{lccccc}
\toprule
Model Variant 
& Thermodynamic and lattice stability (\%) 
& Thermal stability (\%) 
& MSE (\%) 
& SUN (\%) 
& Extrapolated SG \\
\midrule
\textbf{UFO-MGen-Full} 
& 29.5 
& 93.3 
& 27.5 
& \textbf{20.3} 
& 0 \\

No Stage I& 28.7 
& 90.7 
& 26.0 
& 19.8 
& \textbf{52} \\

No Stage II& \textbf{32.0} 
& \textbf{95.8} 
& \textbf{30.7} 
& 6.5 
& 0 \\

No Stage III& 29.2 
& 89.4 
& 26.1 
& 18.8 
& 1$^\textbf{*}$ \\
\bottomrule
\end{tabular}
}
\end{table}
%

% \rev{\subsection{Ablation test for Diffusion and Flow matching}}
% \label{subsec:s8_diffvsFM}

\clearpage

\begin{comment}
\end{comment}

% ============================================================
% SECTION S7: Generation of Complex Crystalline Structures
% ============================================================

%%%%%%%%%%%%%%%%%%%%%%
%fine-tune
%%%%%%%%%%%%%%%%%%%%%%

\section{Inverse Materials Design}

\subsection{Fine-tuning architecture}
\begin{figure}
    \centering
    \includegraphics[width=0.8\linewidth]{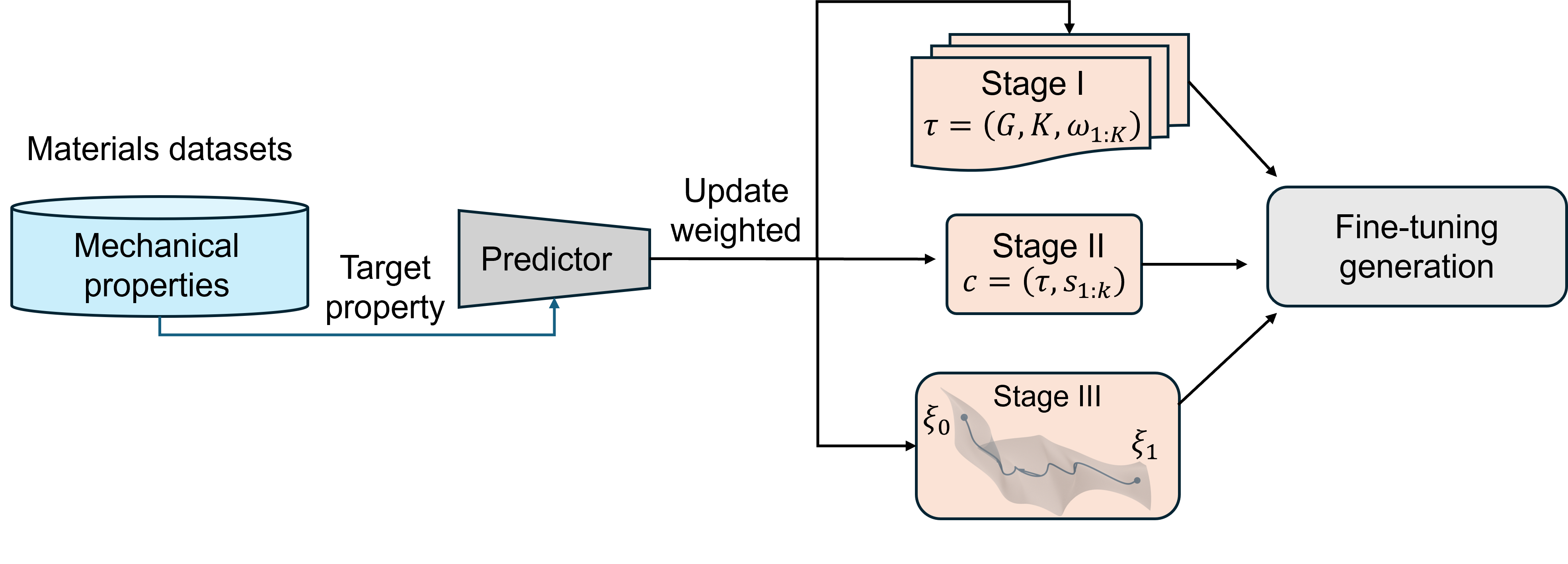}
    \caption{Workflow of fine-tuning architecture in {\model} for predicting materials properties of generated crystal structures with target mechanical properties .}
    \label{fig.s9.1}
\end{figure}

Property-constrained crystal generation is a critical step toward inverse materials design.
To enable this capability, we introduce property-guided tasks into {\model} and fine-tune all three stages toward targeted material properties.
Figure~\ref{fig.s9.1} illustrates the three-stage conditional decomposition of {\model}, in which property guidance is independently incorporated into the selection of crystal topology scaffolds, elemental occupancy, and continuous geometric degrees of freedom.

%
% Before the formal fine-tuning our three-stage {\model}, we used a extreme performance database \rev{(S8.2)} to train a property predictor using \texttt{SimpleImputer--ExtraTreesRegressor}.
% %
% % In the mechanical properties test set, the predictor achieved an MAE of 14.0 GPa for $K$, 11.6 GPa for $G$. 
% %
% The predictor is used only for fast scoring, ranking and generation guidance to accelerate FT-Module's learning of extreme performance, while final validation is confirmed by MatterSim and DFT.
%

%\paragraph{Property-weighted fine-tuning across three stages}
Specifically, property-based fine-tuning is applied throughout all three generation stages by assigning larger training weights to structures with stronger target properties.
In Stage~I, the model learns from predictor high score and the extreme scaffold distributions of the training set labels, guiding the {\model} to a extreme performance generative space. 
In Stage~II, property-weighted training guides element selection and orbit occupancy, while \texttt{COM-ALLOWED} further constrains the sampled chemical space to elements related to extreme properties. 
In Stage~III, the flow-matching objective is weighted by the same property scores, encouraging the generation of crystal geometries that better support the target performance. 

\subsection{Mechanical property datasets}
In this work, we select mechanical properties as the target material properties for fine-tuning {\model}, and we refer this specialized model as UFO-Mech.
The mechanical property predictor is trained on elastic data collected from multiple sources. 
The main dataset is obtained from JARVIS-DFT~\cite{choudhary2020jarvis}, where 8,753 cleaned samples are used for elastic-label training. 
Materials Project~\cite{jain2013materialsproject} elasticity adds 4,933 samples, and the de Jong Scientific Data 2015 dataset~\cite{de2015charting} provides 1,181 samples for external validation. 
After removing duplicates across databases, the final labeled dataset contains 13,686 crystals, which are split into 10,948 training samples, 1,369 validation, and 1,369 test samples. 
The prediction targets include bulk modulus, shear modulus, Young’s modulus, and Poisson’s ratios.
During training, samples exhibiting extreme mechanical properties are assigned higher weights to bias the model toward mechanically robust materials. 
The mechanical-extreme score, $S_\mathbf{mech}$, is defined as jointly considering high bulk modulus, shear modulus, and high Young's modulus:
\begin{equation}
S_\mathbf{mech}=z(log(\textrm{Bulk}))+z(log(\textrm{Shear}))+z(log(\textrm{Young's}))-P_\mathbf{unstable}
\end{equation}
where $z$ is standard normalization, $\log(\textrm{Bulk})$, $\log(\textrm{Shear})$, and $\log(\textrm{Young's})$ are the logarithmically transformed bulk modulus, shear modulus, and Young's modulus, respectively, and $P_\mathbf{unstable}$ is an additional penalty for structural instability.

%\subsection{Results fo Fine-tuning model}

\subsection{DFT validation of UFO-Mech}

To fully evaluate the performance of UFO-Mech, we first use the this fine-tuned model to generate a series of new crystal structures and predict their associated mechanical properties.
Next, DFT calculations are performed to independently validate both the generated crystal structures and their predicted mechanical properties. 
For each fully optimized structure, the elastic constants are calculated using the finite-difference method, in which small lattice distortions are applied and the resulting stress response is used to determine the elastic stiffness tensor, $C_{ij}$, within the linear elastic regime~\cite{de2015charting}.
The calculated elastic constants are expressed as a $6\times6$ stiffness matrix in Voigt notation. The corresponding elastic compliance tensor, $S$, is obtained by inversion of the stiffness tensor:
\begin{equation}
S_{ij} = C_{ij}^{-1}
\end{equation}

The bulk modulus and shear modulus are determined using the Voigt--Reuss--Hill (VRH) approximation~\cite{hill1952elastic}. The Voigt estimates~\cite{voigt1928lehrbuch} are calculated from the elastic stiffness constants as:
\begin{equation}
B_V =
\frac{
C_{11}+C_{22}+C_{33}
+2(C_{12}+C_{13}+C_{23})
}{9},
\end{equation}
\begin{equation}
G_V =
\frac{
C_{11}+C_{22}+C_{33}
-(C_{12}+C_{13}+C_{23})
+3(C_{44}+C_{55}+C_{66})
}{15},
\end{equation}
where $B_V$ and $G_V$ are the Voigt bulk and shear moduli, respectively. The corresponding Reuss estimates~\cite{reuss1929berechnung} are calculated from the elastic compliance constants as:

\begin{equation}
B_R =
\frac{1}{
S_{11}+S_{22}+S_{33}
+2(S_{12}+S_{13}+S_{23})
},
\end{equation}
\begin{equation}
G_R =
\frac{15}{
4(S_{11}+S_{22}+S_{33})
-4(S_{12}+S_{13}+S_{23})
+3(S_{44}+S_{55}+S_{66})
},
\end{equation}
where $B_R$ and $G_R$ are the Reuss bulk and shear moduli, respectively. The final bulk and shear moduli are obtained using the Hill approximation by taking the arithmetic average of the corresponding Voigt and Reuss estimates:
\begin{equation}
\textrm{Bulk modulus} = \frac{B_V+B_R}{2},
\end{equation}
\begin{equation}
\textrm{Shear modulus} = \frac{G_V+G_R}{2}.
\end{equation}

Finally, Young's modulus is calculated from the Hill-averaged bulk and shear moduli according to the following equation:
\begin{equation}
\textrm{Young's modulus} = \frac{9\times\textrm{Bulk modulus}\times\textrm{Shear modulus}}{3\times\textrm{Bulk modulus}+\textrm{Shear modulus}}.
\end{equation}

\begin{figure*}[t]
    \centering
    \includegraphics[width=\textwidth]{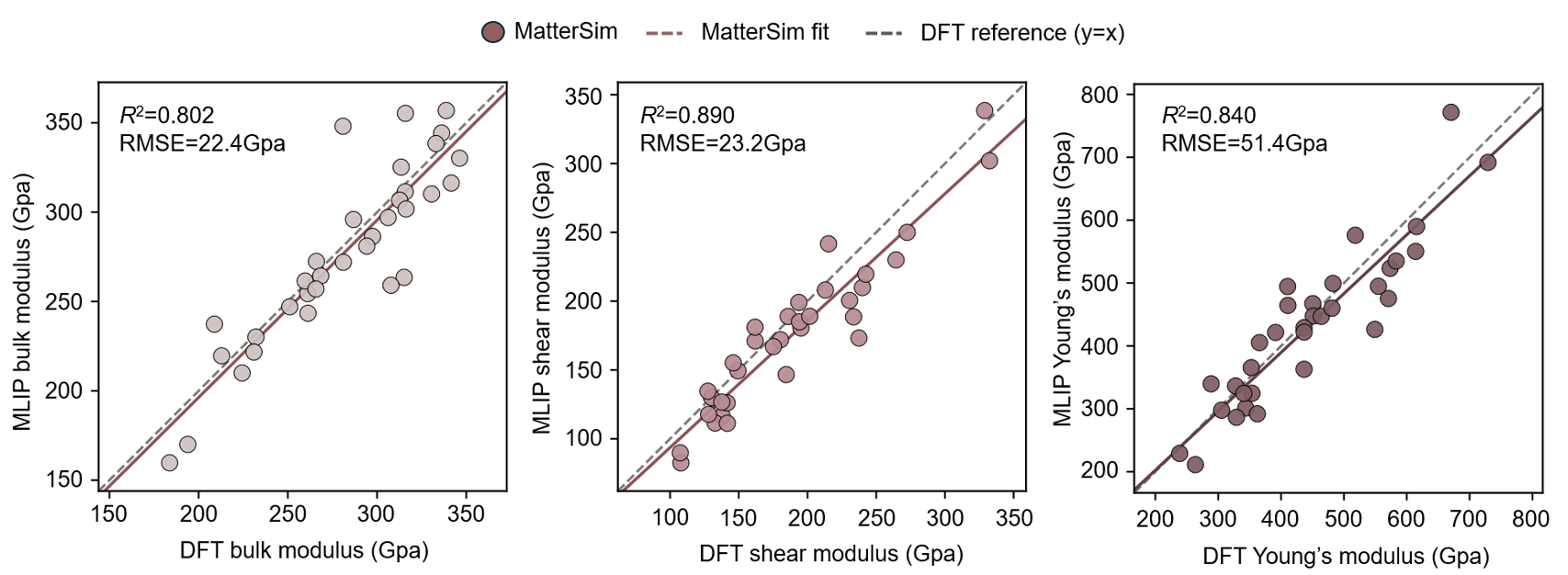}
    \caption{DFT validation of mechanical properties predicted by UFO-Mech.
    }
    \label{fig.s7.2}
\end{figure*}

The resulting DFT-derived bulk, shear, and Young's moduli are used to evaluate the mechanical-property predictions of UFO-Mech.
%
%
%was used to determine the elastic constants for all generated.
%
%The shear, bulk, and Young's moduli of each crystal are then derived from the calculated elastic constants using the \rev{XXX method}.
Figure~\ref{fig.s7.2} presents parity plots comparing the three mechanical properties predicted by UFO-Mech with those calculated by DFT for 33 generated crystal structures.
The strong linear correlations demonstrate the high accuracy of UFO-Mech in predicting these mechanical properties.
Moreover, the structural diversity of the generated candidates indicates that UFO-Mech does not rely on a single preferred structural motif to achieve property optimization.
Instead, it generates diverse stable crystal structures that exhibit exceptional mechanical properties.
%
%Tab.~\ref{tab:s8.1} presents more details and mechanical properties of these verified crystals.

\subsection{Scaffold effect on mechanical properties}
\begin{figure*}[t]
    \centering
    \includegraphics[width=\textwidth]{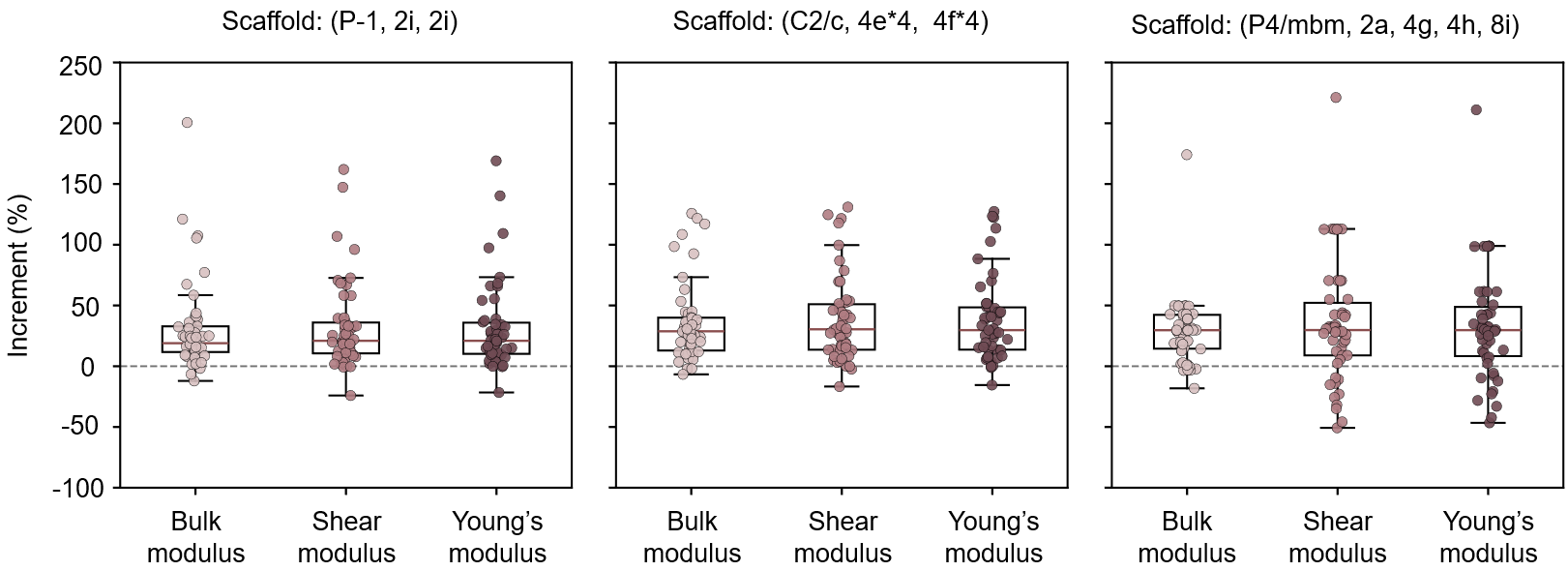}
    \caption{Additional tests of three specific scaffold features identified within the red clusters in main-text Fig. 5(b) to affect the three key mechanical properties of materials.
    %DFT validation of mechanical properties predicted by UFO-Mech.
    }
    \label{fig.s7.3}
\end{figure*}

As discussed in the main-text Section 2.7, scaffold features play a critical role in governing mechanical properties of materials.
For instance, main-text Fig. 5(c) shows that imposing a specific scaffold feature on 50 material systems can significantly enhance their mechanical properties.
To further validate this effect, we select additional three common scaffold features among the crystal structures located in the red cluster in Fig. 5(b), including $c$ = ($P-1, 2i, 2i$), ($C2/c, 4e*4,  4f*4$), and ($P4/mbm, 2a, 4g, 4h, 8i$).
Each scaffold is then imposed on 50 new crystal structures, and their mechanical properties are subsequently calculated and compared with those of the corresponding original structures.
Figure~\ref{fig.s7.3} shows that all three scaffold can significantly improve the mechanical properties of the tested crystals, with improvements of 31.08\%, 37.33\%, and 32.08\%, respectively.
These additional tests further demonstrate the critical roles of crystal scaffolds in governing the mechanical properties of materials.
Since our {\model} can be easily fine-tuned to target other materials properties, we anticipate that scaffold features may also play an important role in governing a broad range of materials properties.
%can affect a broad range of material properties.

%

% Because of {\model}'s powerful ability to capture the latent space of complex structures, UFO-Mech also retains these capabilities and can explicitly improve material properties.
% %
% In the process of guiding the model to generate excellent mechanical properties, UFO-Mech discovered a high frequency topology scaffold ($\tau$ = \texttt{sg62\_4c\_4c\_4c\_4c\_4c\_4c}). These structural candidates have excellent $K$, $G$, $E$.
% %
% The reason lies in its six independently adjustable 4c orbits in the $Pnma$ (SG62), in which 24 symmetrically related sites can form staggered in-plane, inter-layer, and diagonal connections.
% %
% This connection method not only suppresses simple inter-layer slip, but also allows strong bonds to be distributed in multiple directions, thereby improving compressive and shear strength.
% %

% Although mechanical properties are determined by topology scaffold and chemical occupancy, replacing original scaffold with \texttt{sg62\_4c\_4c\_4c\_4c\_4c\_4c} is very limited. Because their elemental composition, bonding network, and geometry are already close to a local optimum.
% %
% In contrast, \texttt{sg62\_4c\_4c\_4c\_4c\_4c\_4c} has a very significant improvement for crystals with the same chemical composition but and poor performance. This greatly improves the diversity and functionality of target material design.

\clearpage

\bibliographystyle{unsrt}
\bibliography{jobname}